\documentclass[a4paper,12pt]{article}

\pdfoutput=1

\usepackage{graphicx} 
\usepackage[a4paper,left=2.73cm,right=2.7cm,top=3cm,bottom=3.5cm]{geometry}
\usepackage[latin2]{inputenc}
\usepackage{epsfig}
\usepackage{amssymb}
\usepackage{amsfonts}
\usepackage{amsmath,euscript,array,mathrsfs}
\usepackage{epsf}
\usepackage{slashed}
\usepackage{comment}
\usepackage{colortbl}
\usepackage{cite}
\usepackage{parskip}
\usepackage{mathtools}
\usepackage{upgreek}
\usepackage{setspace}

\DeclareMathOperator{\diag}{diag}
\usepackage{caption,subcaption,wrapfig}
\usepackage[font={small}]{caption}
\usepackage[colorlinks=true,linktocpage=true,linkcolor=blue,citecolor=blue]{hyperref}

\usepackage{tikz}
\usetikzlibrary{decorations.markings,decorations.pathmorphing}
\numberwithin{equation}{section}

\newcommand{\rnum}[1]{\uppercase\expandafter{\romannumeral #1\relax}}

\allowdisplaybreaks

\begin{document}

 \begin{titlepage}

\thispagestyle{empty}

$\,$

\vspace{40pt} 
	
\begin{center}

{\LARGE \textbf{Non-linear instability of slowly rotating Kerr-AdS black holes}}

\vspace{30pt}
		
{\large \bf Pau Figueras and Lorenzo Rossi}

\vspace{25pt}

{\normalsize 
School of Mathematical Sciences, Queen Mary University of London, Mile End Road, London E1 4NS, United Kingdom}

\vspace{10pt}
\texttt{p.figueras@qmul.ac.uk}, \texttt{l.rossi@qmul.ac.uk}

\vspace{40pt}
				
\abstract{
Generic scalar perturbations on a fixed slowly rotating Kerr-AdS black hole background exhibit stable trapping, that is, the scalar field remains in a region between the exterior of the black hole and the AdS boundary for a very long time, decaying only inverse logarithmically in time. We study this effect employing fully general simulations that take into account the  backreaction of the scalar field on the geometry.
We find that the stable trapping of generic perturbations of Kerr-AdS persists at the non-linear level.
Furthermore, at late times the spacetime settles into a time-dependent and non-axisymmetric black hole which differs from Kerr-AdS. Since our perturbations are suitably small and generic, our results indicate that slowly rotating Kerr-AdS black holes are non-linearly unstable.
}

\end{center}

\end{titlepage}

\thispagestyle{empty}
\setcounter{page}{0}
\tableofcontents

\newpage

\section{Introduction}
\label{sec:conj}

The AdS/CFT correspondence \cite{Maldacena:1997re,Gubser:1998bc,Witten:1998qj} has motivated the studies of the dynamics of gravity in asymptotically anti-de Sitter (AdS) spacetimes. The presence of a timelike boundary at infinity implies that the dynamics of asymptotically AdS spacetimes has to be studied in the context of the initial boundary value problem, with the appropriate boundary conditions. 

While Minkowski space is known to be non-linearly stable under generic perturbations \cite{Christodoulou:1993uv}, Refs. \cite{DafHolz1,DafHolz:1981ff} conjectured that, with reflective boundary conditions, AdS spacetime would be non-linearly unstable to black hole formation.\footnote{On the other hand, AdS is expected to be stable with dissipative boundary conditions \cite{Holzegel:2015swa}.}
Convincing numerical evidence for this conjecture was provided in \cite{Bizon:2011gg}, while \cite{Moschidis:2017llu,Moschidis:2018ruk} rigorously proved it for the Einstein-null dust and Einstein-Vlasov theories respectively. The study of the non-linear stability/instability of AdS and its implications for holography and the thermalization of certain strongly interacting conformal field theories (CFTs) have since led to a vast body of literature.

The study of the dynamics of black holes in AdS also has a long history. Holography boosted the interest in the subject due to the applications to studies of the real time dynamics of certain strongly coupled CFTs at finite temperature and hydrodynamics \cite{Policastro:2001yc,Policastro:2002se,Kovtun:2003wp,Kovtun:2004de,Baier:2007ix,Bhattacharyya:2007vjd}.
In the Poincar\'e patch of AdS and with standard boundary conditions and no matter, black branes appear to be dynamically stable in many situations of interest.\footnote{A notable exception is the turbulent instability of black branes \cite{Carrasco:2012nf,Adams:2013vsa}.} This aspect of black branes has turned out to be important in the context of the AdS/CFT correspondence. Indeed, the linear stability properties of these black holes under perturbations, i.e., the spectrum of quasi-normal modes, capture the linear response of the dual CFT at strong coupling \cite{Kovtun:2005ev,Berti:2009wx}. At the non-linear level, the absence of caustic formation outside apparent horizons in generic physically relevant situations has been exploited, with great success, to study the far-from-equilibrium regime of such strongly interacting CFTs by solving, numerically, the fully non-linear Einstein equations in AdS on dynamical black brane backgrounds using the characteristic formulation \cite{Chesler:2013lia}. 
Note, however, that in the closely related backgrounds that exhibit confinement, localized black holes (known as plasma balls) exist \cite{Aharony:2005bm,Figueras:2014lka}, and other techniques are needed to study their non-linear dynamics \cite{Bantilan:2020pay}.

On the other hand, the full non-linear dynamics of gravity in global AdS is much less understood, arguably because it is technically more difficult to study.
One of the reasons is that the gravitational dynamics in the Poincar\'e patch of AdS is dissipative; on the other hand, global AdS with standard reflective boundary conditions is non-dissipative.\footnote{It is possible to impose fully dissipative boundary conditions in global AdS \cite{Holzegel:2015swa}, but these are, perhaps, less relevant for the AdS/CFT correspondence.} This implies that even if one starts with initial data consisting of small perturbations of a given stationary solution of the equations of motion, upon evolution non-linear effects may build up over sufficiently long times, eventually becoming important and potentially giving rise to new gravitational phenomena  that have no analogues in asymptotically flat spaces or in the Poincar\'e patch of AdS. This is precisely what happens in the non-linear instability of AdS.

The studies of the non-linear dynamics of black holes in global AdS with reflective boundary conditions are still in their infancy because of the technical difficulties of the problem. 
In the general case and without assuming any symmetries, the rigorous mathematical treatment only started rather recently \cite{Holzegel:2009ye,Holzegel:2011qk,Holzegel:2011qj,Holzegel:2011rk,Holzegel:2011uu,Holzegel:2012wt,Holzegel:2013kna}.
On the other hand, the numerical studies of such black holes were made possible by the development of techniques that employ Cauchy evolution to capture the generic gravitational dynamics and can benefit from large scale simulations \cite{Bantilan:2012vu,Bantilan:2020xas}.\footnote{The characteristic approach, which is computationally less costly because it uses spectral methods, was successfully used to study the non-linear dynamics of Kerr-AdS in \cite{Chesler:2018txn,Chesler:2021ehz}. However, the expectation is that caustics will generically form outside the (apparent) horizons of localized black holes, just as in asymptotically flat spaces, and this formulation will eventually break down.}
These techniques rely on computational tools originally developed to solve the black hole binary problem in general relativity \cite{Pretorius:2005gq}.

\subsubsection*{The black hole stability problem in AdS}
\label{sec:backgr}

We consider Einstein's gravity with a negative cosmological constant $\Lambda$ coupled to a real scalar field $\varphi$ with mass parameter $\mu$ in $D=4$ spacetime dimensions:
\begin{eqnarray}
\label{eq:EFE}
  &R_{\mu\nu}-\frac{1}{2}\,R\, g_{\mu\nu}+\Lambda\, g_{\mu\nu}=8\pi\, T_{\mu\nu}, \\
  \label{eq:KG}
  &\left(\Box_g-\mu^2\right)\varphi=0,
\end{eqnarray}
where $\Box_g:=g^{\mu\nu}\nabla_\mu \nabla_\nu$ with $\nabla$ the Levi-Civita covariant derivative associated with the spacetime metric $g$, $R_{\mu\nu}$ is the corresponding Ricci tensor, $R:=g^{\mu\nu}R_{\mu\nu}$ is the Ricci scalar, and
\begin{equation}
\label{eq:KHmomtenscov}
T_{\mu\nu}=(\nabla_\mu \varphi)(\nabla_\nu \varphi) -\tfrac{1}{2}\,g_{\mu\nu}\left[ g^{\rho\sigma} (\nabla_{\rho} \varphi)(\nabla_{\sigma} \varphi) +\mu^2\varphi^2\right]
\end{equation}
is the energy-momentum tensor of the scalar field.
It is convenient to define the AdS radius $L>0$ so that $\Lambda=-\frac{3}{L^2}$.
Without loss of generality, we set $L=1$. We assume that the scalar field mass parameter $\mu$ satisfies the 4-dimensional Breitenlohner-Freedman (BF) bound, $\mu^2\geq-\frac{9}{4}$. 
In this article we consider the coupled Einstein-scalar field system for convenience; as it will become clear from our results, the scalar field does not play a fundamental role in the late time dynamics of the system and it should be possible to obtain qualitatively similar results in the purely gravitational case.

The maximally symmetric solution of this theory in vacuum (i.e., $\varphi=0$) is pure AdS spacetime. The corresponding spacetime manifold is $\mathbb{R}^4$ and its metric is given in spherical coordinates in \eqref{eqn:ads}. This spacetime is conformally compact, i.e., it admits a conformal compactification, with a conformal boundary at null infinity given by the Einstein Static Universe $\mathbb{R}\times S^2$ with metric
\begin{equation}
\label{eq:ESU}
  g_{(0)}=-dt^2+d\Omega^2\,,
\end{equation}
where $d\Omega^2=d\theta^2+\sin^2\theta \,d\phi^2$ is the standard metric on the unit round sphere. The fact that the AdS boundary is timelike implies that pure AdS is not globally hyperbolic.

We are interested in studying conformally compact solutions of the Einstein-scalar field equations that approach pure AdS everywhere near the conformal boundary.
In particular, this implies that the conformal boundary of these spacetimes has the same topology and metric as the boundary of pure AdS and hence we still refer to it as the AdS boundary. We refer to spacetimes defined in this way as asymptotically AdS.\footnote{Strictly speaking, these spacetimes are typically referred to as asymptotically globally AdS. They are a subset of the set of asymptotically locally AdS spacetimes, defined as conformally compact spacetimes that solve the Einstein equations, possibly coupled to matter, but have arbitrary boundary topology and metric. Such spacetimes approach pure AdS only locally near the conformal boundary, and hence the terminology. In this article we will only be concerned with asymptotically globally AdS spacetimes. Thus, we can drop the word ``globally''.\label{fn:asylocalAdS}}
This definition is equivalent to the requirement that $g$ approaches $g^{\text{AdS}}$ and $\varphi$ vanishes near the boundary at some suitable rates.
These boundary conditions are commonly known as reflective, since they cause null waves to be reflected off the AdS boundary. 
The rates at which $g$ and $\varphi$ approach the reference values are determined by the near-boundary expansion of the equations of motion, and can be obtained from the Fefferman-Graham theorem \cite{AST_1985__S131__95_0,deHaro:2000vlm}.
For instance, for $\mu=0$ and employing the asymptotically Cartesian coordinates used in our numerical scheme, one has the rates in \eqref{eq:gadsfalloffs}--\eqref{eq:phiadsfalloffs}.

The  Kerr-AdS spacetime \cite{Carter:1968ks} is an example of  a vacuum ($\varphi=0$) asymptotically AdS spacetime and it is
a generalization of Kerr's stationary, axisymmetric, rotating black hole solution in asymptotically flat space \cite{Kerr:1963ud}. 
It is fully specified by the pair of parameters $(r_+,a)$, which correspond to (roughly) the horizon radius and the rotation parameter respectively (in units of the AdS radius). Given the importance that the stability of Kerr's solution has in the understanding of general relativity in the context of asymptotically flat spacetimes \cite{Press:1973zz,Chandrasekhar:1984siy,Whiting:1988vc,Dafermos:2010hb,Dafermos:2014cua,Klainerman:2019owz,Klainerman:2019uaa,Giorgi:2020oli,Dafermos:2021cbw,Klainerman:2021qzy,Dafermos:2022yzb,giorgi2022wave}, it is natural to ask whether Kerr-AdS is stable within the space of all asymptotically AdS solutions. In this article we consider this problem in the context of the Einstein-scalar field theory, but the coupling to the scalar field should not be essential for the late time dynamics of the system that we shall describe later.

It is known that Kerr-AdS black holes whose event horizon rotates faster than the speed of light, i.e., $\Omega_H>1$ (or, equivalently, $r_+^2<a$), where $\Omega_H$ is the angular velocity of the event horizon, are linearly unstable. 
On these fast-rotating Kerr-AdS black hole backgrounds, there exist single mode solutions of the linearized equations in both the scalar \cite{Cardoso:2004hs,Uchikata:2009zz,Dold:2015cqa} and gravitational sectors \cite{Cardoso:2006wa,Cardoso:2013pza,Green:2015kur} that grow exponentially in time due to superradiance, i.e., the classical amplification of waves scattering off a sufficiently fast-rotating object.\footnote{Presumably there should also be superradiantly unstable spin 1 modes on a fast-rotating Kerr-AdS background.}
Since these modes are compelled to remain in the bulk of the spacetime by the reflective AdS boundary, they undergo amplification multiple times and, at some point, the amplitude of such modes becomes large enough so that neglecting their backreaction onto the Kerr-AdS background is no longer justified.
When the backreaction is taken into account, the original Kerr-AdS black hole transitions to a dynamical black hole with scalar or gravitational hair, depending on the nature of the perturbations.\footnote{Black holes with scalar or gravitational hair and a single, helical Killing vector field have been constructed in \cite{Dias:2011at} and \cite{Dias:2015rxy} respectively. The latter have been dubbed black resonators. These spacetimes have an ergoregion, hence they are unstable \cite{Green:2015kur}.}
Simulations of the early stages of this non-linear process in the case of gravitational perturbations have been performed in \cite{Chesler:2018txn,Chesler:2021ehz} employing the characteristic scheme described in \cite{Chesler:2013lia}.
The end-point of the superradiant instability is currently unknown; Refs.~\cite{Cardoso:2006wa,Niehoff:2015oga} made some heuristic speculations and recently \cite{Kim:2023sig} constructed a new class of solutions of the Einstein equations, called gray galaxies, which they conjectured to be the endpoints of the superradiant instability. 
 We refer the reader to \cite{Brito:2015oca} for a detailed review of the superradiant instability and an extensive list of references.

On the other hand, slowly rotating Kerr-AdS black holes, i.e., those that satisfy the Hawking-Reall bound $\Omega_H<1$, are linearly stable \cite{Hawking:1999dp}. However, the question about the non-linear stability of these black holes remains open. As we will review below, there are reasons to believe that slowly rotating Kerr-AdS black holes may be non-linearly unstable.
The numerical study of the non-linear stability of slowly-rotating Kerr-AdS black holes within the Einstein-scalar field theory under suitably generic small perturbations with reflective boundary conditions is the focus of the program initiated in this work.

We note that Kerr-AdS black holes are known to suffer from other linear instabilities, such as the scalar field condensation near extremality \cite{Dias:2010ma} or the Aretakis instability at extremality \cite{Aretakis:2011ha,Aretakis:2011hc,Aretakis:2011gz,Aretakis:2012ei,Lucietti:2012sf}. These instabilities are of a very different nature compared to the non-linear instability that is the focus of this article, so we will not discuss them any further.

\subsubsection*{The non-linear instability conjecture}
\label{sec:invlogconj}

Valuable insights on the non-linear stability of slowly rotating Kerr-AdS were obtained in \cite{Holzegel:2011rk,Holzegel:2011qk,Holzegel:2011uu,Holzegel:2013kna} from the study of general solutions to the Klein-Gordon equation on a fixed Kerr-AdS background, as we summarize.

We start off by working in the geometric optics approximation, i.e, restricting the attention to highly oscillating modes, which follow null geodesics. In Kerr-AdS, generic null geodesics coming from the boundary encounter a potential barrier that they cannot overcome if their energy is below the barrier's maximum \cite{Hackmann:2010zz}. 
Therefore, these null geodesics are trapped in the region of the spacetime between the potential barrier and the AdS boundary.
Since this phenomenon occurs for a generic set of null geodesics, it is called stable trapping.\footnote{As opposed to unstable trapping, which refers to the case in which only geodesics with finely-tuned trajectories are trapped. One familiar example of unstable trapping is that of null geodesics at the photon sphere of the Schwarzschild(-AdS) black hole.} Notice that even though the potential barrier is also present in the asymptotically flat Kerr geometry, there is no stable trapping phenomenon in that setting. Heuristically and ignoring non-linearities, this can be understood because in asymptotically flat spaces null waves travel indefinitely towards null infinity and thus they are never redirected back into the bulk.

When improving on the geometric optics approximation by considering proper solutions to the Klein-Gordon equation on a fixed Kerr-AdS background, an exponentially small portion of any mode can tunnel through the barrier and fall into the black hole.
Consequently, on linearly stable backgrounds, solutions composed of a finite number of modes are expected to decay exponentially in time.
This is the case, for example, for spherically symmetric perturbations, which can only include the $l=m=0$ mode. 
Such fast decay suggests that in spherical symmetry no instability occurs, even in the full non-linear problem that includes backreaction onto the geometry.
This expectation was confirmed in \cite{Holzegel:2011rk}, which rigorously proved that Schwarzschild-AdS is non-linearly stable to spherically symmetric perturbations.

On the other hand, away from spherical symmetry, solutions to the Klein-Gordon equation can be composed of infinitely many spheroidal harmonic modes and such solutions are expected to decay much more slowly in time. 
In fact, since generic null geodesics in Kerr-AdS exhibit stable trapping, the expectation would be that solutions of the Klein-Gordon equation in such backgrounds and in absence of superradiance decay inversely logarithmically in time.\footnote{In the context of the obstacle problem for the wave equation in Minkowski space, \cite{BurqNicolas1998Ddll} showed that whenever there is stable trapping, the local energy of the solutions decays logarithmically. This result is independent of the geometry of the object causing the trapping.}
In the case of slowly rotating Kerr-AdS black holes, this was confirmed in \cite{Holzegel:2011uu,Holzegel:2013kna}, who proved uniform decay statements for solutions of the Klein-Gordon equation in these backgrounds.
More precisely, \cite{Holzegel:2011uu} showed that the slowest possible decay in time of  solutions of the Klein-Gordon equation on a Kerr-AdS background with $\Omega_H<1$ is inverse logarithmic, whereas \cite{Holzegel:2013kna} showed that the fastest possible decay of an axisymmetric scalar field perturbation on any Kerr-AdS background is also inverse logarithmic.
Combining these two results, it follows that the inverse logarithmic decay in time of solutions of the Klein-Gordon equation on a Kerr-AdS background with $\Omega_H<1$ is sharp. Since non-linear interactions are typically an obstacle to decay, such a slow decay in the linearized problem led \cite{Holzegel:2011uu} to conjecture that, once non-linearities are taken into account, slowly rotating Kerr-AdS black holes would be unstable.\footnote{On the other hand, \cite{Dias:2012tq} argued that for (slowly rotating) black holes in AdS, if the initial data contains sufficiently low power in the modes with large $l$ angular momentum quantum number, then the black holes should be non-linearly stable. In particular, smooth perturbations of black holes should be stable, whereas perturbations with low differentiability in higher dimensions may lead to instabilities, e.g., $C^2$ perturbations may be unstable in $D\geq 12$.\label{ftn:noinst}}
In this article we investigate this conjecture by performing the first numerical simulations of the non-linear evolution of a slowly rotating Kerr-AdS black hole perturbed by a massless scalar field, and we analyze the effects of the trapping mechanism in the linear and non-linear cases.

The rest of this article is organised as follows. In Section~\ref{sec:potbar} we provide an heuristic description of the potential barrier that leads to stable trapping in both Schwarzschild-AdS and Kerr-AdS. Section~\ref{sec:setup} describes our general setup to address the problem, including novel coordinates for Kerr-AdS that are suitable for Cauchy evolution, and the choice of gauge and initial conditions. We present our results in Section~\ref{sec:res}. In subsection~\ref{sec:fixbackgr} we consider the generic evolution of a massless scalar field on a fixed Kerr-AdS background, and in subsection~\ref{sec:back} we consider the general case with backreaction. In Section~\ref{sec:conclusions} we summarize and discuss our findings. Several technical details have been relegated to the appendices. In Appendix~\ref{sec:Causch} we review the Cauchy evolution scheme of \cite{Bantilan:2020xas} for general asymptotically AdS spacetimes. In Appendix~\ref{sec:CFTquants} we provide details of how to extract the boundary CFT observables in our setting. The definitions of the AdS norms that we use in the paper are given in Appendix~\ref{sec:AdSnorms}. Appendix~\ref{sec:appKerrAdS} collects some properties of Kerr-AdS that are relevant for this work, and in Appendix~\ref{sec:Kerr_effective_potential} we provide the details of the construction of the approximate effective potential in Kerr-AdS. Tests of the accuracy of our simulations can be found in Appendix~\ref{sec:conv}.

\section{Potential barrier}
\label{sec:potbar}

In this section we provide a heuristic description of the properties of the trapping potential. In particular, we plot the potential barrier and study how its height and width varies in the parameter space of sub-extremal Kerr-AdS black holes. 

The trapping potential appears in the radial part of the equation of motion of a scalar field on a Kerr-AdS background.
Let $\varphi$ be a scalar field on a fixed Kerr-AdS background with parameters $(r_+,a)$ in Boyer-Lindquist (BL) coordinates $(\tau,r,\Theta,\Phi)$, eq.~\eqref{eq:KerrAdS1}.
Consider a mode solution of the form $e^{-i\omega \tau} \tilde\varphi^{(\omega)}(r,\Theta,\Phi)$, oscillating in time with frequency $\omega$.\footnote{Strictly speaking, since we are studying a real scalar field, we should consider either the real or imaginary part of this complex mode. To simplify the notation, we work with a complex mode; one can straightforwardly adapt the discussion of this section to $\operatorname{Re}(e^{-i\omega \tau} \tilde\varphi^{(\omega)})$ or $\operatorname{Im}(e^{-i\omega \tau} \tilde\varphi^{(\omega)})$.}
For given $\omega$ and $r$, $\tilde\varphi^{(\omega)}(r,\Theta,\Phi)$ can be treated as a smooth function of $(\Theta,\,\Phi)$ on the oblate 2-spheres $S^2_{\tau,r}$ at fixed $(\tau,\,r)$, and thus it can be decomposed into spheroidal harmonics $S^{(\omega)}_{lm}(\Theta,\Phi), l\geq|m|$ (see Appendix~\ref{subsec:sphharm} for a brief review). 
Using \eqref{eq:decsphharm} and extracting the factor $\frac{1}{\sqrt{r^2+a^2}}$ from the coefficient of the spheroidal harmonics gives
\begin{equation}
\tilde \varphi^{(\omega)}(r,\Theta,\Phi)=\frac{1}{\sqrt{r^2+a^2}}\sum_{lm} u_{lm}^{(\omega)}(r) S_{lm}^{(\omega)}(\Theta,\Phi).
\end{equation}
Plugging this ansatz into the Klein-Gordon equation \eqref{eq:KG}, we obtain an ordinary differential equation in $r$ for the radial part $u^{(\omega)}_{lm}(r)$.
Writing the $r$-derivatives in terms of the ``tortoise'' coordinate $r_\ast$,\footnote{This is uniquely defined by $\frac{d r_\ast}{dr}=\frac{r^2+a^2}{\Delta_-}$ and $r_\ast\to\pi/2$ as $r\to+\infty$. Note that $r_\star\to-\infty$ at the event horizon $r=r_+$.} the radial equation reads
\begin{equation}
  \label{eq:radeq}
  \left(-\frac{d^2}{dr_\ast^2}+V_{lm}^{(\omega)}\right)u^{(\omega)}_{lm}=\omega^2 u^{(\omega)}_{lm}\,,
\end{equation}
where the effective potential $V_{lm}^{(\omega)}(r)$ is given by
\begin{equation}
  \label{eq:Vlmomega}
  V_{lm}^{(\omega)}(r)=V_{+,lm}^{(\omega)}(r)+V_{0,lm}^{(\omega)}(r)+V_{\mu}(r)
\end{equation}
with
\begin{align}
  \label{eq:Vlmpart1}
  V_{+,lm}^{(\omega)}(r)\,:=\,&-\Delta^2\frac{3r^2}{(r^2+a^2)^4}+\Delta \frac{5 r^4+3r^2\left(1+a^2\right)-4 M r+a^2}{(r^2+a^2)^2} &\nonumber\\
  \,=\,&\frac{1}{\sqrt{r^2+a^2}}\frac{d^2}{dr_\ast^2}\big(\sqrt{r^2+a^2}\big),&\\
  \label{eq:Vlmpart2}
  V_{0,lm}^{(\omega)}(r)\,:=\,&\frac{\Delta\big(\lambda_{lm}^{(\omega)}+\omega^2 a^2\big)-\Xi^2 a^2 m^2-2 m\,\omega\, a\, \Xi\,\big(\Delta-(r^2+a^2)\big)}{\left(r^2+a^2\right)^2},&\\
  \label{eq:Vlmpart3}
  V_{\mu}(r)\,:=\,&\mu^2\frac{\Delta}{\left(r^2+a^2\right)^2}\left(r^2+\chi(\mu^2)a^2\right),& 
\end{align}
and $\chi(\mu^2)=1$ for $-\frac{9}{4}\leq\mu^2<0$ and $\chi(\mu^2)=0$ for $\mu^2\geq 0$.
Here, $\lambda_{lm}^{(\omega)}$, called separation constant, is the eigenvalue of the operator \eqref{eq:sphharmop} associated with the spheroidal harmonic $S^{(\omega)}_{lm}(\Theta,\Phi)$.
Imposing reflective boundary conditions at infinity, i.e., $u^{(\omega)}_{lm}=0$ for $r\to+\infty$, and ingoing-wave boundary conditions at the horizon, i.e., $u^{(\omega)}_{lm}\sim e^{-i\omega r_\ast}$ for $r_\ast\to-\infty$, one obtains an eigenvalue problem for a Schr\"{o}dinger-like operator. 
For any Kerr-AdS parameters $(r_+,a)$ and spheroidal harmonic indices $(l,m)$, the solutions to this problem are the quasi-normal modes $u^{(\omega)}_{lm}$ with the associated (complex) quasi-normal frequencies $\omega$.
Each solution has a different number of radial nodes $n$, called overtones. 
Kerr-AdS black holes that satisfy the Hawking-Reall bound $\Omega_H<1$ only admit decaying ($\operatorname{Im}\omega<0$) modes \cite{Holzegel:2009ye,Holzegel:2012wt}, with the lowest overtone mode being the least damped one.
On the other hand, for any Kerr-AdS black hole with $\Omega_H>1$, there exist exponentially growing linear modes ($\operatorname{Im}\omega>0$) \cite{Dold:2015cqa}. These modes are responsible for the superradiant instability mentioned in the Introduction.

Our goal in the rest of this section is to develop an intuition for the properties of the potential barrier in the family of sub-extremal Kerr-AdS black holes, which can guide the choice of parameters in the numerical studies of the trapping instability.
For simplicity we consider a massless scalar field $\mu=0$;
for other non-zero values of $\mu^2$, the features of the potential barrier are qualitatively the same as those discussed in Sections \ref{subsec:potbarschads} and \ref{subsec:potbarkerrads}.
The mass term dominates the potential at large $r$, away from the barrier, where $V_{lm}^{(\omega)}\sim +r^2$ if $\mu^2>-2$, $V_{lm}^{(\omega)}\sim \lambda_{lm}^{(\omega)}$ if $\mu^2=-2$, and $V_{lm}^{(\omega)}\sim -r^2$ if $-\frac{9}{4}<\mu^2<-2$.

\subsection{Schwarzschild-AdS}
\label{subsec:potbarschads}

In the case of Schwarzschild-AdS, $a=0$, the eigenvalue problem simplifies considerably.
First, the potential does not depend on $\omega$ (hence we say that the eigenvalue problem is linear) nor on the index $m$. Second, the spheroidal harmonics $S_{lm}^{(\omega)}(\Theta,\Phi)$ reduce to the usual spherical harmonics $Y_{lm}(\Theta,\Phi)$, and the corresponding eigenvalues $\lambda_{lm}^{(\omega)}$ reduce to $l(l+1)$.
Therefore, \eqref{eq:radeq} and \eqref{eq:Vlmomega} take the simpler form
\begin{equation}
  \label{eq:radeqschads}
  \left(-\frac{d^2}{dr_\ast^2}+V_{l}\right)u_{lm}=\omega^2 u_{lm}\,,
\end{equation}
with
\begin{equation}
  \label{eq:Vl}
  V_{l}(r)=\left(1-\frac{2M}{r}+r^2\right)\left(\frac{l(l+1)}{r^2}+\frac{2 M}{r^3}+2\right).
\end{equation}
The eigenvalue problem for Schwarzschild-AdS was investigated using analytical and numerical techniques in several articles \cite{Horowitz:1999jd,Konoplya:2002zu,Berti:2003ud,Cardoso:2003cj,Cardoso:2004up,Festuccia:2008zx,Dias:2012tq,Berti:2009wx,Gannot:2012pb}.

\begin{figure*}[t!]
    \centering
    \includegraphics[scale=0.6]{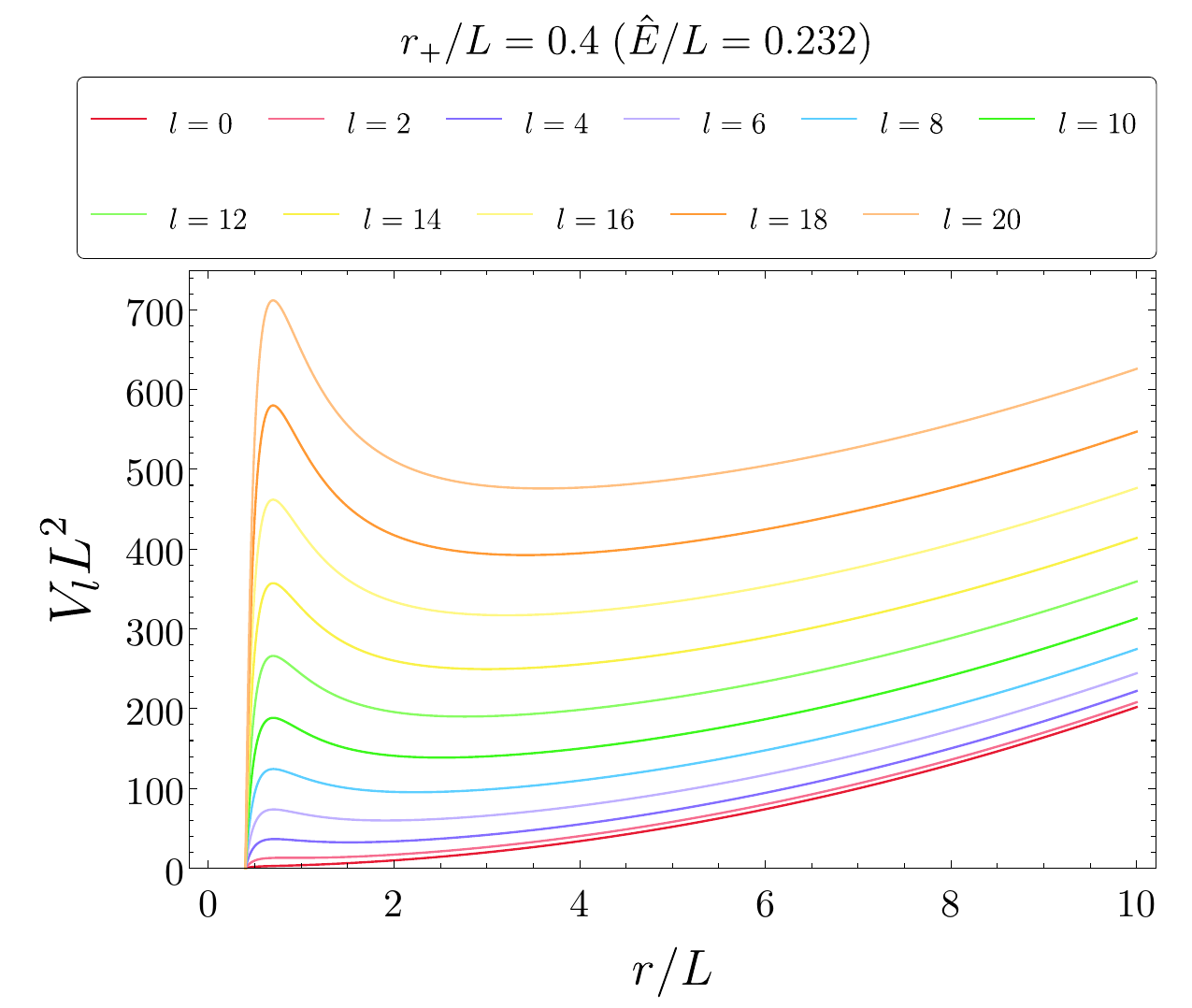}
\parbox{5.0in}{\vspace{0.1cm}\caption{Effective potential of Schwarzschild-AdS with $r_+/L=0.4$ for several values of $l$. Units of the AdS radius $L$ are restored.
    }\label{fig:barrierschads}}
\end{figure*}

In Fig.~\ref{fig:barrierschads} we plot $V_l(r)$ for different values of $l$ at fixed horizon radius $r_+=0.4$.
We notice the existence of a potential barrier for $l\geq 2$ modes. It is this barrier that, together with the reflective boundary conditions at the AdS boundary, can stably trap single mode solutions in the exterior of the black hole, only allowing for leaking into the black hole through a classical tunnelling effect. 
Comparing the height $\Delta V_{l}:=V_{l,\text{max}}-V_{l,\text{min}}$, and the width, $\Delta r_{l}:=|r_{l,\text{max}}-r_{l,\text{min}}|$, of the potential barrier as functions of $l$ for fixed $r_+$, one sees that both quantities increase with $l$.
Therefore, for a given Schwarzschild-AdS black hole, high $l$ modes are more trapped than the low $l$ ones.
Indeed, in the $l\to+\infty$ limit, the height and width of the barrier diverge, while the location of the maximum tends to $r_{+\infty,\text{max}}=\frac{3}{2}r_+(1+r_+^2)$.
Furthermore, for any value of $r_+$, one can look for the lowest value of $l$ for which $V_l(r)$ exhibits a barrier outside of the horizon. It turns out that this ``minimum trapped $l$'' increases with $r_+$ more slowly than an exponential, which implies that stable trapping of a sufficiently high $l$ mode can occur for Schwarzschild-AdS black holes of arbitrary size. 

\subsection{Kerr-AdS}
\label{subsec:potbarkerrads}

We now turn to the rotating case, $0 < a < 1$, for which the eigenvalue problem is more complicated.
In this case, the potential depends on both the eigenvalue $\omega$ (so the eigenvalue problem is non-linear) and the separation constant $\lambda^{(\omega)}_{lm}$. 
As a consequence, the potential $V_{lm}^{(\omega)}$ is not known analytically.
Nevertheless, a mix of analytic and numerical methods have been successfully used to obtain the solutions to this problem; for instance,
analytic expressions of quasi-normal frequencies in the regime $a\ll r_+\ll 1$ were found in \cite{Cardoso:2004hs}, whilst \cite{Uchikata:2009zz} determined them numerically in the range $r_+\lesssim 1$ for some of the modes.

\begin{figure*}[t!]
    \centering
    \includegraphics[width=6in,clip=true]{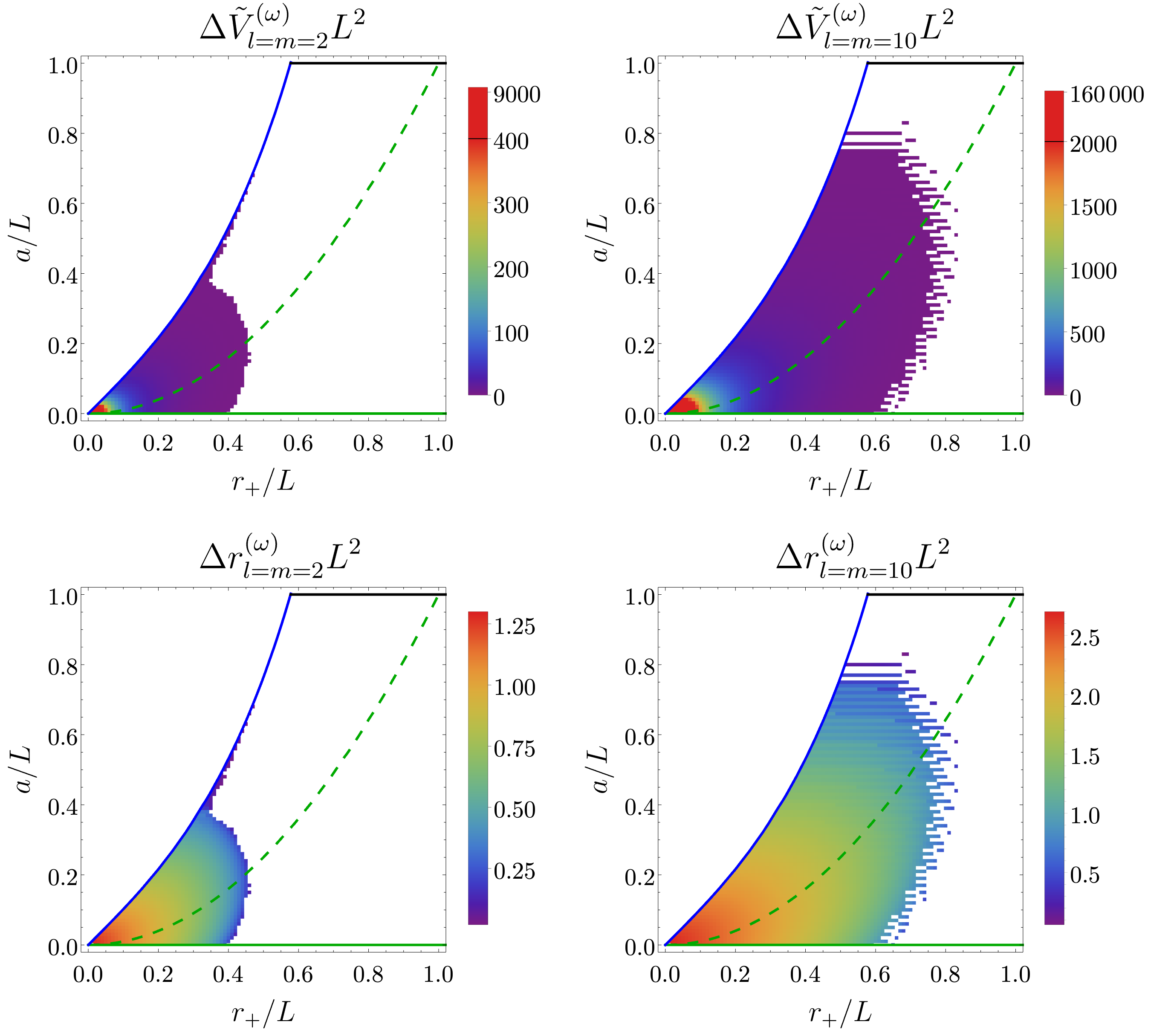}
\parbox{5.0in}{
\vspace{0.15cm}\caption{Height (top) and width (bottom) of the approximate potential barrier $\tilde V_{l=m}^{(\omega)}$ over the sub-extremal Kerr-AdS parameter space for $l=2$ (left) and $l=10$ (right). The black line is the boundary of the region containing regular Kerr-AdS black holes, the blue line represents extremal Kerr-AdS black holes, and the continuous green line represents Schwarzschild-AdS black holes. The dashed green line represents Kerr-AdS black holes with $\Omega_H=1$; superradiantly unstable black holes are above this line. In the top panels, the color function is constant for values larger than the threshold displayed in the bar legends (400 for $l=2$ and 2000 for $l=10$). Units of the AdS radius $L$ are restored.}
 \label{fig:heighwidthkads} 
}
\end{figure*}

Our goal in this subsection is to provide a reasonably accurate qualitative picture of the potential barrier for the entire range of sub-extremal Kerr-AdS black holes. One can achieve this by considering an approximate effective potential, obtained through the argument detailed in Appendix~\ref{sec:Kerr_effective_potential}. This is given  by
\begin{equation}
\label{eq:lowboVlm}
  \tilde{V}_{lm}^{(\omega)}:=V_{+,lm}^{(\omega)}+\tilde V_{0,lm}^{(\omega)}
\end{equation}
with $V_{+,lm}^{(\omega)}$ as in \eqref{eq:Vlmpart1}, 
\begin{equation}
\label{eq:lowboV0lm}
  \tilde V_{0,lm}^{(\omega)}:=\frac{\Delta \Xi^2 l(l+1)-\Xi^2 a^2 m^2-2 m\omega a \Xi\left(\Delta-\left(r^2+a^2\right)\right)}{\left(r^2+a^2\right)^2}\,,
\end{equation} 
and $\omega\in \mathbb{R}$. Our approximation is convenient because, unlike the actual potential $V_{lm}^{(\omega)}$, $\tilde{V}_{lm}^{(\omega)}$ is known analytically and hence it is straightforward to plot. 
In the rest of this discussion we focus on the $l=m$ case, since this is the most interesting one for our study.
In fact, among all modes with given $l$, the $l = m$ ones are those that ``feel'' the highest and widest potential barrier, as shown by \cite{Dias:2012tq} in the large $l$ limit. Therefore, these are the most trapped modes, that is, the most long-lived modes in the exterior of the horizon and hence the most likely to cause the non-linear instability. 
We note that the $l=m$ case is the one for which our approximation of the potential is expected to be most accurate; see Appendix~\ref{sec:Kerr_effective_potential}.

To depict $\tilde{V}_{l=m}^{(\omega)}$ in the entire range of the sub-extremal Kerr-AdS parameters $(r_+,a)$, we fix $l$ and search for a value of $\omega\in\mathbb{R}$ such that 1) $\tilde{V}_{l=m}^{(\omega)}$ has a local maximum $\tilde{V}_{l=m,\text{max}}^{(\omega)}$ at $r^{(\omega)}_{l=m,\text{max}}$ and a local minimum $\tilde{V}_{l=m,\text{min}}^{(\omega)}$ at $r^{(\omega)}_{l=m,\text{min}}$ with $r_+<r^{(\omega)}_{l=m,\text{max}}<r^{(\omega)}_{l=m,\text{min}}$, and 2) $\tilde{V}_{l=m,\text{min}}^{(\omega)}\leq \omega^2<\tilde{V}_{l=m,\text{max}}^{(\omega)}$. We carry out the search by specifying a grid in $(r_+,a,\omega)$; then, for each $(r_+,a)$-grid point, we start from $\omega=0$ and increase $\omega$ until a value is found that satisfies criteria 1) and 2). This value is then used to compute $\tilde V_{l=m}^{(\omega)}$, and thus the height $\Delta\tilde V_{l=m}^{(\omega)}$ and width $\Delta r_{l=m}^{(\omega)}$.\footnote{We use $\Delta r_+=\Delta a=\Delta \omega=0.01$ for the $l=2$ case, and $\Delta r_+=\Delta a=0.01, \Delta \omega=0.2$ for the $l=10$ case.} 

The result is shown in Fig.~\ref{fig:heighwidthkads}. This figure depicts the height (top panels) and width (bottom panels) of the approximate potential $\tilde{V}_{l=m}^{(\omega)}$ for the representative values of $l=2,10$. 
First, we note that the boundaries in these plots are an artefact of our discretization in $(r_+,a,\omega)$. Secondly, comparing the left and right panels, we notice that for given $(r_+,a)$, larger $l$ modes ``feel'' a higher and wider potential barrier, therefore they are more trapped. 
Finally, we see that the region of the parameter space in which stable trapping can occur in the Kerr-AdS family of black holes is larger for larger $l$. 
Armed with this qualitative picture of the potential, we can proceed to study the non-linear evolution of generic scalar perturbations of a slowly rotating Kerr-AdS black hole.

\section{Setup}
\label{sec:setup}

In this section we provide the details of our numerical methods. 
We essentially follow  \cite{Bantilan:2020xas}, which evolves a minimally coupled massless scalar field in global AdS by solving the Cauchy problem without symmetry assumptions. 
In the present study we generalize this scheme by replacing the background AdS spacetime with Kerr-AdS.
This only requires minimal adaptations of the techniques of \cite{Bantilan:2020xas}, which we explain in this section. For completeness, we review the scheme of  \cite{Bantilan:2020xas} in Appendix~\ref{sec:Causch}; more details can be found in this reference and also in \cite{Rossi:2022nao}.

\subsection{Kerr-AdS in suitable coordinates}
\label{sec:horpenbackgr}

To evolve a perturbed Kerr-AdS spacetime with our numerical scheme, we need to write the Kerr-AdS metric in coordinates that are asymptotically Cartesian. 
This is easy to do once we have coordinates adapted to the asymptotic geometry in the sense explained in Appendix~\ref{sec:appKerrAdS}. Using such coordinates, we then simply need to define a compactified radial coordinate and  apply the usual transformation that maps spherical coordinates to Cartesian coordinates.
In this section, we construct suitable asymptotically Cartesian, horizon-penetrating coordinates, which we use in our numerical scheme.\footnote{The coordinates introduced by Hawking, Hunter and Taylor-Robinson (HHT) in \cite{Hawking:1998kw} (see Appendix~\ref{sec:appKerrAdS}) are adapted to the asymptotic geometry; however, they are not horizon-penetrating and thus not suitable for numerical implementation.}

We start with the Kerr-Schild coordinates $x^{\alpha'}=(t,r,\Theta,\phi)\in(-\infty,+\infty)\times(0,+\infty)\times(0,\pi)\times(0,2\pi)$ proposed in \cite{Gibbons:2004uw}.
They are related to the usual BL coordinates $(\tau,r,\Theta,\Phi)$, by 
\begin{equation}
\label{eq:BLtoKS}
dt=d\tau+\frac{2M}{(1+r^2)(V-2M)}dr, \quad d\phi=d\Phi-ad\tau+\frac{2aM}{(r^2+a^2)(V-2M)}dr,
\end{equation}
where $
V=\frac{1}{r}(1+r^2)(r^2+a^2)$.
The Kerr-AdS metric in these coordinates reads
\begin{equation}
\label{eq:KSKAdS}
\hat g=g^{\text{AdS}}+\frac{2M}{U}\big(\lambda_{\alpha'} dx^{\alpha'} \big)^2,
\end{equation}
where $U=\frac{1}{r}(r^2+a^2\cos^2\Theta)$, $g^{\text{AdS}}$ is the pure AdS metric, 
\begin{equation}
g^{\text{AdS}}=-\frac{(1+r^2)\Delta_\Theta}{\Xi}dt^2+\frac{\Sigma^2}{(1+r^2)(r^2+a^2)}dr^2+\frac{\Sigma^2}{\Delta_\Theta}d\Theta^2+\frac{(r^2+a^2)}{\Xi}\sin^2\Theta d\phi^2,
\end{equation}
and $\lambda=\lambda_{\alpha'} dx^{\alpha'}$ is a null covector given by
\begin{equation}
\lambda_{\alpha'} dx^{\alpha'}=\frac{\Delta_\Theta}{\Xi}dt+\frac{\Sigma^2}{(1+r^2)(r^2+a^2)}dr-\frac{a}{\Xi}\sin^2\Theta d\phi.
\end{equation}
The functions $\Delta_\Theta$, $\Sigma$, $\Xi$ can be found in \eqref{eq:fnrule}.
It is straightforward to verify that Kerr-Schild coordinates are horizon-penetrating but not adapted to the asymptotic geometry.
We can transform them into coordinates that satisfy both requirements by defining new $(R,\theta)$ coordinates, related to $(r,\Theta)$ by the same transformation that relates the $(r,\Theta)$ BL coordinates to the $(\mathcal{R},\chi)$ HHT coordinates, 
\begin{equation}
\begin{aligned}
R^2\cos^2\theta&=r^2\cos^2\Theta,\\
\Xi\, R^2\sin^2\theta&=(r^2+a^2)\sin^2\Theta.
\end{aligned}
\end{equation}
We refer to these new coordinates as Kerr-Schild-HHT, and we denote them by $x^{\hat{\alpha}}=(t,R,\theta,\phi)\in(-\infty,+\infty)\times(0,+\infty)\times(0,\pi)\times(0,2\pi)$.
The explicit expressions for the transformation $\left(R(r,\Theta), \theta(r,\Theta)\right)$, as well as its inverse $\left(r(R,\theta), \Theta(R,\theta)\right)$, can be obtained from \eqref{eq:nonrotnonhorpenexpl} and \eqref{eq:nonrotnonhorpenexplinv} respectively.
The location of the event horizon in these coordinates, $R_+(\theta)$, can be obtained from \eqref{eq:Rpluschi}.
We do not explicitly write the Kerr-AdS metric in the $(t,R,\theta,\phi)$ coordinates because it is too cumbersome and not particularly informative.
We just note that the components of the tensor $h\equiv\hat g-g^{\text{AdS}}$, describing the deviation of Kerr-AdS from pure AdS, have the following fall-offs near the AdS boundary, $R\to\infty$:
\begin{align}
\label{eq:sphKSfo}
&h_{tt}=O(R^{-1}), &h_{tR}=O(R^{-3}), &\qquad h_{t\theta}=O(R^{-2}), &h_{t\phi}=O(R^{-1}), \nonumber\\
&h_{RR}=O(R^{-5}), &h_{R\theta}=O(R^{-4}), &\qquad h_{R\phi}=O(R^{-3}),\nonumber\\
&h_{\theta\theta}=O(R^{-3}), &h_{\theta\phi}=O(R^{-2}), \nonumber\\
&h_{\phi\phi}=O(R^{-1}) 
\end{align}

It is now easy to obtain the desired horizon-penetrating, asymptotically Cartesian coordinates $x^\mu=(t,x,y,z)$ for Kerr-AdS.
First, define the compactified radial coordinate 
\begin{equation}
  \label{eq:comprad}
  \rho=\frac{-1+\sqrt{1+R^2}}{R},\quad \rho\in(0,1)\,,
\end{equation}
thus obtaining asymptotically spherical coordinates, $x^\alpha=(t,\rho,\theta,\phi)$.
The asymptotics of the deviation tensor $h$ near the AdS boundary $\rho\to 1$ in these coordinates are
\begin{equation}
\label{eq:sphbounconh}
h_{\alpha\beta}=O((1-\rho)).
\end{equation}
Finally, suitable compact asymptotically Cartesian coordinates $(t,x,y,z)$ are given by
\begin{equation}\label{eqn:txyz}
\begin{aligned}
x&=\rho \sin\theta \cos\phi\,, \\
y&=\rho \sin\theta \sin\phi\,,\\
z&=\rho \cos\theta\,,
\end{aligned}
\end{equation}
with $x,y,z\in(-1,1)$ and $\sqrt{x^2+y^2+z^2}<1$. 
It is straightforward to verify that the Kerr-AdS metric components $\hat{g}_{\mu\nu}$ and the corresponding source functions $\hat H_{\mu}$ asymptote to their pure AdS values near the boundary $\rho\to 1$ with the fall-offs \eqref{eq:gadsfalloffs} and \eqref{eq:Hadsfalloffs} respectively.  
Therefore, it follows that the numerical scheme of \cite{Bantilan:2020xas} can be straightforwardly used to evolve non-linear deformations of Kerr-AdS spacetimes.

For later use in Section~\ref{sec:indat}, we also define an uncompactified version of the asymptotically Cartesian coordinates, which we denote by $x^M=(t,X,Y,Z)\in\mathbb{R}^4$. They are obtained from the Kerr-Schild-HHT coordinates by
\begin{align}\label{eqn:txyzunc}
X&=R \sin\theta \cos\phi\,, \nonumber \\
Y&=R \sin\theta \sin\phi\,,\\
Z&=R \cos\theta\,. \nonumber
\end{align}

\subsection{Evolving deviations from Kerr-AdS}
\label{sec:KAdSasbackgr}

The scheme employed in \cite{Bantilan:2020xas} solves the initial boundary value problem in AdS in terms of quantities, $\bar g_{\mu\nu},\bar\varphi, \bar H_{\mu}$, which are in essence the deviations of the full spacetime quantities $g_{\mu\nu},\varphi$ and $H_\mu$ from their pure AdS values (see Appendix~\ref{sec:Causch} for the precise definition).
When applying the scheme to evolve deformations of Kerr-AdS, it is more efficient to directly evolve the deviations from the Kerr-AdS background, as this significantly reduces numerical errors.\footnote{A similar approach was used in \cite{East:2013iwa} to evolve extreme mass ratio binaries in general relativity.} Therefore, in this work we solve the initial-boundary value problem in AdS in terms of the following evolution variables:
\begin{align}
\tilde{g}_{\mu\nu}&=g_{\mu\nu}-\hat g_{\mu\nu}\,,\label{eqn:evogkads}\\
\tilde\varphi&=\frac{\varphi}{(1-\rho^2)^2}\label{eqn:evophikads}\,,\\
\tilde H_\mu&=\frac{H_\mu - \hat H_\mu}{1-\rho^2}\label{eqn:evoHkads}\,,
\end{align}
where $\hat g_{\mu\nu}$ and $\hat H_\mu$ are the pure Kerr-AdS quantities defined in the previous section.
We have verified that this background subtraction technique significantly improves the numerical stability and accuracy of the simulations. 
The factors of $(1-\rho)$ in \eqref{eqn:evogkads}--\eqref{eqn:evoHkads} have been chosen so that the evolution variables vanish as $(1-\rho)$ near the AdS boundary. Hence, the reflective boundary conditions \eqref{eq:gadsfalloffs}--\eqref{eq:phiadsfalloffs} take the simple Dirichlet form
\begin{equation}
\label{eq:dirrefl}
  \tilde{g}_{\mu\nu}|_{\rho=1}=0\,,\quad \tilde{\varphi}|_{\rho=1}=0\,,
\end{equation}
which can be readily implemented with our finite difference scheme as explained in \cite{Bantilan:2020xas}.

\subsection{Stable gauge choice}
\label{sec:gaukads}

We can now obtain a stable gauge for evolving deviations from a Kerr-AdS background by straightforwardly applying the prescription of Appendix~\ref{sec:Causch} to the evolution variables.
In terms of the coefficients of the near-boundary expansions of the evolution variables and the sources, 
\begin{align}
\label{eqn:qexpgtild}
\tilde g_{\mu \nu} &= \tilde g_{(1) \mu \nu} (1-\rho) + \tilde g_{(2) \mu \nu} (1-\rho)^2 + O\left((1-\rho)^3\right), \\
\label{eqn:qexpphitild}
\tilde \varphi &= \tilde \varphi_{(1)} (1-\rho) + \tilde \varphi_{(2)} (1-\rho)^2 + O\left((1-\rho)^3\right),\\
\label{eqn:qexpHtild}
\tilde H_{\mu} &=\tilde H_{(1) \mu} (1-\rho) + \tilde H_{(2) \mu} (1-\rho)^2 + O\left((1-\rho)^3\right),
\end{align}
the near-boundary stable gauge is given by expressions analogous to those in \eqref{eqn:target_gauge_txyz}:
\begin{equation}
\begin{aligned}
\tilde H_{(1)t}&=\frac{3}{2\sqrt{x^2+y^2+z^2}}(x \, \tilde g_{(1)tx}+y\, \tilde g_{(1)ty}+z\, \tilde g_{(1)tz}),\\
\tilde H_{(1)x}&=\frac{3}{2\sqrt{x^2+y^2+z^2}}(x \, \tilde g_{(1)xx}+y\, \tilde g_{(1)xy}+z\, \tilde g_{(1)xz}), \\
\tilde H_{(1)y}&=\frac{3}{2\sqrt{x^2+y^2+z^2}}(x \, \tilde g_{(1)xy}+y\, \tilde g_{(1)yy}+z\, \tilde g_{(1)yz}), \\
\tilde H_{(1)z}&=\frac{3}{2\sqrt{x^2+y^2+z^2}}(x \, \tilde g_{(1)xz}+y\, \tilde g_{(1)yz}+z\, \tilde g_{(1)zz}).
\end{aligned}
\label{eqn:target_gauge_txyztild}
\end{equation}
This is our choice of source functions near the boundary. Deep in the bulk, the evolved source functions $\tilde H_{\mu}$ smoothly transition to zero. 
This smoothly varying spatial configuration is reached after few time steps through a smooth transition in time, starting from the initial configuration $\tilde H_\mu|_{t=0}$ specified in Section~\ref{sec:indat}.
The details of the implementation of the spatial and time transitions can be found in Appendix E of \cite{Bantilan:2020xas} (see also \cite{Bantilan:2012vu}), after replacing the barred variables in this reference with the tilded variables.
Using the same notation and conventions as in \cite{Bantilan:2020xas}, in the simulations of Section~\ref{sec:back} the parameters in the transition function are set to $\rho_{1a}=0.05,\rho_{1b}=0.95,\rho_{0a}=0.0,\rho_{0b}=0.95,\xi_1=0.1,\xi_2=0.0$.

\subsection{Initial data}
\label{sec:indat}

In this section we detail our choice of initial data for the simulations in Section~\ref{sec:res}. In our evolution scheme we have to provide initial data for the deviations from the Kerr-AdS background, $\tilde{\varphi}|_{t=0}$, $\tilde{g}_{ij}|_{t=0}$, $\partial_t\tilde{\varphi}|_{t=0} $ and $\partial_t\tilde{g}_{ij}|_{t=0}$, as well as the source functions $\tilde{H}_\mu$ at all times $t\geq 0$.
Here $i,j=1,2,3$ are indices associated with the spatial coordinates $x^i=(x,y,z)$.
The choice of source functions for $t>0$ is discussed in Section~\ref{sec:gaukads}; 
 at $t=0$ we specify $\tilde{g}_{t\mu}|_{t=0}$, and $\partial_t\tilde{g}_{t\mu}|_{t=0}$, from which we can compute $\tilde{H}_\mu|_{t=0}$ using equation \eqref{eq:GHcoords}.

As initial data for the metric, we use the pure Kerr-AdS values in asymptotically Cartesian coordinates, which implies $\tilde{g}_{\mu\nu}|_{t=0}=0$, $\partial_t\tilde{g}_{\mu\nu}|_{t=0}=0$. This choice corresponds to setting the source functions at $t=0$ to their pure Kerr-AdS values, i.e., $\tilde{H}_\mu|_{t=0}=0$.
As initial data for the scalar field, we use a Lorentz-boosted Gaussian profile, constructed as follows.
Since at $t=0$ the geometry is that of pure Kerr-AdS, we can work in the coordinates for Kerr-AdS defined in Section~\ref{sec:horpenbackgr}.
In the asymptotically Cartesian $x^\mu=(t,x,y,z)$, we consider a Gaussian profile of amplitude $A$, width $\delta$, centred at $(x_0,y_0,z_0)$ and distorted along each Cartesian direction with eccentricities $(e_x,e_y,e_z)$:
\begin{equation}
\label{eq:scaGaupro}
\begin{split}
f(x,y,z)&=A \exp\left[-\left(\frac{d(x,y,z)}{\delta}\right)^2\right],\\
d(x,y,z)&:=\sqrt{ (x-x_0)^2(1-e_x^2)+ (y-y_0)^2(1-e_y^2)+ (z-z_0)^2(1-e_z^2)}.
\end{split}
\end{equation}
To obtain the boosted version of $f$ we proceed as follows. In the uncompactified coordinates $x^M=(t,X,Y,Z)$, a Lorentz boost can be straightforwardly implemented by a transformation, $(t,X,Y,Z)\to \Lambda_{\vec{v}}(t,X,Y,Z)$, where $\Lambda_{\vec{v}}(t,X,Y,Z)$ is a standard Lorentz transformation with 3-velocity $\vec v=(v_X,v_Y,v_Z)$. Therefore, to implement the Lorentz boost in the compactified coordinates $x^\mu$, we simply transform first to the uncompactified coordinates $x^M$, implement the Lorentz boost, and tranform back to the compactified coordinates. Schematically, $(t,x,y,z)\to \left(T_{X\to x}\, \circ \, \Lambda_{\vec{v}}\,\circ\, T_{x\to X}\right)(t,x,y,z)$. Denoting the operator that implements these steps by $\Lambda^{\text{compact}}_{\vec{v}}:=T_{X\to x}\, \circ \, \Lambda_{\vec{v}}\,\circ\, T_{x\to X}$, the inverse boost is implemented by $\left(\Lambda^{\text{compact}}_{\vec{v}}\right)^{-1}=\Lambda^{\text{compact}}_{-\vec{v}}$ since $\left(\Lambda_{\vec{v}}\right)^{-1}=\Lambda_{-\vec{v}}$. In this way, the action of a Lorentz boost on the Gaussian profile in \eqref{eq:scaGaupro} is:
\begin{equation}
  f(x)\to \tilde f(x)=f\big((\Lambda^{\text{compact}}_{\vec{v}})^{-1}(x)\big).
\end{equation}
Notice that $\tilde f(x)$ depends on $t$ (unless $\vec{v}=0$), even though $f(x)$ does not.
Then, the boosted initial data for the scalar field is given by
\begin{equation}
  \tilde\varphi\big|_{t=0}=\tilde f\big|_{t=0},\quad\quad
  \partial_t\tilde\varphi\big|_{t=0}=\partial_t \tilde f\big|_{t=0}.
\end{equation}

It should be emphasized that this choice of initial data does not solve the Hamiltonian and momentum constraints of the coupled Einstein-scalar field theory. 
However, if the constraint violations at $t=0$ are sufficiently small, the constraint-damping terms in the evolution equations can ensure that a solution of the equations of motion is recovered on a timescale that is faster than any physical timescale in the system (see the discussion below eq.~\eqref{eq:GHconstr}).
In our work, the initial constraint violations can be made  small enough by choosing a sufficiently small amplitude $A$ and boost velocity $|\vec v|$ of the initial scalar field profile.
This is precisely the regime that we are interested in to study the non-linear instability of slowly rotating Kerr-AdS black holes because for small $A$ and $|\vec v|$, the initial data can be understood as a suitably small perturbation of the stationary Kerr-AdS background solution.
In practice, we verify that the constraint violations are kept under control at all times in the non-linear evolutions by checking that the conserved charges (see Appendix~\ref{sec:CFTquants}) are indeed conserved, to a very good accuracy, in our simulations, as expected for a solution of the equations of motion \cite{Fischetti:2012rd}; see Fig.~\ref{fig:cons_charges} in Appendix~\ref{sec:conv}. 
Furthermore, in the non-linear case, i.e., with backreaction from the scalar field, we verify that the initial data is ``small'' by checking that the conserved charges of the initial spacetime are a small deviation from those of the background stationary black hole and they depend quadratically on the small parameters characterizing the initial scalar field configuration, namely  $A$ and $|\vec v|$.\footnote{Note that in our initial data we only modify the scalar field $\tilde\varphi$ from its background value. Since the latter enters quadratically in the equations of motion through the stress tensor, for small ``data'' one should expect that the deviations of the spacetime geometry from its stationary background values scale quadratically with the amplitude of the scalar field. We have verified that this is indeed the case for our choice of initial data.}
We leave the construction of constraint-preserving initial data with arbitrary amplitude $A$ and initial velocity $|\vec v|$ of the scalar field for future work.

\section{Results}
\label{sec:res}

We used the numerical scheme reviewed in Appendix~\ref{sec:Causch}, with the enhancements described in Section~\ref{sec:setup}, to simulate the trapping mechanism in a slowly-rotating Kerr-AdS spacetime. In subsection~\ref{sec:fixbackgr} we consider the evolution of the scalar field $\varphi$ on a fixed Kerr-AdS background. This warm-up exercise is useful to ensure that we identify a class of initial conditions for the scalar field that exhibits logarithmic decay in time \cite{Holzegel:2011uu,Holzegel:2013kna}. Then, in subsection~\ref{sec:back}, we consider the same initial data for the scalar field but we take into account the backreaction of the scalar field $\varphi$ on the metric $g$ and solve the full non-linear Einstein-scalar field system of PDEs, eqs.~\eqref{eq:EFE}--\eqref{eq:KG}. For both the simulations on a fixed background and those including full backreaction on the geometry, we consider a Kerr-AdS black hole with parameters $M=0.4$ and $a=0.2$ or, equivalently, $r_+\approx 0.544$ and $\Omega_H\approx0.772$, which correspond to an energy $\hat E\approx0.434$ and angular momentum $\hat J\approx0.087$.  
We emphasize that these parameters are chosen so that the initial Kerr-AdS black hole is slowly rotating (i.e., $\Omega_H<1$), but their are otherwise arbitrary. Therefore, one should expect that the dynamics that we describe below is a generic property of all slowly rotating Kerr-AdS black holes, including Schwarzschild-AdS.

Before describing our results we collect some technical details of our runs.
At any time step, the numerical solution is given by the values of the variables $\tilde g_{\mu\nu}$, $\tilde\varphi$ and $\tilde H_\mu$, defined in eqs.~\eqref{eqn:evogkads}--\eqref{eqn:evoHkads}, on a grid whose points are located at discretised, equally spaced values of the compact, asymptotically Cartesian coordinates $(x,y,z)$. We use the PAMR/AMRD libraries \cite{PAMR} to solve \eqref{eq:EFE} (with $\mu=0$) on a single Cartesian grid. For the results presented in this section, we considered a grid with $N=217$ points along each Cartesian direction and a Courant-Friedrichs-Lewy (CFL) factor of $0.3$.
In our coordinates, the Kerr-AdS black hole rotates on the $x$-$y$ plane. 
As is customary in numerical simulations, on any slice at fixed coordinate time $t$, we use the location of the apparent horizon (AH) as a proxy for the location of the event horizon on that slice.
In our simulations, the  $S^2$ of the AH is covered by $N_\theta=33$ equally spaced points along the polar direction $\theta$ and $N_\phi=65$ points along the azimuthal direction $\phi$, so the spacing in the two angular directions is the same. 
In the case where the background geometry is fixed, the location of the AH is set to the exact value of the location of the Kerr-AdS event horizon, i.e., $R=R_+(\theta)$ in Kerr-Schild-HHT coordinates. On the other hand, in the simulations with full backreaction, we locate the AH at each time step using the flow method. In both cases we excise $60\%$ of the coordinate region inside the AH.
The details of the implementation of these techniques can be found in \cite{Choptuik:2003qd,Pretorius:2004jg,Bantilan:2012vu,Bantilan:2020xas}.
Finally, we use the bulk solution to obtain the boundary CFT observables given in Appendix~\ref{sec:CFTquants} via  extrapolation of the bulk quantities as described in Appendix F of \cite{Bantilan:2020xas}, with the marginal adaptations explained in Appendix E of \cite{Rossi:2022nao}.

\subsection{Fixed background}
\label{sec:fixbackgr}

As a non-trivial test of the accuracy of the scheme and to check that we have identified a suitable class of initial data for the scalar field on a fixed Kerr-AdS background, we verify that at late times the scalar field exhibits a (uniform) inverse logarithmic decay in time, consistent with the theorems in \cite{Holzegel:2011uu,Holzegel:2013kna}.\footnote{We note that these references do not prove the existence of solutions to the Klein-Gordon equation on a fixed Kerr-AdS background that asymptotically decay logarithmically in time (let alone the fact that a generic solution asymptotically decays logarithmically in time). We thank the anonymous referee for pointing this out to us.} To obtain this logarithmic decay at late times we do not need to fine tune any of the parameters that specify the initial scalar field configuration.  This numerical results suggest that generic (in a suitable sense) solutions of the Klein-Gordon equation on a fixed slowly rotating Kerr-AdS blackground asymptotically decay logarithmically in time. 

\begin{figure*}[t!]
    \centering
    \includegraphics[scale=0.75]{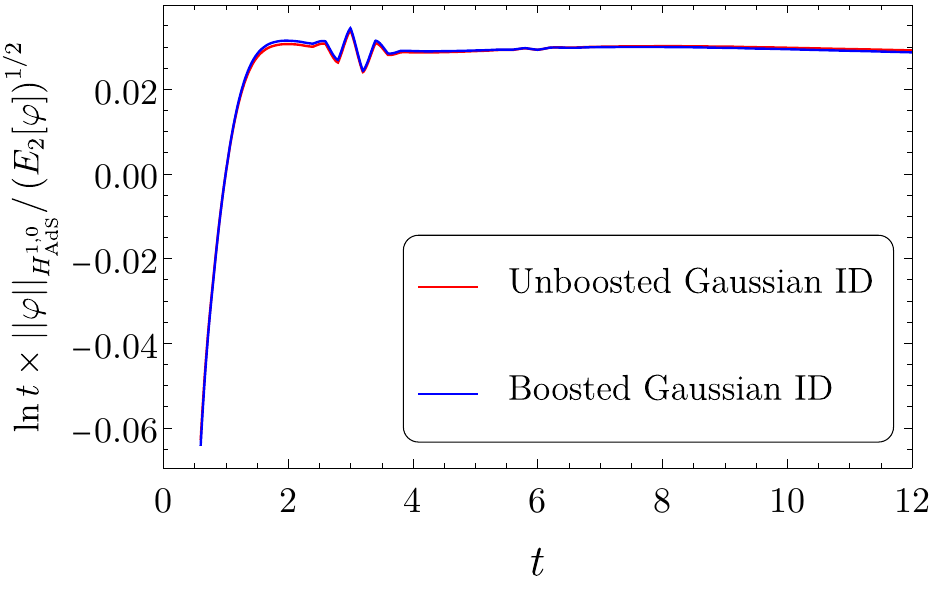}
\parbox{5.0in}{\vspace{0.1cm}\caption{Norm $||\varphi||_{H^{1,0}_{\text{AdS}}}$ of the scalar field on a fixed Kerr-AdS background for boosted (blue) and unboosted (red) Gaussian initial data. This norm exhibits inverse logarithmic decay in time, with a proportionality constant that only depends on the parameters of the background, in consistent with the theorems of \cite{Holzegel:2011uu,Holzegel:2013kna}.
}\label{fig:fix_bkg_init_dat}}
\end{figure*}

We considered two types of initial data for the scalar field. Firstly, we evolve the deformed Gaussian scalar field profile defined in Section~\ref{sec:indat} with parameters $A=20$, $\delta=0.2$, $(e_x,e_y,e_z)=(0.1,0.5,0.9)$, no boost, and centred at $(x_0,y_0,z_0)=(0.7,0.0,0.0)$. 
Fig.~\ref{fig:fix_bkg_init_dat} shows that, after a sufficiently long time, the norm of the scalar field behaves as
\begin{equation}
  ||\varphi||_{H^{1,0}_{\text{AdS}}}(t)=\frac{1}{\ln t}C E_2[\varphi]^{\frac{1}{2}},
  \label{eq:linear_log_decay}
\end{equation}
where $C$ is a positive constant that should only depend on the Kerr-AdS background parameters, namely $r_+,a$, and the scalar field mass $\mu$ (in units of the AdS radius $L$) \cite{Holzegel:2011uu,Holzegel:2013kna}. 
The quantities $||\varphi||_{H^{1,0}_{\text{AdS}}}$ and $E_2[\varphi]$ are defined in Appendix~\ref{sec:AdSnorms}.
For our choice of initial data and background parameters, we find that $C\approx 0.03$. The oscillation that can been seen on this figure at around $t\sim 3$ corresponds to the first time that the scalar field hits the AdS boundary; another, much smaller, wiggle  can be seen at around $t\sim 6$, and it corresponds to the second reflection of the scalar field off the AdS boundary. The interval between these two oscillations is consistent, roughly, with the light crossing time in the spacetime. Later oscillations cannot be seen on the scale of this plot. 
We have repeated this test for different values of the scalar field amplitude, namely $A=10,15$, and we have obtained the same late time profile as in Fig.~\ref{fig:fix_bkg_init_dat}; in particular, the value of the constant $C$ that we extract from our data is the same for the various runs (within the numerical errors). 

We have carried out a similar run with a boosted Gaussian profile but with a smaller amplitude, $A=0.1$, same velocity, less extreme eccentricities, $(e_x,e_y,e_z)=(0.1,0.3,0.6)$, and centred at $(x_0,y_0,z_0)=(0.6,0.0,0.0)$. 
The reason is that this is exactly the same initial scalar field configuration that we use in the simulations with full backreaction (see Section~\ref{sec:back}). 
However, with such a small amplitude it is hard to accurately extract the late time logarithmic decay and infer the value of the constant $C$. 
This simulation with a small amplitude and no backreaction is useful because it allows us to compare with the one with backreaction. 
In particular, we can easily obtain the decay rates of the different modes of scalar field and assess the impact of the non-linearities. 
We defer the detailed discussion of the latter to Section~\ref{sec:back}.

\begin{table}[t!]
  \centering
  \begin{tabular}{c|ccccccccccc}
  \hline
  \hline
   $l=m $ & 0 & 1 & 2 & 3 & 4 \\
  \hline
   $- \operatorname{Im}\omega$ & 1.734 & 1.128 & 0.620 & 0.399 & 0.266 \\
   \hline
   \hline
   $l=m $ & 5 & 6 & 7 & 8 & 9 & 10 \\
  \hline
   $- \operatorname{Im}\omega$ & 0.170 & 0.096 & 0.051 & 0.026 & 0.015 & 0.011 \\
   \hline
   \hline 
  \end{tabular}
  \caption{Decay rates of the first few spherical harmonic modes with  $l=m$ of the expectation value of the operator dual to the scalar field in the bulk, $\langle \mathcal{O} \rangle_{\text{CFT}}$. These decay rates are compatible with those in \cite{Uchikata:2009zz}.}
  \label{tab:decay_linear}
\end{table}

The amplitudes and the corresponding decay rates of the scalar field modes can be extracted in a simple way by decomposing the expectation value of the operator dual to the bulk scalar field, $\langle \mathcal{O}\rangle_{\text{CFT}}$, in  spherical harmonics.
The simulations without backreaction also have the obvious advantage of being computationally much cheaper; by comparing the results obtained with different resolutions we can get a quantitative understanding of which modes can be resolved for a given number of grid points. The results are depicted in Fig.~\ref{fig:scalopdecay} with dashed lines.
This figure shows the evolution of the amplitudes of the first few $l=m$ spherical harmonic modes  of $\langle \mathcal{O}\rangle_{\text{CFT}}$. At the linear level (dahsed lines), the plot shows that the various modes are decoupled and decay exponentially, as expected.
This figure also shows the well-known fact that the decay rates of the individual modes decrease as $l$ increases. 
In Table \ref{tab:decay_linear} we present the  decay rates of the first few scalar harmonics extracted from this simulation without backreaction.\footnote{We do not quote the decay rates for the $l=m=11,\,12$ modes because the errors in the decay rates for these modes are too large.} To confirm the correctness of our decay rates, we should compare the results in Table \ref{tab:decay_linear} with the direct calculation of the scalar quasi-normal frequencies of Kerr-AdS. Unfortunately, we are not aware of any results in the literature for the background parameters that we are considering; the closest ones that we could find are those in \cite{Uchikata:2009zz}. This reference studied scalar perturbations of Kerr-AdS black holes for different values of the background parameters; of particular relevance for us is the case with $r_+=0.5$, $a=0.2$ and $l=m=1$ (see their Fig.~4). Our results are compatible with the results in this reference.

\subsection{Backreaction}
\label{sec:back}

In this subsection we present the results of the simulations with backreaction. As initial data, we consider the scalar field Gaussian profile of the previous subsection, with amplitude $A=0.1$, eccentricities $(e_x,e_y,e_z)=(0.1,0.3,0.6)$, centred at $(x_0,y_0,z_0)=(0.6,0,0)$ and boosted in the $X$-direction with $(v_X,v_Y,v_Z)=(0.2,0.0,0.0)$. 
Such a small amplitude of the scalar field is necessary to guarantee that the initial data is suitably small and,
as discussed in Section~\ref{sec:indat},  to ensure that the constraint damping terms in the equations can effectively dissipate the initial constraint violations on a timescale that is faster than any other timescale in the problem.

\subsubsection{Bulk quantities}
\label{sec:bulkqts}

In this subsection, we describe the evolution of various bulk quantities. We first consider the evolution of the scalar field and then we move on to characterize the evolution of the geometry. 

\begin{figure*}[t!]
    \centering
    \includegraphics[scale=0.2,clip=true]{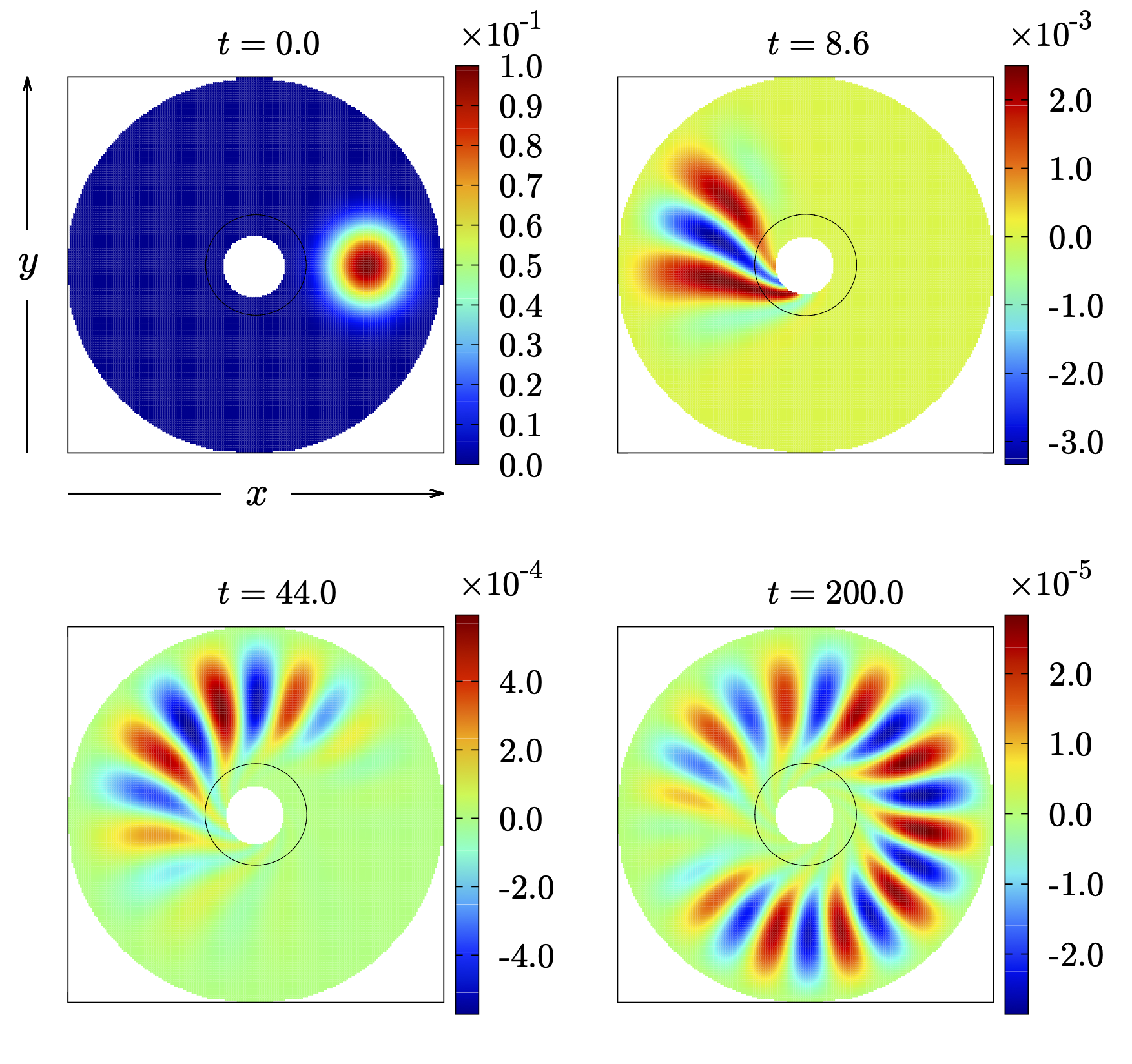}
\parbox{5.0in}{\vspace{0.1cm}\caption{Snapshots of the bulk scalar field $\tilde{\varphi}$ on the rotation plane of the black hole. The black line denotes the location of the AH and the blank region at the centre of each plot corresponds to the excised region. The scalar field is dragged by the rotating black hole and it rotates with an angular velocity $\sim 2\pi/5.0$. Low $l$ modes decay faster than the higher $l$ modes, which dominate at late times.
    }\label{fig:phievo}}
\end{figure*}

\subsubsection*{Scalar field}

In Fig.~\ref{fig:phievo} we display four representative snapshots of the evolved scalar field $\tilde\varphi$ on the rotation plane, $z=0$, of the black hole.\footnote{For animations of the bulk scalar field, see \url{https://youtu.be/aVXCCBesyGU} and \url{https://youtu.be/Wk8N0d6qfJE}.} The location of the AH is shown with a solid black line and the blank region at the centre of each plot corresponds to the excised region. This sequence of plots shows that a significant amount of the scalar field falls into the black hole during the initial stages of the evolution. However, a small amount of scalar field remains in the exterior of the black hole and is dragged along by the rotation as the evolution proceeds. The plots also show that the total amplitude of the remnant scalar field decreases with time as the scalar field continues to fall into the black hole.

Initially, the low $l$ modes dominate the dynamics but  they decay fast; in the later stages of the evolution, small distance structure, i.e., high $l$ modes, in the scalar field profile becomes visible in the region between the AH and the AdS boundary. 
As we argue below, the reason for this is the stable trapping of the higher $l$ scalar harmonic modes, which leads to a slow decay of the scalar field; we have seen in Section \ref{sec:fixbackgr} that the decay rates in the linear problem are smaller for the higher $l$ harmonics and it remains true in the full non-linear problem. 

From Fig.~\ref{fig:phievo} we see that the low $l$ modes that dominate the early dynamics of the scalar field  are supported outside as well as in the region near and inside the black hole. For instance, the second panel in Fig.~\ref{fig:phievo} shows that the scalar field profile has two clear maxima and a minimum, and both in the second and third panels of this figure one can see that the scalar field profile penetrates the AH.  Hence the low $l$ modes rapidly fall into the black hole; in fact, in Section~\ref{sec:bdyobs} we show that they decay exponentially and do not exhibit non-linearities. On the other hand, at late times the evolution of the scalar field is dominated by the high $l$ modes.  The last panel in Fig.~\ref{fig:phievo} shows that in the late stages the scalar field profile has many peaks and troughs;  these modes with high $l$ barely have any support in the near horizon region, which implies that they decay very slowly, resulting in very different dynamics. This can be clearly seen in the last panel of Fig.~\ref{fig:phievo}, which shows that the high $l$ modes of the scalar field are supported in a region between the AdS boundary and a potential barrier, and they remain in this region for a very long time. This is  the  stable trapping mechanism and it results in non-linear effects as a consequence of the interactions between the different scalar modes and with the metric.  Below we discuss these non-linear effects on the spacetime geometry.

\subsubsection*{Spacetime geometry}

\begin{figure}[t!]
  \centering
  \includegraphics[scale=0.65,clip=true]{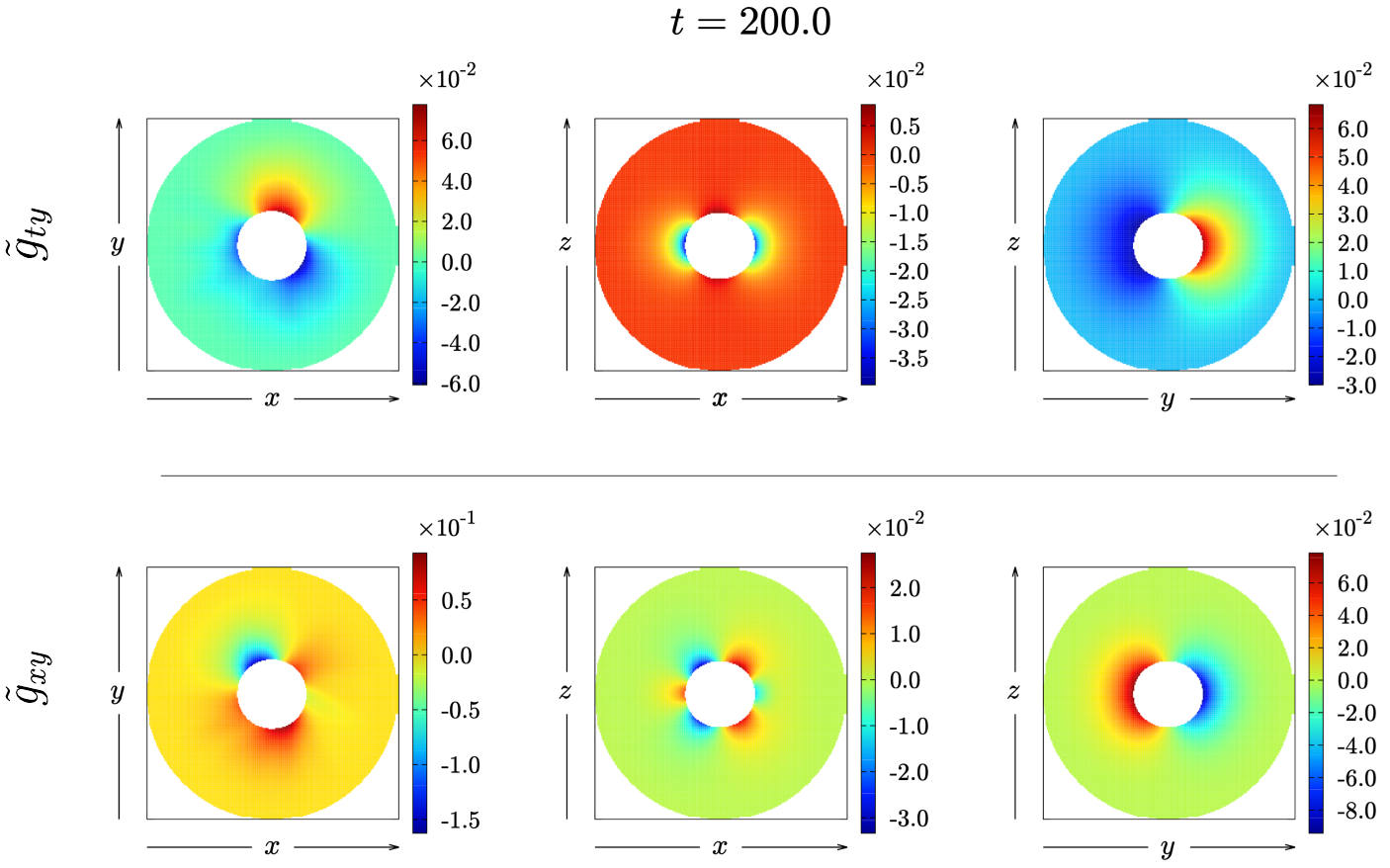}
  \parbox{5.0in}{\vspace{0.4cm}\caption{Evolved metric components $\tilde g_{ty}$ (top) and $\tilde g_{xy}$ (bottom) on various slices of the computational domain in the final snapshot of our simulation. The metric deviations from Kerr-AdS are non-axisymmetric and time-dependent, and their changes are very small (on the scale of this plot) from $t\sim 20$ until the end of our simulations, $t=200$. } \label{fig:metric_deviations}}
\end{figure}

The area of the AH increases by a few percent in the very early stages of the evolution due to the fact that most of the scalar field is absorbed when it first collides with the black hole. 
The small increase of the AH area is consistent with the amplitude of the scalar in the initial data and the latter being a small perturbation of the Kerr-AdS background. 
After a short ($\Delta t\sim 2$) period of adjustment, the AH area remains essentially constant for the rest of the simulation because the amplitude of the leftover scalar field is too small to have a detectable effect on this quantity.

The scalar field backreacts onto the geometry giving rise to a time-dependent spacetime. There is an initial period of rapid dynamics that is triggered by our initial data since the scalar field is localized on the rotation plane, away from the black hole and moving towards it with some small velocity until they collide. 
The scalar field  backreacts on the geometry, which results in a significant evolution of the spacetime metric.\footnote{This initial real dynamics is clearly distinguished from the initial gauge adjustment and constraint damping period, which happens on an even faster time scale.}
With our choice of initial conditions, by $t\sim 20$ the absolute value of the scalar field amplitude is already $O(10^{-3})$. From then onwards, on large scales, the spacetime geometry appears to have settled into what seems to be a periodic state, with some small scale dynamics due to the fact that the scalar field continues to slowly decay for a very long time. 
In Fig.~\ref{fig:metric_deviations} we display two representative components of the evolved metric, namely $\tilde g_{ty}$ and $\tilde g_{xy}$, in the final snapshot of our simulations. Superficially, they change very little from $t\sim 20$ until $t\sim 200$. These deformations of the metric have an imprint on the geometry of the AH, which is also  deformed away from axisymmetry. Note that from Fig.~\ref{fig:metric_deviations} we see that the metric deviations $\tilde g_{\mu\nu}$ are $O(10^{-2})$ while the background Kerr-AdS metric components $\hat g_{\mu\nu}$ are $O(1)$; this implies that the local deformations in the shape of the AH in these stages of the simulation are also at the percent level.

\begin{figure}[t]
  \centering
   \includegraphics[scale=0.6]{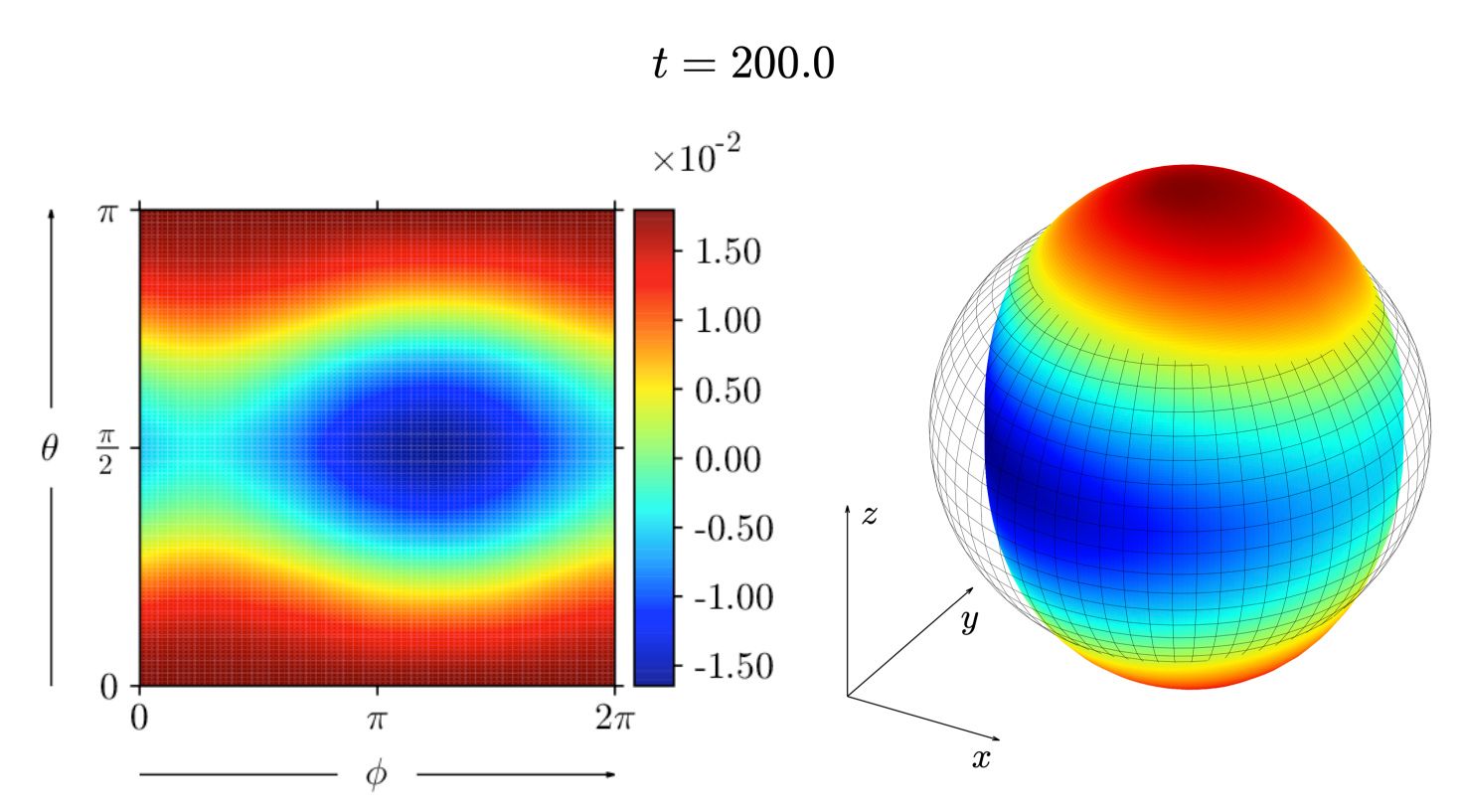}
  \parbox{5.0in}{\vspace{0.1cm}\caption{Left: Relative difference of the radial function, $\rho_{AH}/\rho_+ -1$, that determines the location of the AH at $t=200$, where $\rho_+$ is the location of the event horizon of the background Kerr-AdS black hole. Right: the same quantity is displayed on a  sphere with radius $1+10(\rho_{AH}/\rho_+ -1)$. The factor of 10 is used to enhance the deformation of the unit sphere and make it visible. To guide the eye, we superpose, in light gray, a sphere of radius $\rho_+(\theta)/(\rho_+|_{\theta=0})$ as a way to visualize the deformation to the unit round sphere induced by the radial function in Kerr-AdS with the same background parameters used in the simulations.}
  \label{fig:radius_AH}}
\end{figure}

In Fig.~\ref{fig:radius_AH} (left) we plot the relative difference between the radius of the AH in our dynamical spacetime at the final time of our simulation, $\rho_{\text{AH}}$, and the radius of the initial Kerr-AdS event horizon, $\rho_+$; the latter is obtained by plugging the Kerr-Schild-HHT horizon radius $R_+(\theta)$ in the expression for the compactified radial coordinate $\rho$, eq.~\eqref{eq:comprad}.
Even though this is a coordinate-dependent quantity, it provides some intuition about the deformation of the AH. 
Fig.~\ref{fig:radius_AH} (left) shows that the radius of the horizon $S^2$ of our dynamical spacetime is larger at the poles and smaller around the equator than that of Kerr-AdS with $M=0.4$ and $a=0.2$; this type of deformation suggests that the horizon of the dynamical black hole is rotating slower than that of the background Kerr-AdS black hole. This is consistent with the fact that in the time-dependent spacetime, some of the mass and angular momentum are carried by the gravitational field and the remnant scalar field outside the black hole. This figure also shows that the horizon $S^2$ of the dynamical black hole is non-axisymmetric; there is a blob localized around the equator that breaks the axisymmetry of the spacetime. This blob rotates around the black hole with some periodicity and persists all the way to the end of our simulation. We will come back to this point later. 

To get some more intuition about the shape of the AH, in Fig.~\ref{fig:radius_AH} (right) we plot the same deformation function, $\rho_{AH}/\rho_+ -1$, on the unit sphere multiplied by a factor of 10 to make it visible. To guide the eye, we also plot, in light gray,  the deformation of the unit sphere  induced by  $\rho_+$ along the polar angle, namely $\rho_+(\theta)/(\rho_+|_{\theta=0})$. We emphasize that this plot is not meant to be an accurate depiction of the deformed AH geometry but it should provide some intuition of its most salient features.

To  further assess the non-axisymmetric deformations of AH, we compute the $m=1$ mode, ``bar mode'' ($m=2$) \cite{Shibata:2010wz} and ``quadrupole mode'' ($m=4$) deformations, defined as follows.
\begin{align}
  \eta_{m=1} &:= \frac{2\sqrt{(\bar\ell_0-\bar\ell_\pi)^2+(\bar\ell_{\pi/2}-\bar\ell_{3\pi/2})^2}}{\bar\ell_0 + \bar\ell_\pi}\,, \label{eq:AH_m=1}\\
  \eta_\text{bar} &:= \frac{2\sqrt{(\ell_0-\ell_{\pi/2})^2+(\ell_{\pi/4}-\ell_{3\pi/4})^2}}{\ell_0+\ell_{\pi/2}}\,, \label{eq:AH_m=2}\\
  \eta_\text{quad} &:= \frac{2\sqrt{(\ell_0-\ell_{\pi/4})^2 + (\ell_{\pi/8}-\ell_{3\pi/8})^2+(\ell_{\pi/2}-\ell_{3\pi/4})^2+(\ell_{5\pi/8}-\ell_{7\pi/8})^2}}{\ell_0 + \ell_{\pi/4} + \ell_{\pi/2} + \ell_{3\pi/4}}\,, \label{eq:AH_m=4}
\end{align}
where $\bar\ell_\phi$ is the proper length of a meridian at a given value of the azimuthal angle $\phi$ on the $S^2$ of the AH, 
\begin{equation}
  \bar\ell_\phi := 2\int_0^{\pi/2}d\theta\sqrt{\sigma_{\theta\theta}(\phi)}\,,\quad \ell_\phi:= \bar\ell_\phi + \bar\ell_{\phi+\pi}\,,
\end{equation}
$\sigma_{AB}$ is the induced metric on the AH and $\theta$ is the polar angle.\footnote{Note that with our choice of initial conditions, the spacetime is symmetric under reflections about the equatorial plane of the background Kerr-AdS black hole. However, we do not impose this symmetry in our simulations.} 

\begin{figure}[t!]
  \centering
  \includegraphics[scale=0.55]{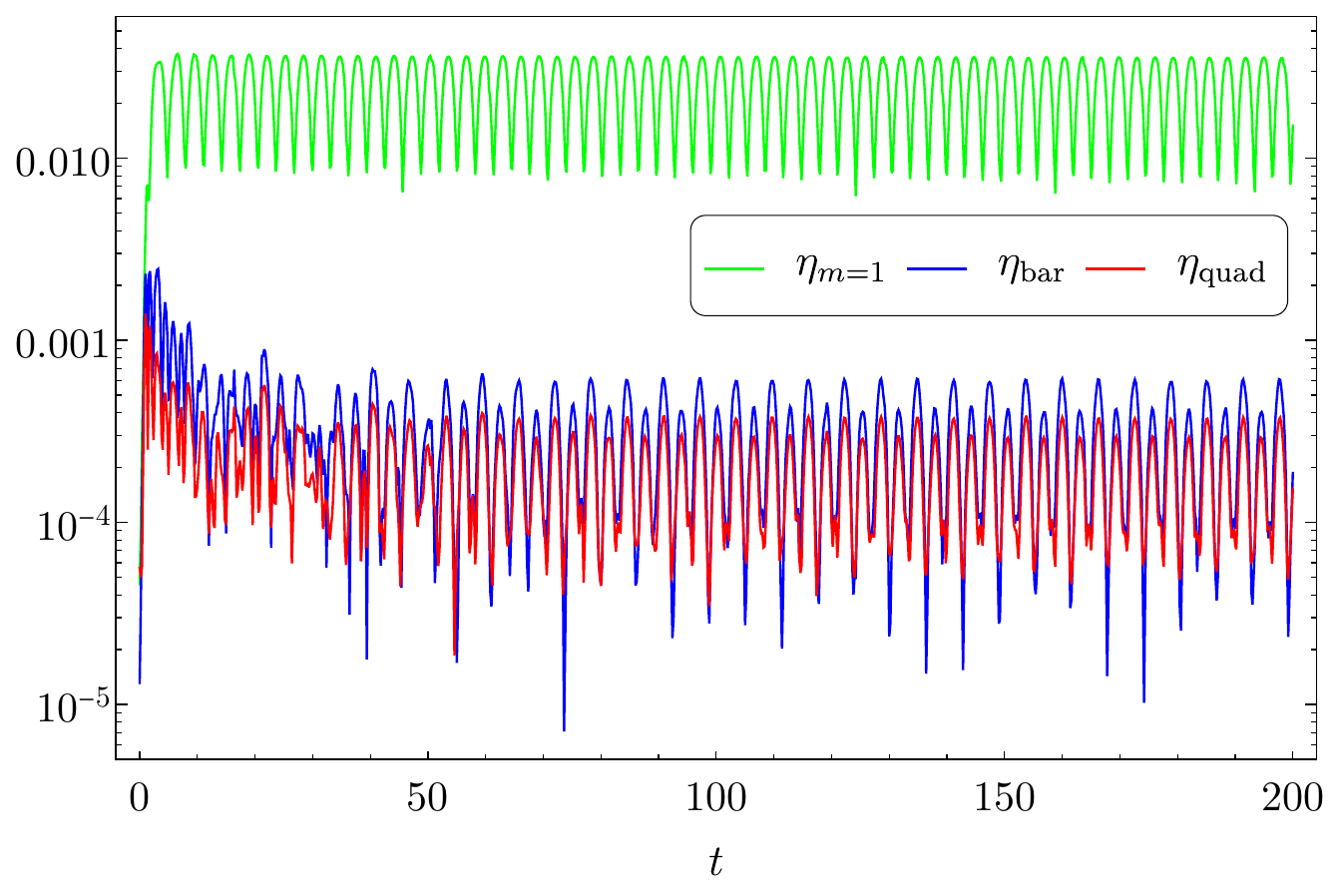}
  \parbox{5.0in}{\vspace{0.1cm}\caption{$m=1$, ``bar mode'' and quadrupole deformations of the $S^2$ of the AH as functions of time. The final black hole is non-axisymmetric and evolves on two time scales.}
  \label{fig:AH_deformations}}
\end{figure}

In Fig.~\ref{fig:AH_deformations} we display the deformations as measured by these modes as functions of time. The figure shows that the $m=1$ deformation dominates, as the relative radius function of the AH suggests, see Fig.~\ref{fig:radius_AH}. 
Furthermore, the $m=1$ deformation is imprinted onto the black hole by the initial collision with the incoming  scalar field and, unlike the higher mode deformations, it reaches its final state shortly after this initial interaction between the black hole and the scalar field. Fig.~\ref{fig:AH_deformations} shows that the end state of the $m=1$ deformation $\eta_{m=1}$ is given by an oscillating function that does not decay. On the other hand, this plot shows that the higher mode deformations of the AH undergo an initial period of adjustment and reach their asymptotic final states at $t\sim 20$--$40$; from then on, they oscillate with two characteristic frequencies. The fast frequency of the higher $m$ modes is the same as that of the $m=1$ mode. Therefore, from this plot we conclude that the black hole settles into a time depdendant and non-axisymmetric spacetime that evolves on two time scales. As we shall see shortly, the deformation of the AH and the mode decomposition of the energy density of the dual CFT exhibit similar features, as expected.

\begin{figure*}[t!]
    \centering
    \includegraphics[scale=0.2,clip=true]
    {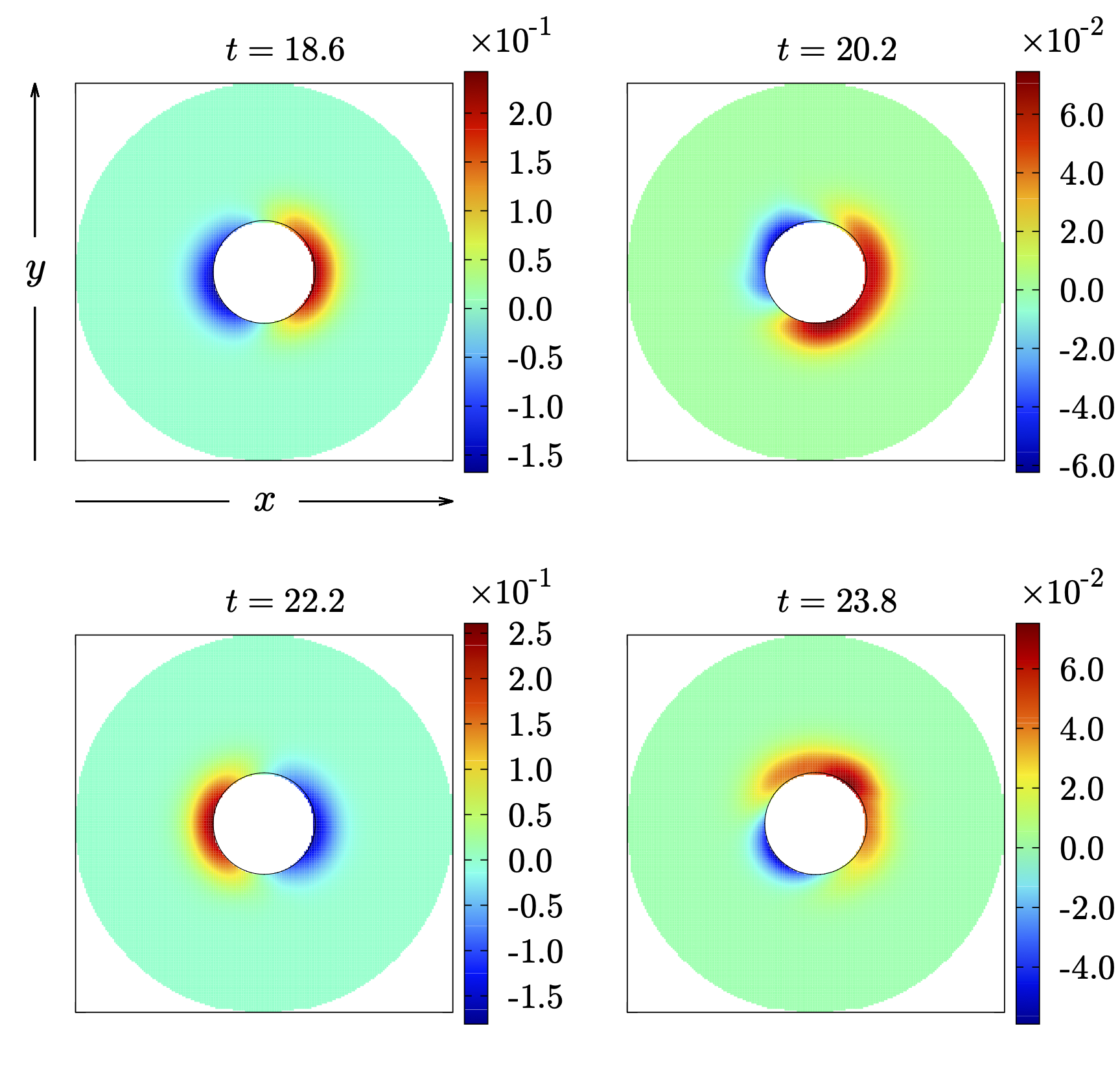}
\parbox{5.0in}{\vspace{0.1cm}\caption{Snapshots of the relative (to Kerr-AdS) Kretschmann scalar on the rotation plane. The black line denotes the location of the apparent horizon.
    }\label{fig:relkretschevo}}
\end{figure*}

To get a better (and coordinate-invariant) understanding of the spacetime geometry, we consider the evolution of the Kretschmann scalar, and we compare it with the Kretschmann scalar of the background Kerr-AdS black hole. In Fig.~\ref{fig:relkretschevo} we display four representative snapshots of the evolution of the relative Kretschmann scalar. After an initial period in which the geometry reacts very quickly to the scalar perturbation, we observe that the relative Kretschmann settles to its (essentially) final state, which is time-dependent and oscillating around a central value. We have observed such behavior in the evolved metric components and in the geometry of the AH; the fact that the Kretschmann scalar exhibits the same dynamics confirms that this is effect is physical, not gauge-dependent.

The size of the local deviations from Kerr-AdS as measured from the relative Kretschmann scalar is of a few percent, consistent with the initial amplitude of the scalar perturbation and the metric deviations. We note that the Kretschmann scalar rotates in the opposite direction with respect to the scalar field.\footnote{For animations of the relative Kretschmann scalar, see \url{https://youtu.be/PEG8l7NnqG0} and \url{https://youtu.be/izNYPwSohig}.} 
The local maximum and minimum of the relative Kretschmann spend a time $\Delta t\sim 1.6$ close to the horizontal $y=0$ line (the top-left panel); the rotation appears to be interrupted at this time; it then resumes and it takes another $\Delta t\sim 1.6$ to complete half a cycle and return close to the $y=0$ line, but now with the local maximum and minimum on opposite sides of the black hole (compare the top and bottom left panels of this figure). These times are compatible with the light-crossing time of the spacetime and the shape of the relative Kretschmann is consistent with the deformation of the geometry being dominated by an $m=1$ mode. On top of this global rotation of the relative Kretschmann, we observe that the maximum and minimum also oscillate. This evolution pattern of the relative Kretschmann repeats itself and the amplitude of these oscillations does not decrease with time.
Therefore, our simulations provide evidence that the relative Kretschmann exhibits periodic dynamics, which implies that the spacetime itself is also periodic. As we shall discuss below, the two periodic time scales that we observe in the evolution of the relative Kretschmann are the same as in the spacetime metric and in the deformations of the AH. Clearly, since Kerr-AdS is stationary and axisymmetric, our dynamical spacetime is not settling down to another member of the Kerr-AdS family of black holes.

\begin{figure*}[t!]
    \centering
    \includegraphics[scale=0.6]{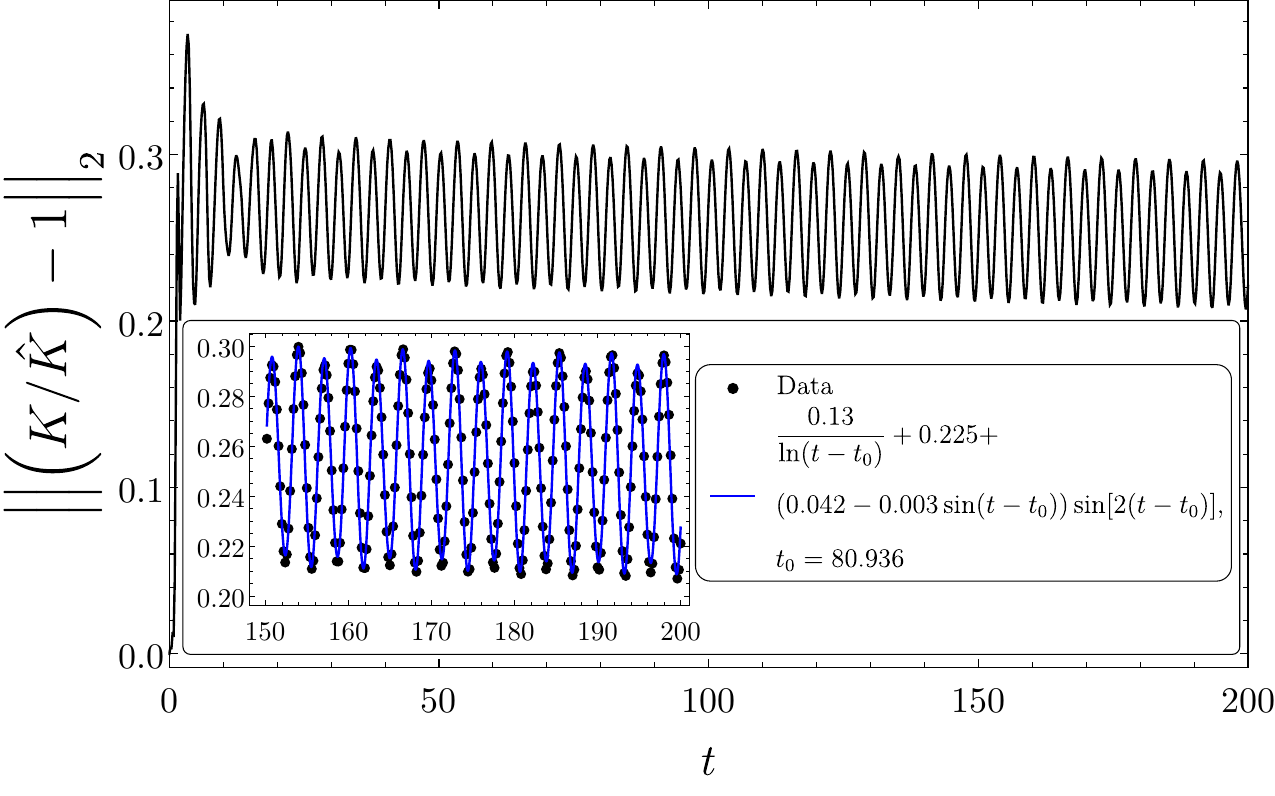}
\parbox{5.0in}{\vspace{0.1cm}\caption{$L^2$ norm of Kretschmann scalar relative to the Kretschmann scalar $\hat K$ of Kerr-AdS with $M=0.4$ and $a=0.2$ as a function of time. At each time step, the $L^2$ norm is computed over the region between the AdS boundary and the AH. The inset shows a fit of the data set at late times with a function oscillating on two time scales.
    }\label{fig:L2norm-relkretsch}}
\end{figure*}

To provide further evidence, in Fig.~\ref{fig:L2norm-relkretsch} we display the time evolution of the $L^2$ norm of the relative Kretschmann scalar. At every time step, this norm is computed over the region between the AH and the AdS boundary, and hence it provides a global measure of the relative difference between the two Kretschmann scalars. As this plot shows, for the duration of our simulation, the deviation from pure Kerr-AdS as measured by the $L^2$ norm is at around 25\%, even though the local differences remain at the percent level throughout the evolution. Furthermore, the fact that the relative Kretschmann does not appear to grow over time in our simulations can be interpreted as evidence that, if some kind of singularity eventually forms, it will have to happen on a very long (possibly infinite) timescale. 

Our simulations indicate that the deviation from Kerr-AdS will eventually settle down to a finite value, with oscillations of finite amplitude on top. In the inset in Fig.~\ref{fig:relkretschevo} we show that the evolution of the relative Kretschmann can be accurately captured by a fit to a function of the form
\begin{equation}
  f(t)=\tfrac{a_1}{\ln(t-t_0)}+[a_2+a_3\sin(t-t_0)]\sin[2(t-t_0)]+a_4\,,
  \label{eq:K_fit}
\end{equation}
where $t_0$ is chosen to be a sufficiently large ``late'' time and such that the fitting function and the dataset have the same phase shift; we use $t_0\sim 80.9$ but any other value differing by an integer multiple of the light-crossing time provides an equally good fit. 
The first term in \eqref{eq:K_fit} can be replaced by any other slowly decaying function and the fit is equally good; for instance, $e^{-a_1(t-t_0)}$ and $a_1/(t-t_0)$ work just as well.
This term  can be interpreted as accounting for the remnant scalar field in the spacetime that is slowly decaying.
The two periodic functions in \eqref{eq:K_fit} capture the two timescales in the evolution of the relative Kretschmann that can be seen in Fig.~\ref{fig:L2norm-relkretsch}, namely the rotation of the relative Kretschmann around the central black hole (see Fig.~\ref{fig:relkretschevo}) and the oscillation of the amplitude. 
More importantly, the constant term $a_4$ in the fit is pretty much insensitive to the choice of the slowly decaying function in the first term in \eqref{eq:K_fit}; the possible choices outlined above all give $a_4\sim 0.23-0.25$. This term is the central value that the $L^2$ norm of the relative Kretschmann is oscillating about at late times, thus suggesting that the relative Kretschmann will not decay at all. 
Furthermore, the evolution of the $m=2$ and $m=4$ deformations of the AH shown in Fig.~\ref{fig:AH_deformations} can be also be fitted by functions of the form \eqref{eq:K_fit} (obviously with different fitting coefficients); in particular with the same two frequencies and with a constant term. This shows that the metric itself is evolving on these two periodic timescales, which confirms the previous conclusion that the spacetime is not going to settle down to another member of the Kerr-AdS family of black holes.
As we will show in the next subsection, other geometric invariants that measure the deviations from Kerr-AdS do not decay either, which supports this observation.

Given that the duration of our simulations is necessarily finite, one cannot be absolutely certain that the endpoint of the evolution is a periodic spacetime, as the fit \eqref{eq:K_fit} would suggest. Neither we can rule out the possibility that the geometry experiences an energy cascade to the UV ending at a singularity on a very long (possibly infinite) timescale that our simulations do not capture. Similar conclusions were reached from the numerical simulations of the superradiant instability of Kerr-AdS \cite{Chesler:2018txn,Chesler:2021ehz}. A third possibility that cannot be discarded is that the spacetime eventually does decay to another Kerr-AdS black hole but on a very long timescale. However, there is no evidence in our simulations that favors this last possibility.

Finally, we note that we have perturbed a slowly rotating Kerr-AdS black hole by adding a scalar field with a small amplitude to the background spacetime. The interaction between the scalar field and the geometry induces deformations of the spacetime metric away from Kerr-AdS and in this subsection we have provided evidence that the latter do not decay in time. On the other hand, the scalar field continues to decay throughout the evolution and does not play a role in the late time dynamics of the system. Therefore, it should be possible to induce qualitatively similar deformations to the Kerr-AdS geometry by a judicious choice of initial conditions without the need to introduce matter and we expect that such deformations of the metric would not decay at late times.

\subsubsection{Boundary observables}
\label{sec:bdyobs}

\begin{figure*}[t!]
    \centering
    \includegraphics[scale=0.2,clip=true]{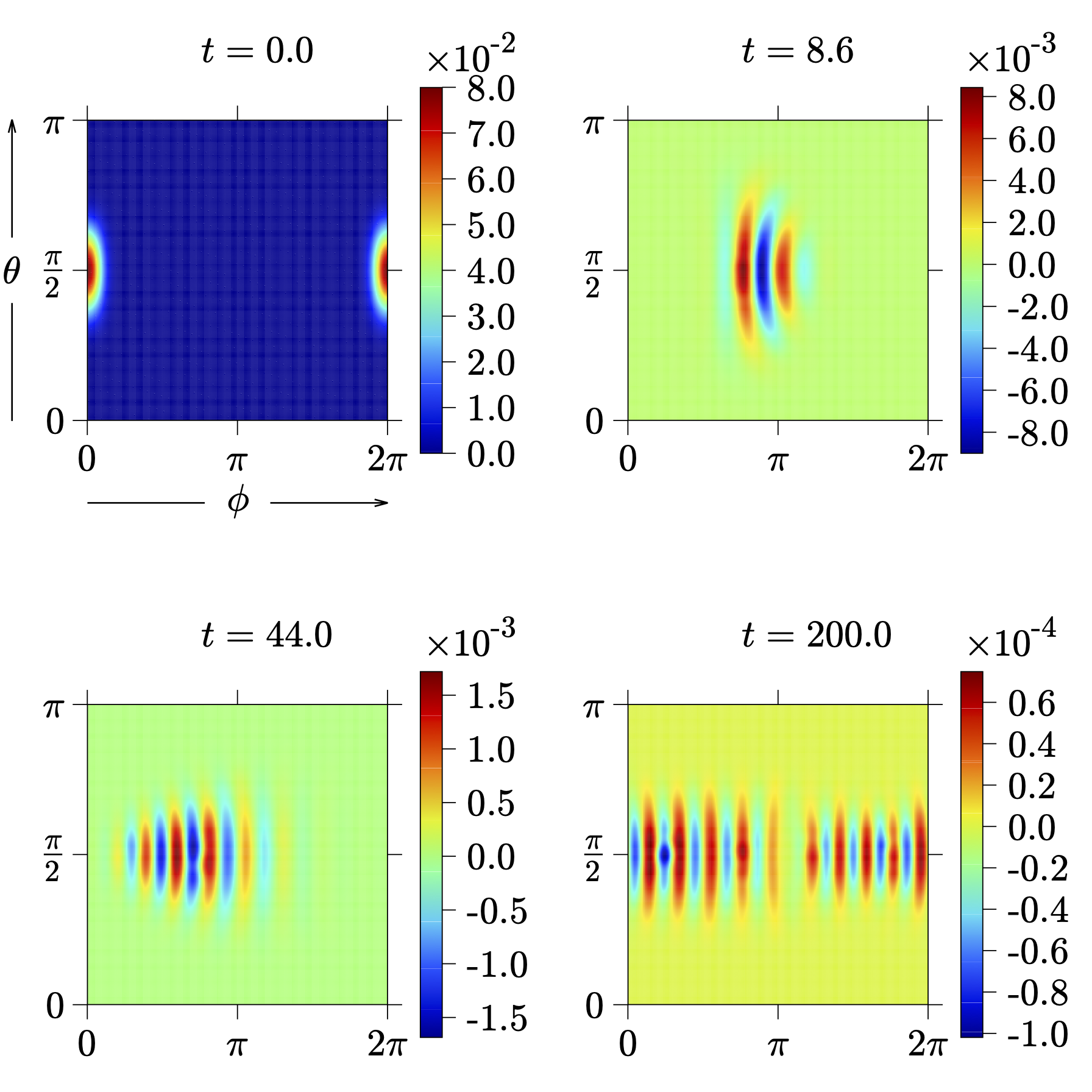}
\parbox{5.0in}{\vspace{0.1cm}\caption{Evolution of the expectation value of the  operator dual to the bulk scalar field, $\langle \mathcal{O}\rangle_{\text{CFT}}$. The low $l$ modes decay quickly, but the higher $l$ ones survive until very late times due to the stable trapping in the bulk.
}\label{fig:sphereevobdyscalop}}
\end{figure*}

We now turn to the discussion of the observables of the dual CFT. These are useful because they also provide a gauge invariant characterization of the bulk dynamics. The expressions of the boundary quantities in terms of the bulk solution are given in Appendix~\ref{sec:CFTquants}.

Fig.~\ref{fig:sphereevobdyscalop} shows four snapshots of the expectation value of the operator dual to the bulk scalar field, $\langle \mathcal{O}\rangle_{\text{CFT}}$, and is the boundary translation of Fig.~\ref{fig:phievo}.\footnote{For animations of the expectation value of the boundary scalar operator, see \url{https://youtu.be/Sfha75Fmt9I} and \url{https://youtu.be/oeGBwwa8rtE}.} 
In the bulk, the scalar field is localized on the rotation plane of the black hole and it is dragged along by the rotation; from the boundary point of view, the scalar operator $\langle \mathcal{O}\rangle_{\text{CFT}}$ is localized at the equator of the boundary sphere. Fig.~\ref{fig:sphereevobdyscalop} shows that low $l$ modes decay rapidly, while at late times large $l$ modes are still visible. In these snapshots, the harmonic quantum number $l$ can be identified by the number of peaks and troughs. At the end of our simulation we see that the $l=m=10$ and $l=m=11$ modes are the dominant ones, with the $l=m=10$ mode having a marginally larger amplitude.  Note that the last snapshots in Figs.~\ref{fig:phievo} and \ref{fig:sphereevobdyscalop} are consistent with each other: both figures show eleven local maxima (as well as eleven local minima)  in the profile of the scalar field; ten of these are large enough to be clearly visible, while the last one is small and barely visible.

\begin{figure*}[t!]
    \centering
    \includegraphics[width=6.0in,clip=true]{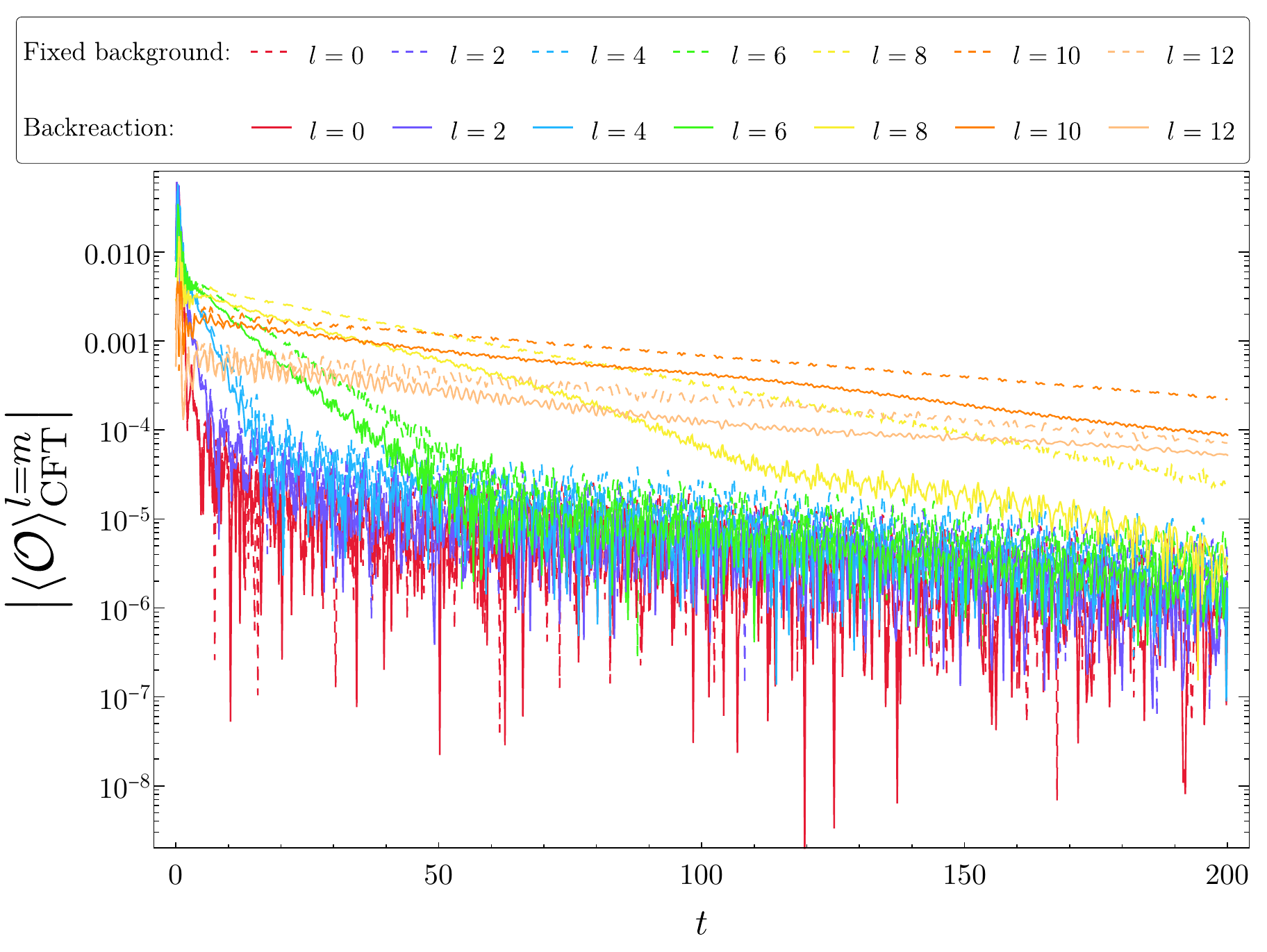}
\parbox{5.0in}{\vspace{0.1cm}\caption{Decay of the first few $l=m$ modes of $\langle\mathcal{O}\rangle_{\text{CFT}}$ in the linear (dashed lines) and backreacting (solid lines) cases respectively. Modes with $l\gtrsim 8$ exhibit significant differences from the linear behavior at late times.
    }\label{fig:scalopdecay}}
\end{figure*}

We perform a decomposition of $\langle \mathcal{O}\rangle_{\text{CFT}}$ into spherical harmonics and we monitor the amplitude of each mode as a function of time. The result of this decomposition is shown in Fig.~\ref{fig:scalopdecay} for the first few $l=m$ modes. In this figure we compare the decay of the various modes in the linear (dashed lines) and fully non-linear (solid lines) cases respectively.\footnote{We note that the oscillations of each mode in Fig.~\ref{fig:scalopdecay} are separated approximately by a light-crossing time, and they can be interpreted as arising when the given scalar mode reaches the AdS boundary.}
We are not displaying modes with $l>12$ since we cannot accurately resolve them with the computational resources at our disposal. The solid lines in Fig.~\ref{fig:scalopdecay} clearly show that, for modes with $l\gtrsim 8$, the decay rates change with time, deviating from the linear behaviour (dashed lines) quite significantly at late times. For instance, initially the $l=8$ mode has the same decay rate as in the linear theory but at around $t\sim 75$ the decay rate becomes faster, indicating the onset of the non-linearities for this mode. 
Other higher $l$ modes exhibit the same qualitative dynamics, with the deviation from linear behavior occurring at later times for higher $l$'s. On the other hand, the decay rates of those modes with $l\lesssim 6$ are essentially the same as in the linear theory. Recall that such modes in the bulk have a strong overlap with the black hole horizon, which implies that modes with low $l$ are not stably trapped for long enough to enter the non-linear regime. We will argue below that the effect of the non-linearities in the high $l$ modes is to transfer energy to the metric. 

\begin{figure*}[t!]
    \centering
    \includegraphics[scale=0.2,clip=true]{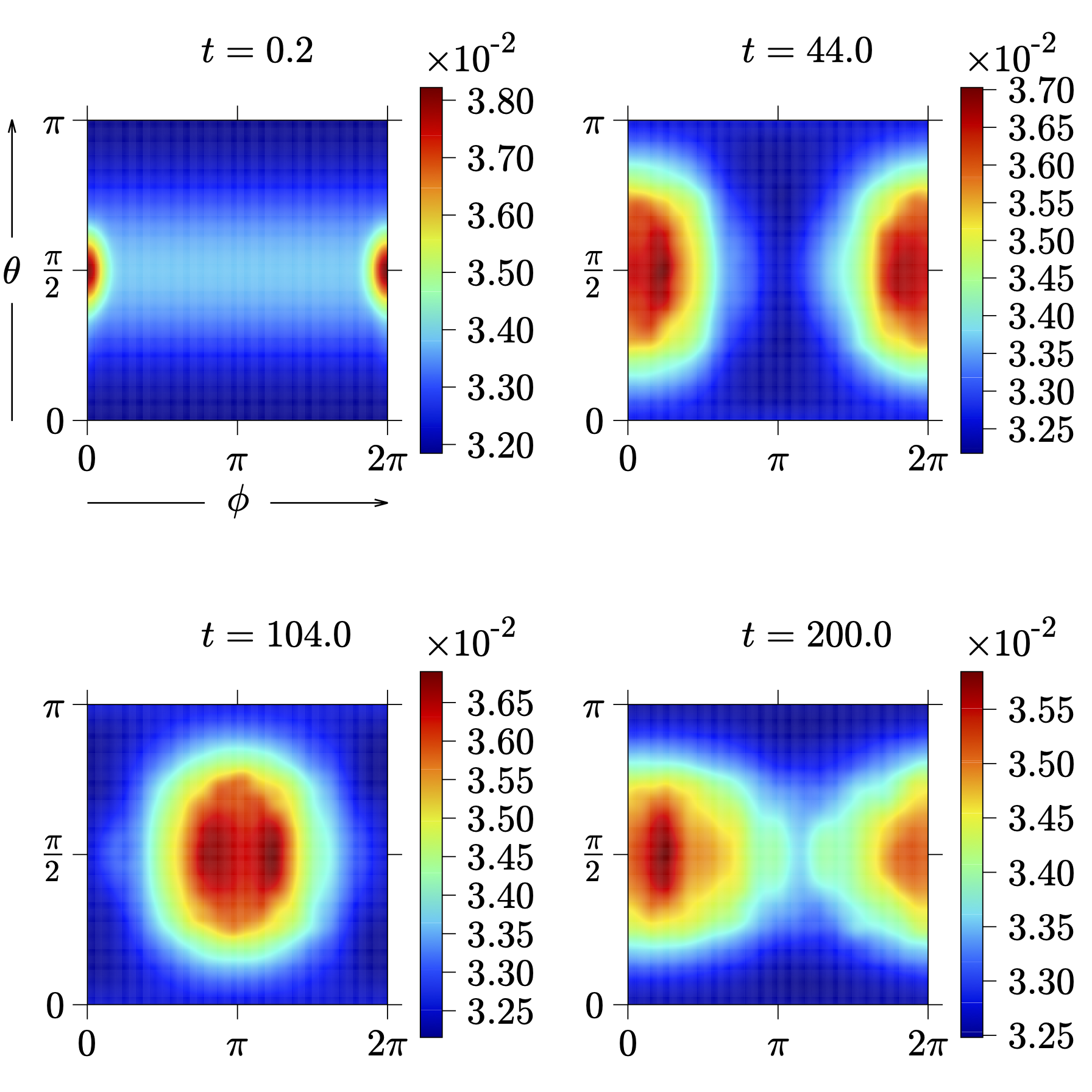}
\parbox{5.0in}{\vspace{0.1cm}\caption{Evolution of boundary energy density $\epsilon_{\text{CFT}}$.
This quantity has a clear dependence on the azimuthal angle on the boundary $S^2$, which is not present in Kerr-AdS.}
    \label{fig:bdy_energy_density_plane_plot}}
\end{figure*}

In Fig.~\ref{fig:bdy_energy_density_plane_plot} we show the profile of the boundary energy density of the CFT, $\epsilon_\text{CFT}$, at four representative times.
Note that $\epsilon_\text{CFT}$ is a particularly relevant quantity for us because it is a direct boundary measure of properties of the bulk geometry.\footnote{Recall that $\epsilon_\text{CFT}$ is defined as (minus) the eigenvalue corresponding to the unique timelike eigenvector of the boundary stress-energy tensor, which is dual to the bulk metric; we could have equivalently displayed $-\langle T^t_{\phantom{t}t}\rangle_{\text{CFT}}>0$ but the two quantities look qualitatively the same because the angular velocity of the black hole is small compared to the speed of light.}
The first panel at $t\sim 0.2$ provides the boundary picture of our choice of initial conditions. One can see a perturbation localized at the equator of the boundary sphere at around $\phi\sim 0$ with amplitude $O(10^{-2})$ on top of a $\phi$--independent (yet dependent on the polar angle $\theta$) background; the latter corresponds to the energy density of the unperturbed background Kerr-AdS black hole.
At subsequent times, the evolution of $\epsilon_\text{CFT}$ is reminiscent of the evolution of the relative Kretschmann scalar described above: the profile rotates in the opposite direction as the expectation value of the boundary scalar operator $\langle \mathcal{O}\rangle_{\text{CFT}}$ around the equator, and it displays the same periodic dynamics as the bulk geometry.\footnote{For animations of the boundary energy density, see \url{https://youtu.be/9BOLqTBPRHE} and \url{https://youtu.be/OT9dYXHOO8c}.} 
Fig.~\ref{fig:bdy_energy_density_plane_plot} shows that $\epsilon_{\text{CFT}}$ has a non-trivial dependence on the azimuthal direction on the boundary $S^2$ at the percent level. Furthermore, this dependence exhibits periodic dynamics and it remains pretty much unaltered from $t\sim 20$, i.e., it does not decay away at late times, at least on the timescale of the simulation. Therefore, the boundary observables unambiguously also show that  there is a persistent, non-axisymmetric deviation from the corresponding Kerr-AdS quantities.

\begin{figure*}[t!]
    \centering
    \includegraphics[width=6.0in,clip=true]{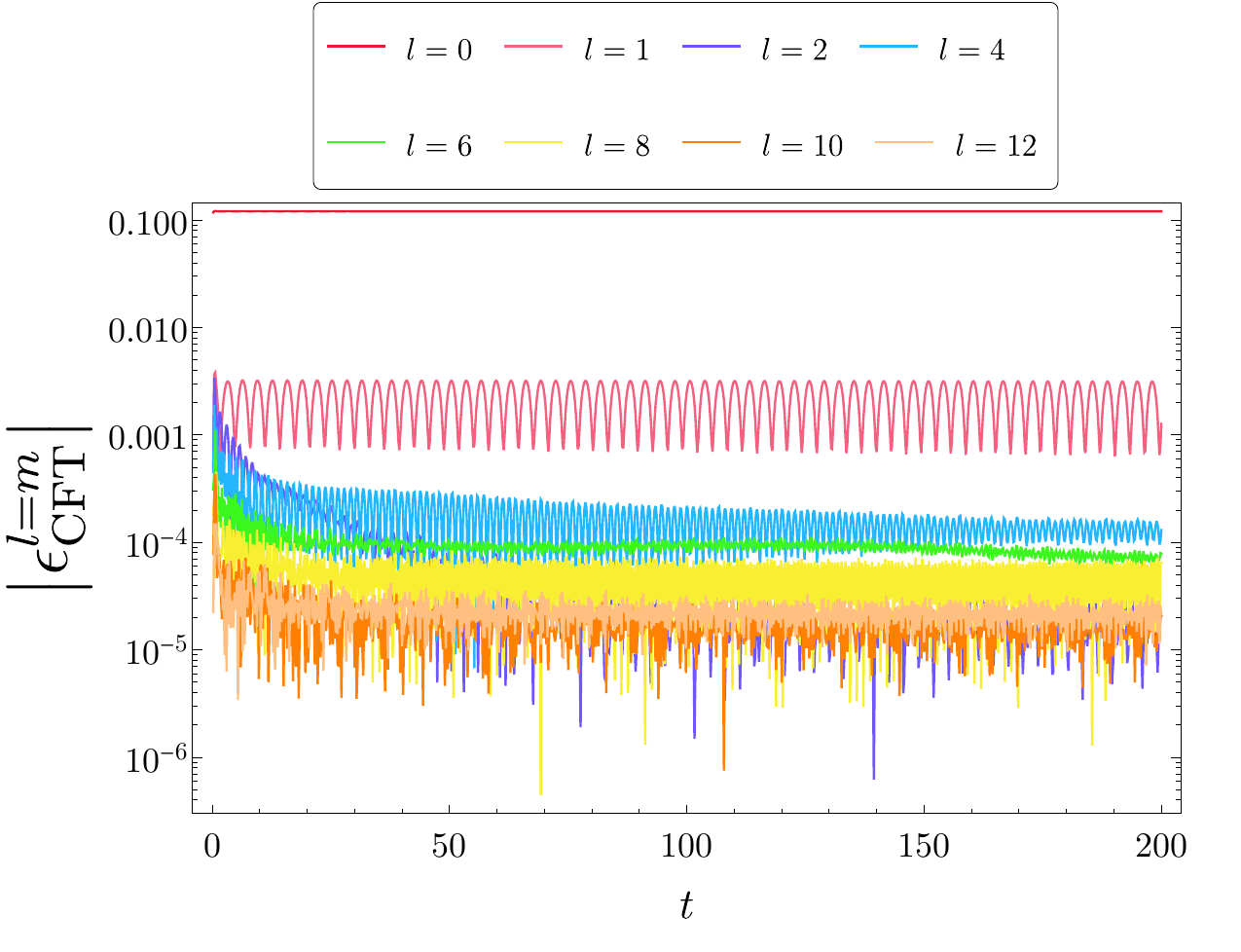}
\parbox{5.0in}{\vspace{0.1cm}\caption{Mode decomposition of boundary energy density $\epsilon_{\text{CFT}}$. The $l=m=1,4,6,8$ modes remain above the level of numerical error and do not decay on the timescale of the simulation, signalling a persistent time-dependent, non-axisymmetric deviation from Kerr-AdS. 
  }\label{fig:bdyendensdecaybackreact}}
\end{figure*}

In Fig.~\ref{fig:bdyendensdecaybackreact} we display the amplitudes, as functions of time, of certain $l=m$ modes in the decomposition of the boundary energy density $\epsilon_\text{CFT}$. Except for $l=m=0$, the modes shown in this figure are not present in Kerr-AdS and hence their amplitudes constitute a direct measurement of the deviations from Kerr-AdS. 
Fig.~\ref{fig:bdyendensdecaybackreact} shows that, with our numerical resolution, the scale of the numerical error in the mode decomposition of the boundary energy density is at around $\sim2\times 10^{-5}$. 
On the scale of this plot, some modes (e.g., $l=m=1,\,4,\,6,\,8$) are well above the numerical noise levels and do not decay on the timescale of the simulation. 
The $l=m=0$ mode exhibits a very slow decay (not visible on the scale of the plot). The slow decay of this mode is consistent with the fact that there is a remnant scalar field in the bulk that is slowly decaying. Since the total energy is conserved (see Fig.~\ref{fig:cons_charges} in Appendix~\ref{sec:conv}), there must be a transfer of energy from this mode to the higher $l$ modes, but this effect appears to be too small to be clearly seen. The $l=m=1$ mode is clearly the dominant non-axisymmetric mode of the boundary energy density, and it exhibits the same behavior as the analogous mode in the deformation of the AH. That is, this mode is excited as soon as the scalar field interacts with the black hole, and its behavior remains essentially unchanged thereafter. On the other hand, the higher $l=m$ modes undergo an initial period of adjustment before they settle to their final, non-zero, state. In particular, the amplitude of the oscillations of $l=m=4$ mode decays very slowly with time and only towards the end of the simulation it seems that this mode has settled. 
It is interesting to note that the $l=m=6$ mode exhibits some non-trivial dynamics at intermediate times, which could be interpreted as a sign of mode interactions. Likewise, there is a hint in our data that the $l=m=8$ is starting to grow at late times. Finally, we note that whilst all boundary scalar modes eventually decay, that is not quite the case for the modes that characterize the energy density (and hence the bulk geometry). In the latter case, some $l=m>0$ modes do not decay at all, which reflects the permanent, non-axisymmetric deviations of the bulk geometry from Kerr-AdS.

\section{Conclusions}
\label{sec:conclusions}

In this article we have studied, using numerical methods, the non-linear evolution of massless scalar field perturbations of a slowly rotating Kerr-AdS black hole as an initial boundary value problem. This was achieved by the use of novel coordinates on Kerr-AdS that are both horizon-penetrating and adapted to the asymptotic geometry.
For concreteness we have considered a Kerr-AdS black hole with parameters $M=0.4$ and $a=0.2$. This black hole rotates slowly in the sense that its angular velocity satisfies $\Omega_H<1$, and hence it is below the Hawking-Reall bound \cite{Hawking:1999dp}. Such black holes are linearly stable. However,~\cite{Holzegel:2011uu,Holzegel:2013kna} showed that scalar perturbations of slowly rotating black holes in AdS decay only logarithmically in time, which led these authors to conjecture that such black holes should all be non-linearly unstable. In this work, we have investigated this conjecture and provided evidence that it is correct.

The main results of the paper are the following. First, we have identified a class of initial conditions for a scalar field on the background of a slowly rotating Kerr-AdS black hole such that, upon evolution,  decay  logarithmically in time at late times, eq.~\eqref{eq:linear_log_decay}. Furthermore, we have explicitly checked that the constant in this equation only depends on the background parameters and not on the parameters that specify the initial scalar field configuration, which is consistent with the theorems of \cite{Holzegel:2011uu,Holzegel:2013kna}. The fact that we do not make any symmetry assumptions  nor we need to fine-tune any parameters in the initial data, suggest that the observed asymptotic logarithmic decay in time may be a feature of generic solutions of the massless Klein-Gordon equation on a fixed slowly rotating Kerr-AdS background. After performing a decomposition of the scalar perturbations in spherical harmonics, the linear decay rates that we compute from our simulations are compatible with the results in the literature \cite{Uchikata:2009zz}. The main conclusion from our numerical studies of the linear problem is that our initial conditions are ``generic'' in the sense that the asymptotic decay of the scalar field that we observe is compatible with the uniform decay estimates proven in  \cite{Holzegel:2011uu,Holzegel:2013kna}.

In our simulations of the full non-linear problem, turning on the backreaction of the scalar field onto the geometry, we have managed to probe timescales of the order $t\sim200 L$, which are roughly two orders of magnitude larger than the AdS light-crossing time and much larger than the timescale set by the black hole mass. 
Including non-linearities, we find that the dynamics of $l=m$ scalar modes  with $l\lesssim 6$ is similar to the linear problem, which supports the fact that our initial data is suitably small. On the other hand, modes with $l\gtrsim 8$ are stably trapped between a certain region of the exterior of the black hole and the AdS boundary, and their late time dynamics differs from the linear problem. 
The precise location and size of this region depends on the harmonic under consideration and it changes with time as the low $l$ modes eventually decay.
The modes that experience stable trapping are long lived; in fact, we have verified that the lifetime of each individual mode grows exponentially with $l$. 
The fact that many modes remain in the same region of the spacetime for sufficiently long times leads  to non-linear effects when we turn on the interactions. We have unambiguously observed two types of non-linearities. First,  for individual scalar modes with $l\gtrsim 8$, the decay  rates deviate from the linear behavior at late times, becoming faster. The time when this happens depends on the mode $l$ under consideration; the larger the $l$, the later the time.  This phenomenon suggests that there is transfer of energy to the higher $l$ modes as well as to the spacetime geometry. Second, the interaction between the scalar field and the metric results in a time-dependent and non-axisymmetric spacetime geometry that differs from that of the Kerr-AdS black hole. 

To study the geometry of the spacetime, we have first considered the deformations of the AH away from Kerr-AdS. We have shown that the main deformation of the horizon $S^2$ is due to an $m=1$ mode, but higher $m$ modes are also present even though they are two orders of magnitude smaller. Interestingly, whilst the evolution of the $m=1$ is governed by only one timescale, the dynamics of the higher $m$ deformations is characterized by two timescales. These time-depdendant, non-axisymmetric deviations from Kerr-AdS are not a gauge artefact, as confirmed by the evolution of relative Kretschmann scalar as a function of time. The $L^2$ norm of this quantity oscillates around a central, finite value with the same two timescales as the deformations of the AH. Hence, we conclude that the geometry is not decaying to a nearby slowly rotating Kerr-AdS black hole. Since there is nothing special about our choice of parameters of the background black hole other than it is slowly rotating and our initial data can be regarded as a small perturbation, this result implies that slowly rotating Kerr-AdS  black holes are generically non-linearly unstable to non-axisymmetric deformations. 

From the boundary perspective, we measure the deviation from the Kerr-AdS geometry by performing a harmonic decomposition of the expectation value of the operator dual to the bulk scalar field, $\langle \mathcal{O}\rangle_{\text{CFT}}$, and of the boundary energy density, $\epsilon_{\text{CFT}}$. We observe that the $l=m$ modes of $\langle \mathcal{O}\rangle_{\text{CFT}}$ with $l\gtrsim 8$ exhibit non-linear behavior at late times. Moreover, all scalar modes that we could accurately extract exhibit decay. On the other hand,  the energy density of the perturbed spacetime is characterized by modes with $l=m\neq0$ that are not present in Kerr-AdS, such as the dominant $l=m=1$ mode, and these modes do not decay with time. This is further evidence that the bulk geometry is not going to settle down to another nearby member of the Kerr-AdS family of black holes. The amplitudes of the $l=m\neq 0$ modes  of the boundary energy density are consistent with the bulk deviations from Kerr-AdS.

The evolution of the perturbed slowly rotating Kerr-AdS black hole presents some similarities with the late time evolution of the superradiant instability of Kerr-AdS. 
Refs.~\cite{Chesler:2018txn,Chesler:2021ehz} showed that the latter proceeds through a sequence of spacetimes that are close to the known time-periodic black resonators \cite{Dias:2015rxy}, developing structure at shorter scales in the region far from the central black hole on increasingly longer timescales. Assuming that in the slowly rotating case the evolution proceeds in this way ad infinitum, it would imply that it takes an infinite amount of time to produce a naked singularity \cite{Niehoff:2015oga}.\footnote{This would constitute a violation of the weak cosmic censorship conjecture, at least in spirit, since arbitrarily large curvatures would become accessible to asymptotic observers that wait long enough.} This is qualitatively similar to what happens when one perturbs three-dimensional AdS with initial data whose total energy is below the mass of the first BTZ black hole: in a suitable norm, the spacetime becomes rougher as the evolution proceeds but it takes an infinite classical time to form a singularity \cite{Bizon:2013xha}.\footnote{This should be contrasted with other instabilities, such as the Gregory-Laflamme instability \cite{Gregory:1993vy,Lehner:2010pn,Figueras:2015hkb,Figueras:2017zwa,Bantilan:2019bvf,Andrade:2020dgc,Figueras:2022zkg}, which leads to the formation of a naked singularity in finite asymptotic time.} Eventually, quantum effects would become important and their backreaction on the geometry could no longer be ignored. Recently, \cite{Kim:2023sig} has proposed a possible end state of the superradiant instability of Kerr-AdS black holes that consists of a central black hole with horizon angular velocity $\Omega_H=1$, co-rotating at the speed of light with a thermal gas of gravitons that is localized on a thin equatorial disk. These geometries were dubbed `gray galaxies'.
The results of our simulations do not rule out the possibility that the evolution of the perturbed slowly rotating Kerr-AdS black holes proceeds in a qualitatively similar manner as the superradiant instability, possibly on even longer timescales. So it may be possible that there exist other types of gray galaxies with $\Omega_H<1$, although they would not be thermodynamically preferred (in the grand canonical ensemble). 

In our case, we have not been able to identify a suitable norm that grows with time and that could characterize a further instability. 
While there appears to be a transfer of energy from the scalar field to the metric towards small scales, we do not see an unambiguous sign of a strong UV cascade within the geometry itself in the region outside the black hole.
However, there are signs that the cascade might occur on longer timescales, which we could not probe with our limited computational resources. 
For instance, we do observe non-trivial dynamics in the modes that characterise the boundary energy density, and there are hints in our data that the amplitude of the highest modes (e.g., $l=m=8$) may be growing at late times. Furthermore, we find no evidence that a naked singularity may form in the timescales that we have been able to study.

Another possibility that we cannot rule out at this stage is that the spacetime settles down to a new type of slowly rotating black resonator that would be stable. Our simulations indicate that the evolution of the bulk geometry is governed by two frequencies, so one could expect that, generically, these new solutions would also involve multiple frequencies, similar to the (spherically symmetric) multi-oscillating boson stars that appear to be non-linearly stable \cite{Choptuik:2019zji} and the multi-oscillating black resonators in five spacetime dimensions \cite{Ishii:2021xmn,Ishii:2020muv}. These putative new black holes are presently unknown, and would be genuinely different from the black resonators constructed in \cite{Dias:2015rxy} since their horizon angular velocity would satisfy $\Omega_H<1$. 

There are many interesting extensions of the present work. First, one would need to run longer and more accurate simulations to conclusively determine whether the geometry settles down to a new type of black resonator or it exhibits a runaway cascade to the UV. The latter should translate into a certain norm that is sensitive to high enough derivatives growing with time. Ideally one should also use constraint-preserving initial data, which would allow for larger scalar field amplitudes and velocities.
A large amplitude can be expected to lead to ``faster" and  ``stronger'' dynamics, with significant focusing and potentially caustic formation outside of the apparent horizon.
Our scheme is well-suited to handle such situations, as opposed to the characteristic scheme of \cite{Chesler:2013lia}.
Ref. \cite{Dias:2012tq} argued that smooth perturbations of slowly rotating black holes in AdS should be non-linearly stable but perturbations of low differentiability in sufficiently high dimensions could be unstable. The initial data that we used in the simulations presented in this paper is smooth and yet the evolution of the spacetime does not seem to settle down to Kerr-AdS. 
We have identified $C^2$ initial data for the scalar field that exhibits logarithmic decay at the linear level and we hope to study its non-linear evolution in future work. Preliminary results suggest that the non-linear evolution of the $C^2$ initial data is qualitatively similar to that of the smooth initial data presented in this paper, which should not be surprising since for us $D=4$ while \cite{Dias:2012tq} requires $D\geq 12$; see footnote~\ref{ftn:noinst}. 

Finally, we have studied  a scalar field coupled to gravity because it is the system considered in \cite{Holzegel:2011uu,Holzegel:2013kna} and it is convenient to induce small deformations of the spacetime geometry. However, our simulations show that the late dynamics of the system is dominated by the metric while the scalar field decays asymptotically. Therefore, it should be possible to directly observe that suitable non-axisymmetric metric perturbations of slowly rotating Kerr-AdS black holes evolve into a new type of dynamical black hole and do not decay at late times, without the need of introducing matter fields that induce such metric perturbations.

\subsection*{Acknowledgements}

We would like to thank Gustav Holzegel for reading an earlier version of this paper and for providing critical comments. PF would like to thank Bob Wald for discussions. PF would also like to thank the Enrico Fermi Institute and the Department of Physics of the University of Chicago for their hospitality in the final stages of this work. PF and LR were supported by a Royal Society Fellowship URF\textbackslash R\textbackslash201026 and
RF\textbackslash ERE\textbackslash 210291. PF is supported by the STFC Consolidated Grant ST/X000931/1.
This work used the Young Tier 2 HPC cluster at UCL; we are grateful to the UK Materials and Molecular Modelling Hub for computational resources, which is partially funded by EPSRC
(EP/P020194/1 and EP/T022213/1).
Calculations were also performed using the Sulis Tier 2 HPC platform hosted by the Scientific Computing Research Technology Platform at the University of Warwick. Sulis is funded by EPSRC Grant EP/T022108/1 and the HPC Midlands+ consortium. 
PF also acknowledges the use of ARCHER2 UK National Supercomputing Service\footnote{https://www.archer2.ac.uk} under the EPSRC HPC project no. E775 and the CSD3 cluster in Cambridge under Project No. DP128. The Cambridge Service for Data Driven Discovery (CSD3), partially operated by the University of Cambridge Research Computing on behalf of the STFC DiRAC HPC Facility. The DiRAC component of CSD3 is funded by BEIS capital via STFC capital Grants No. ST/P002307/1 and No. ST/ R002452/1 and STFC operations Grant No. ST/R00689X/1. 
This work also used the DiRAC@Durham facility managed by the Institute for Computational Cosmology on behalf of the STFC DiRAC HPC Facility\footnote{www.dirac.ac.uk}. The equipment was funded by BEIS capital funding via STFC capital grants ST/P002293/1, ST/R002371/1 and ST/S002502/1, Durham University and STFC operations grant ST/R000832/1. DiRAC is part of the National e-Infrastructure. 
This research also utilised Queen Mary's Apocrita HPC facility, supported by QMUL Research-IT \cite{apocrita}.
For the purpose of Open Access, the author has applied a CC BY public copyright licence to any Author Accepted Manuscript version arising from this submission.

\appendix
\setcounter{tocdepth}{1}
\numberwithin{equation}{section}

\section{Review of Cauchy evolution scheme}
\label{sec:Causch}

In this appendix we recall the salient points about the scheme of \cite{Bantilan:2020xas} to evolve asymptotically AdS spacetimes as an initial boundary value problem in complete generality. 
The adaptation of this scheme to the evolution of generic deviations from Kerr-AdS, satisfying the same set of boundary conditions, is discussed in Section~\ref{sec:setup}. For simplicity, in the following we consider 3+1 dimensional spacetimes, but the generalization to higher dimensions is straightforward.

We work with the generalised harmonic (GH) formulation of the Einstein equations.
This well-posed formulation was originally employed to prove the first theorems about existence and uniqueness of solutions to the Einstein equations in asymptotically flat spaces \cite{Foures-Bruhat:1952grw,Choquet-Bruhat:1969ywq}.
It was then used in the first breakthrough simulations of black hole binary mergers in asymptotically flat spacetimes \cite{Pretorius:2004jg}, and later in asymptotically AdS spacetimes \cite{Bantilan:2012vu,Bantilan:2017kok,Bantilan:2020pay,Bantilan:2020xas}.
In this formulation one prescribes wave equations with sources to evolve the spacetime coordinates $x^\mu$, 
\begin{equation}
\label{eq:GHcoords}
  \Box_g x^\mu=H^\mu\,,
\end{equation}
where $\Box_g$ is the scalar wave operator. The choice of source functions $H^\mu$ corresponds to the choice of coordinates.\footnote{Notice that, in spite of the notation, $H^\mu$ are not the components of a tensor.}

In the case of coupling to a massless scalar field, the equations of motion of the Einstein-scalar field theory in GH coordinates are
\begin{align}
\hspace{2cm}-\tfrac{1}{2}g^{\sigma\delta}g_{\mu\nu,\sigma\delta}&-g^{\sigma\delta}_{\phantom\sigma\phantom\delta,(\mu}g_{\nu)\sigma,\delta}-H_{(\mu,\nu)}+H_\sigma\Gamma^\sigma_{\phantom\sigma\mu\nu}-\Gamma^\sigma_{\phantom\sigma\delta\mu}\Gamma^\delta_{\phantom\delta\sigma\nu}&\nonumber\\
&-\kappa\left(2n_{(\mu}C_{\nu)}-(1+P)n^\sigma C_\sigma\right)&\nonumber\\
&=\Lambda g_{\mu\nu}+8\pi\left(T_{\mu\nu}-\tfrac{1}{2}T g_{\mu\nu}\right)\,, &\label{eq:evoeqg}\\
&  g^{\mu\nu} \partial_{\mu} \partial_{\nu} \varphi -g^{\mu\nu} \Gamma^\sigma_{\phantom\sigma\nu\mu}\partial_\sigma\varphi= 0\,, &\label{eq:evoeqphi}
\end{align}
where $\kappa$ and $P$ are constants, $n^\mu$ is the future-pointing unit normal to hypersurfaces of constant $x^0$, and
\begin{equation}
  C^\mu:=\Box_g x^\mu-H^\mu=0\,
  \label{eq:GHconstr}
\end{equation} 
are the constraints.
In the continuum limit and for initial data that satisfies the Einstein constraints together with $C_\mu=0$ initially, the constraints continue to be zero at all later times. 
However, that is not the case in a numerical simulation due to the unavoidable truncation errors. 
Following the idea of \cite{Gundlach:2005eh}, constraint-damping terms have been added in the second line of \eqref{eq:evoeqg} to ensure that the constraint violations are exponentially suppressed during the evolution. In our simulations, we choose $\kappa=-10$ and $P=-1$.

When applying these ideas to evolve asymptotically AdS spacetimes, one also needs to specify boundary conditions at null infinity throughout the evolution, thus turning the Cauchy problem into an initial-boundary value problem. 
We consider the reflective boundary conditions mentioned in Section~\ref{sec:setup}.
In this setting, it turns out that not every choice of source functions $H^\mu$ leads to long-time stable numerical evolutions, as originally noted in \cite{Bantilan:2012vu}. 
A prescription to obtain a stable gauge choice was proposed in \cite{Bantilan:2020xas} and 
it can be applied to all asymptotically AdS spacetimes with reflective boundary conditions, without imposing symmetries on the solution.
For concreteness, the prescription is presented for a class of coordinates $x^\mu=(t,x,y,z)$, called asymptotically Cartesian;
these take values in $\mathbb{R}\times (-1,\,1)^3$ with the constraint $\rho(x,y,z):=\sqrt{x^2+y^2+z^2}<1$, where $\rho=1$ corresponds to the location of the AdS boundary.
These coordinates are uniquely defined near the AdS boundary as the coordinates in which any solution to the Einstein-scalar field equations with reflective boundary conditions approaches pure AdS as
\begin{align}
\label{eq:gadsfalloffs}
  g_{\mu\nu}&=g^{\text{AdS}}_{\mu\nu}+O\left(1-\rho\right),\\
  \label{eq:phiadsfalloffs}
  \varphi&=O\left((1-\rho)^3\right).
\end{align}
Here $g^{\text{AdS}}_{\mu\nu}=\frac{4}{(1-\rho^2)^2
}\diag(-\frac{(1+\rho^2)^2}{4},1,1,1)$ is the metric of pure AdS. 
The near-boundary source functions that ensure $C_\mu=0$ to leading order in asymptotically Cartesian coordinates are then given by
\begin{equation}
  \label{eq:Hadsfalloffs}
  H_\mu=H^{\text{AdS}}_\mu+O\left((1-\rho)^2\right),
\end{equation}
where $H^{\text{AdS}}_\mu=\frac{2(3+\rho^2)}{1-\rho^4}(0,x,y,z)$ are the pure AdS source functions.

The prescription introduced in \cite{Bantilan:2020xas} for a choice of source functions $H_\mu$ at the next order in the $1-\rho$ expansion that leads to stable evolutions proceeds as follows. Define barred variables $\bar{g}_{\mu\nu}$, $\bar\varphi$, and $\bar{H}_\mu$ 
as the leading order deviations of $g_{\mu\nu}$, $\varphi$ and $H_\mu$, respectively, from their pure AdS values, and strip the result of as many factors of $(1-\rho^2)$ as needed so that each quantity falls off linearly in $(1-\rho)$ near the AdS boundary.
From the fall-offs \eqref{eq:gadsfalloffs}--\eqref{eq:Hadsfalloffs}, we see that the barred variables in asymptotically Cartesian coordinates are given by
\begin{align}
\bar{g}_{\mu\nu}&=g_{\mu\nu}-g^{\text{AdS}}_{\mu\nu},\label{eqn:evogads}\\
\bar\varphi&=\frac{\varphi}{(1-\rho^2)^2}\label{eqn:evophiads},\\
\bar H_\mu&=\frac{H_\mu - H^{\text{AdS}}_\mu}{1-\rho^2}\label{eqn:evoHads}.
\end{align}
In terms of these variables, the reflective boundary conditions become simple Dirichlet boundary conditions:
\begin{equation}
\label{eq:dirreflcondAdS}
  \bar{g}_{\mu\nu}|_{\rho=1}=0\,,\quad \bar{\varphi}|_{\rho=1}=0.
\end{equation} 

Next, consider the near-boundary expansion of the barred quantities, $\bar{g}_{\mu\nu}$, $\bar\varphi$, and $\bar{H}_\mu$. By construction, we have
\begin{align}
\label{eqn:qexpg}
\bar g_{\mu \nu} &= \bar g_{(1) \mu \nu} (1-\rho) + \bar g_{(2) \mu \nu} (1-\rho)^2 + O\left((1-\rho)^3\right), \\
\label{eqn:qexpphi}
\bar \varphi &= \bar \varphi_{(1)} (1-\rho) + \bar \varphi_{(2)} (1-\rho)^2 + O\left((1-\rho)^3\right),\\
\label{eqn:qexpH}
\bar H_{\mu} &=\bar H_{(1) \mu} (1-\rho) + \bar H_{(2) \mu} (1-\rho)^2 + O\left((1-\rho)^3\right).
\end{align}
We use these to compute the near-boundary expansion of the evolution equations \eqref{eq:evoeqg}--\eqref{eq:evoeqphi} and the GH constraints \eqref{eq:GHconstr}.
The leading order of \eqref{eq:evoeqg} and \eqref{eq:GHconstr} give a set of linear algebraic equations involving $\bar g_{(1)\mu\nu}$ and $\bar H_{(1)\mu}$.\footnote{$\bar \varphi_{(1)}$ appears at higher orders in these expansions, as a consequence of the fact that our reflective boundary conditions set the slowly-decaying scalar solution to zero, and allow only for a fastly-decaying solution.}
The stable gauge prescription consists, in essence, of choosing $\bar H_{(1)\mu}$ by solving this system of equations.
The result for asymptotically Cartesian coordinates is
\begin{equation}\label{eqn:target_gauge_txyz}
\begin{aligned}
\bar H_{(1)t}&=\frac{3}{2\sqrt{x^2+y^2+z^2}}(x \, \bar g_{(1)tx}+y \, \bar g_{(1)ty}+z \, \bar g_{(1)tz}),\\
\bar H_{(1)x}&=\frac{3}{2\sqrt{x^2+y^2+z^2}}(x \, \bar g_{(1)xx}+y \, \bar g_{(1)xy}+z \, \bar g_{(1)xz}),  \\
\bar H_{(1)y}&=\frac{3}{2\sqrt{x^2+y^2+z^2}}(x \, \bar g_{(1)xy}+y \, \bar g_{(1)yy}+z \, \bar g_{(1)yz}), \\
\bar H_{(1)z}&=\frac{3}{2\sqrt{x^2+y^2+z^2}}(x \, \bar g_{(1)xz}+y \, \bar g_{(1)yz}+z \, \bar g_{(1)zz}).
\end{aligned}
\end{equation}
The remaining equations and their first derivatives ensure tracelessness and conservation of the energy-momentum tensor of the boundary CFT.
The values of the higher order coefficients in \eqref{eqn:qexpH}, and essentially the value of $H_\mu$ deeper into the bulk, are, in principle, arbitrary. In practice we set $H_\mu$ to smoothly transition to zero deep in the bulk, so we effectively impose harmonic gauge there.

Given this choice of $\bar H_{\mu}$, the evolution equations \eqref{eq:evoeqg}--\eqref{eq:evoeqphi}, expressed in terms of the barred variables, can be solved with the reflective boundary conditions \eqref{eq:dirreflcondAdS} for $\bar{g}_{\mu\nu}$ and $\bar\varphi$.
The full metric $g_{\mu\nu}$, scalar field $\varphi$ and source functions $H_\mu$ can then be readily obtained by inverting the relations \eqref{eqn:evogads},\eqref{eqn:evophiads} and \eqref{eqn:evoHads}, respectively.

Although there is no rigorous proof, the empirical fact that this prescription leads to stable evolutions can be intuitively understood as follows.
Notice that we have essentially chosen $\bar H_{(1)\mu}$ in such a way that the evolution equation for the metric \eqref{eq:evoeqg} is analytically satisfied to leading order in the near-boundary expansion once all the constraints have been accounted for.
Now, the Fefferman-Graham theorem states that the solution at any order near the boundary is fully determined in terms of the boundary data, specified by the boundary conditions, and the data at order $(1-\rho)$ in the metric, thus order $(1-\rho)^2$ in the source functions, and at order $(1-\rho)^3$ in the massless scalar field.\footnote{To see this explicitly, one needs to convert the asymptotic form of the metric \eqref{eq:gadsfalloffs} and scalar field \eqref{eq:phiadsfalloffs} from asymptotically Cartesian coordinates to Fefferman-Graham coordinates, in terms of which the Fefferman-Graham theorem is typically stated.
This change of coordinates is discussed in detail in \cite{Bantilan:2020xas}.}
The data for $g_{\mu\nu}$ and $H_\mu$ at these orders, i.e., $\bar g_{(1)\mu\nu}$ and $\bar H_{(1)\mu}$ respectively, are precisely the quantities that appear in the leading order of \eqref{eq:evoeqg}.
It is therefore not surprising that, once that order of \eqref{eq:evoeqg} is solved analytically as per our prescription, the solution at higher orders can be determined numerically without stability issues.

\section{Dual CFT observables, energy and angular momentum}
\label{sec:CFTquants}

According to the AdS/CFT dictionary \cite{Maldacena:1997re,Gubser:1998bc,Witten:1998qj}, the bulk massless scalar field $\varphi$ is dual to a scalar operator of scaling dimension $\Delta_+=3$, and the metric $g$ is dual to the energy-momentum tensor of the CFT.
At each time step $t=t_n$ in our simulations, we compute some of the observables that characterise the physics of the boundary CFT, such as the expectation value of the dual scalar operator and the expectation value of the boundary energy-momentum tensor.
We also compute the total energy and angular momentum of the spacetime from the boundary observables.
In this appendix we give the analytic expressions of these quantities.

To present these expressions, it is convenient to use coordinates adapted to the asymptotic geometry in the sense of Appendix~\ref{sec:Causch}.
Specifically, we consider the class of coordinates $x^\alpha=(t,\rho,\theta,\phi)\in(-\infty,+\infty)\times(0,1)\times(0,\pi)\times(0,2\pi)$ for which the AdS boundary is at $\rho=1$, and any solution of the Einstein-scalar field equations satisfying reflective boundary conditions approaches pure AdS near the boundary as
\begin{align}
\label{eq:gadsfalloffssph}
g_{\alpha\beta}&=g^{\text{AdS}}_{\alpha\beta}+O\left(1-\rho\right),\\
  \label{eq:phiadsfalloffssph}
  \varphi&=O\left((1-\rho)^3\right)\,.
\end{align}
Here $g^{\text{AdS}}=\frac{4}{(1-\rho^2)^2}\left(-\frac{(1+\rho^2)^2}{4},1,\rho^2,\rho^2\sin^2\theta\right)$ is the metric of pure AdS in the compactified spherical coordinates.
We refer to $x^\alpha$ as asymptotically spherical coordinates.\footnote{Notice that the definition of asymptotically spherical coordinates used here differs from the definition of \cite{Bantilan:2020xas}, as can be easily seen by comparing \eqref{eq:gadsfalloffssph} and eq. (II.9) of \cite{Bantilan:2020xas}.}
Given any set of coordinates belonging to the class of asymptotically Cartesian coordinates $x^\mu=(t,x,y,z)$ defined in Appendix~\ref{sec:Causch}, a set of asymptotically spherical coordinates can be obtained from the transformation $x=\rho\sin\theta\cos\phi,y=\rho\sin\theta\sin\phi, z=\rho \cos\theta$.
The frames $(t,\rho,\theta,\phi)$ and $(\mathcal{T},\zeta,\chi,\psi)$ in Kerr-AdS defined in Section~\ref{sec:horpenbackgr} and Appendix~\ref{sec:setKAdS} respectively, are two examples of asymptotically spherical coordinates.
Let us also denote by $x^a=(t,\theta,\phi)$ the coordinates on hypersurfaces $\partial \mathcal{M}_q$ at fixed $q:=1-\rho$ for small $q>0$. The AdS boundary coincides with $\partial \mathcal{M}_0$, and the conformal boundary metric $g_{(0)}$ in coordinates $x^a$ is given by \eqref{eq:ESU}.
Since the metric and scalar field in coordinates $x^\alpha$ approach pure AdS at the same rate as in asymptotically Cartesian coordinates $x^\mu$, the appropriate barred variables for asymptotically spherical coordinates, $\bar g_{\alpha\beta}$, are defined by \eqref{eqn:evogkads}--\eqref{eqn:evoHkads} after replacing the Cartesian indices $\mu,\nu,\dots$ with the spherical indices $\alpha,\beta,\dots$.

In order to obtain the numerical values of the boundary observables, given below in terms of $\bar g_{\alpha\beta}$, we will need to obtain $\bar g_{\alpha\beta}$ in terms of the variables $\tilde g_{\mu\nu}$ of our numerical scheme in Section~\ref{sec:KAdSasbackgr}.
This can be done in two simple steps.
First, we obtain the Cartesian barred variables as $\bar g_{\mu\nu}=\tilde g_{\mu\nu}+h_{\mu\nu}$, where $h_{\mu\nu}:=\hat g_{\mu\nu}-g^{\text{AdS}}_{\mu\nu}$ is known analytically.
Then, we observe that $\bar g_{\alpha\beta}$ and $\bar g_{\mu\nu}$ are the components of the tensor $g-g^{\text{AdS}}$, hence they are related by the usual transformation law
\begin{equation}
\label{eq:bargcarttosph}
  \bar g_{\alpha\beta}=\frac{\partial x^\mu}{\partial x^\alpha}\frac{\partial x^\nu}{\partial x^\beta}\, \bar g_{\mu\nu}.
\end{equation}

We are now ready to present the expressions for the boundary quantities.
For the expectation value of the dual scalar operator, following the prescription of \cite{deHaro:2000vlm}, we have
\begin{equation}\label{eq:CFTscal}
\langle \mathcal{O}\rangle_{\text{CFT}}=(2\Delta_+ - 3) \bar\varphi_{(1)},
\end{equation}
where $\bar\varphi_{(1)}$ is defined by the near-boundary expression of $\bar\varphi$ in \eqref{eqn:qexpphi}.
We often refer to $\langle \mathcal{O}\rangle_{\text{CFT}}$ as the boundary scalar operator for simplicity.

To obtain the expectation value of the energy-momentum tensor of the boundary CFT,  $\langle T_{ab}\rangle_{\text{CFT}}$, we follow the prescription of \cite{Balasubramanian:1999re}.\footnote{When considering a solution of the Einstein(-scalar field) equations, this prescription gives the same result as the prescription of \cite{deHaro:2000vlm}, as it should (see the explicit proof in Appendix B.4 of \cite{Bantilan:2020xas}).}
We first compute the quasi-local energy-momentum tensor $^{(q)}T_{\alpha\beta}$ at an arbitrary timelike hypersurface $\partial \mathcal{M}_q$ at some fixed $q=1-\rho$,
\begin{equation}
\label{eq:qslocset}
^{(q)}T_{\alpha\beta}=\frac{1}{8\pi}\biggl(\; \Theta_{\alpha\beta}-\Theta \;\omega_{\alpha\beta}-2\omega_{\alpha\beta}+ G_{\alpha\beta} \biggr),
\end{equation}
where $\Theta_{\alpha\beta}=-\omega^\gamma_{\alpha}\omega^\delta_\beta\nabla_{\gamma}S_{\delta}$ is the extrinsic curvature of $\partial \mathcal{M}_q$, $\omega_{\alpha\beta}=g_{\alpha\beta}-S_\alpha S_\beta$ is the induced metric on $\partial \mathcal{M}_q$, $S^\alpha$ is the spacelike, outward pointing unit vector normal to $\partial \mathcal{M}_q$ and $G_{\alpha\beta}$ is the Einstein tensor of $\partial \mathcal{M}_q$.\footnote{Notice the different sign in the last term of \eqref{eq:qslocset} with respect to \cite{Balasubramanian:1999re}.} 
Restricting to the indices corresponding to the coordinates along the boundary, $x^a$, we obtain the expectation value of the boundary energy-momentum tensor as
\begin{equation}
\label{eq:BalKrabdyset}
\langle T_{ab}\rangle_{\text{CFT}}=\lim_{q\to0}\frac{1}{q} \;^{(q)}T_{ab}\,.
\end{equation}

The boundary energy-momentum tensor can be expressed in terms of the leading order coefficients of the near-boundary expansion of $\bar{g}_{\alpha\beta}$ in powers of $(1-\rho)$, denoted by $\bar{g}_{(1)\alpha\beta}$. We find
\begin{equation}
\label{eq:set_explicit}
\begin{aligned}
\langle T_{tt}\rangle_{\text{CFT}}&=\frac{1}{16\pi}\, \left(2\,\bar{g}_{(1)\rho\rho}+3\,\bar{g}_{(1)\theta\theta}+\frac{3}{\sin^2\theta}\,\bar{g}_{(1)\phi\phi}\right),  \\
\langle T_{t\theta}\rangle_{\text{CFT}}&=\frac{3}{16\pi}\,\bar{g}_{(1)t\theta},  \\
\langle T_{t\phi}\rangle_{\text{CFT}}&=\frac{3}{16\pi}\,\bar{g}_{(1)t\phi},  \\
\langle T_{\theta\theta}\rangle_{\text{CFT}}&=\frac{1}{16\pi}\,\left(3\,\bar{g}_{(1)tt}-2\,\bar{g}_{(1)\rho\rho}-\frac{3}{\sin^2\theta}\,\bar{g}_{(1)\phi\phi}\right),  \\
\langle T_{\theta\phi}\rangle_{\text{CFT}}&=\frac{3}{16\pi}\,\bar{g}_{(1)\theta\phi}, \\
\langle T_{\phi\phi}\rangle_{\text{CFT}}&=\frac{\sin^2\theta}{16\pi}\, \left(3\,\bar{g}_{(1)tt}-2\,\bar{g}_{(1)\rho\rho}-3\,\bar{g}_{(1)\theta\theta}\right).
\end{aligned}
\end{equation}

From these we compute the trace of the energy-momentum tensor, $\langle \text{tr}T\rangle_{\text{CFT}}:=g_{(0)}^{ab} \langle T_{ab}\rangle_{\text{CFT}}$:
\begin{equation}
\label{eq:settrace}
\langle \text{tr}T\rangle_{\text{CFT}}=\frac{3}{8\pi}\left(\bar{g}_{(1)tt}-\bar{g}_{(1)\rho\rho}-\bar{g}_{(1)\theta\theta}-\frac{1}{\sin^2\theta}\bar{g}_{(1)\phi\phi}\,\right).
\end{equation}
Given that the CFT is defined on an odd-dimensional spacetime, $\langle \text{tr}T\rangle_{\text{CFT}}$ should vanish for solutions of the equations of motion.
In addition, the boundary energy-momentum tensor is conserved, i.e., $\mathcal{D}_{(0)}^{\bar{a}}\langle T_{\bar{a}\bar{b}}\rangle_{\text{CFT}}=0$ where $\mathcal{D}_{(0)}$ is the Levi-Civita covariant derivative associated with $g_{(0)}$.

We compute a notion of energy density of the boundary CFT as follows. Assuming that $\langle T_{ab}\rangle_{\text{CFT}}$ satisfies the weak energy condition, we solve the eigenvalue problem, 
\begin{equation}
  \langle {T^a}_{b}\rangle_{\text{CFT}} v^b=\lambda_v v^a
\end{equation}
everywhere on the boundary $S^2$ at a given time, and identify the energy density of the boundary CFT, $\epsilon_{\text{CFT}}$, as minus the eigenvalue associated to the unique (up to rescaling) timelike eigenvector.\footnote{The weak energy condition for an energy-momentum tensor $T_{ab}$ requires that $T_{ab}V^a V^b\geq 0$ for any causal vector $V^a$, at any point. If $\pm \langle T_{ab}\rangle_{\text{CFT}}$ fails to satisfy the weak energy condition, the $L^2$-norm of $\langle T_{ab}\rangle_{\text{CFT}}$, $||\langle T_{ab}\rangle_{\text{CFT}}||_2$, can have complex conjugate pairs of eigenvalues and no real timelike eigenvector, as pointed out in footnote 9 of \cite{Chesler:2013lia}.} Similarly, the boundary anisotropy is given by $\Delta p_{\text{CFT}}:=|p_1-p_2|$, where $p_1$ and $p_2$ are the eigenvalues associated with the remaining two spacelike eigenvectors. We will not discuss the evolution of $\Delta p_{\text{CFT}}$ in this article, as it remains close to its Kerr-AdS value and thus is not particularly informative.

Following \cite{Balasubramanian:1999re}, we can also compute the conserved charges associated with any vector field, $\xi$, generator of the asymptotic symmetry group SO(3,2). 
At any given time $t$, we consider a spacelike two-dimensional surface $\mathcal{S}_q$ on a timelike hypersurface $\partial \mathcal{M}_q$, with induced metric $\sigma_{ab}=\omega_{ab}+u_a u_b$, where $u_a=-N(dt)_a$ is the future pointing unit 1-form normal to $\mathcal{S}_q$ in $\partial \mathcal{M}_q$, and $N$ is the lapse of $\mathcal{S}_q$ regarded as a surface of $\partial \mathcal{M}_q$. The charge associated with $\xi$ is given by
\begin{equation}
\label{eq:charge2}
Q[\xi]=\lim_{q\to0}\int_{\mathcal{S}_q} d\theta d\phi \sqrt{\sigma}\left( ^{(q)}T_{ab} u^a \xi^b\right)\,,
\end{equation}
and is conserved for a solution of the equations of motion \cite{Fischetti:2012rd}.

We can also express the conserved charges in terms of $\bar{g}_{(1)\alpha\beta}$. Here, we are interested in the energy and angular momenta. These are the conserved charges associated, respectively, with the generator of time translations, $k=\frac{\partial}{\partial t}$, and the generators of $SO(3)$ rotations, $m_1=\frac{\partial}{\partial \phi}$, $m_2=-\sin\phi\frac{\partial}{\partial\theta}-\cot\theta\cos\phi \frac{\partial}{\partial \phi}$, $m_3=\cos\phi\frac{\partial}{\partial\theta}-\cot\theta\sin\phi \frac{\partial}{\partial \phi}$.
Plugging these in \eqref{eq:charge2}, we obtain
\begin{equation}
\label{eq:AdSmasscalc}
E:=Q[k]=\int_0^\pi d\theta \int_0^{2\pi}d\phi\,\frac{\sin\theta}{16\pi} \left(2\,\bar{g}_{(1)\rho\rho}+3\,\bar{g}_{(1)\theta\theta}+\frac{3}{\sin^2\theta}\,\bar{g}_{(1)\phi\phi}\right),
\end{equation}
and
\begin{equation}
\label{eq:AdSangmom}
\begin{aligned}
J_1:=-Q[m_1]&=-\int_0^\pi d\theta \int_0^{2\pi}d\phi\, \frac{3\sin\theta}{16\pi}\,\bar {g}_{(1)t\phi}\,,\\\
J_2:=-Q[m_2]&=\int_0^\pi d\theta \int_0^{2\pi}d\phi\,\frac{3\sin\theta}{16\pi}\,\left(\sin\theta\,\sin\phi\,\bar {g}_{(1)t\theta}+\cos\theta\,\cos\phi\, \bar {g}_{(1)t\phi} \right),\\
J_3:=-Q[m_3]&=\int_0^\pi d\theta \int_0^{2\pi}d\phi\,\frac{3\sin\theta}{16\pi}\,\left(-\sin\theta\,\cos\phi\,\bar {g}_{(1)t\theta}+\cos\theta\,\sin\phi\, \bar {g}_{(1)t\phi} \right).
\end{aligned}
\end{equation}

Given that in general the AdS boundary does not lie on points of the Cartesian grid, the numerical values of the boundary observables (including the densities of the charges) at points of the boundary $S^2$ are obtained via extrapolation from the numerical solution at bulk grid points. We employ the third-order extrapolation technique presented in Appendix F of \cite{Bantilan:2020xas}, with some minor adaptations detailed in Appendix E of \cite{Rossi:2022nao}.

\section{AdS norms}
\label{sec:AdSnorms}

Ref.~\cite{Holzegel:2011qj} defines various norms of the scalar field on an arbitrary asymptotically AdS spacetime in terms of coordinates adapted to the geometry near the boundary.
These integrals provide a global measure of the size of the scalar field on a given time slice.
Similar norms, playing the same role, were defined in \cite{Holzegel:2011uu,Holzegel:2013kna} by employing the expressions of \cite{Holzegel:2011qj} and simply replacing the original coordinates with certain horizon-penetrating coordinates that are not adapted to the asymptotic geometry.
These norms were then used to express the global estimates on the decay of a scalar field on a Kerr-AdS background with $\Omega_H<1$.
In this appendix we present the version of these norms that we compute in our simulations to measure the ``size'' of the numerical solution.

Starting from the solution in asymptotically Cartesian coordinates $x^\mu$, we transform it to compactified asymptotically spherical coordinates $x^\alpha=(t,\rho,\theta,\phi)$, defined in Appendix~\ref{sec:CFTquants}. 
Let us denote the location of the apparent horizon on a given $t=\text{const}.$ slice $\Sigma_t$ in these coordinates by $\rho_{\text{AH}}(\theta,\phi)$. The exterior of the apparent horizon is given by $\Sigma_t^{\text{ext}}:=\Sigma_t\cap\left\{\rho>\rho_{\text{AH}}\right\}$.
Let $\sigma_{AB}$ be the metric on an oblate 2-sphere $S^2_{t,\rho}$ at fixed $t$ and $\rho$ in coordinates $x^A=(\theta,\phi)$, and $D_A$ the corresponding Levi-Civita covariant derivative.
For any $s\in\mathbb{R}$, the norms that we employ are given by the following integrals,
\begin{align}
  \label{eq:H0AdS}
  \left|\left|\varphi\right|\right|^2_{H_{\text{AdS}}^{0,s}}(t):=&\int_{\Sigma_t^{\text{ext}}} dR \sin\theta d\theta d\phi\, R^s\varphi^2 R^2 ,&\\
  \label{eq:H1AdS}
   \left|\left|\varphi\right|\right|^2_{H_{\text{AdS}}^{1,s}}(t):=&\int_{\Sigma_t^{\text{ext}}}d R \sin\theta d\theta d\phi\, R^s\left[ R^2 (\partial_R\varphi)^2+\sigma^{AB}D_A\varphi D_B\varphi +\varphi^2 \right]R^2 ,&\\
   \label{eq:H2AdS}
   \left|\left|\varphi\right|\right|^2_{H_{\text{AdS}}^{2,s}}(t):=&\left|\left|\varphi\right|\right|^2_{H_{\text{AdS}}^{1,s}}(t)\nonumber\\
   &+\int_{\Sigma_t^{\text{ext}}} d R \sin\theta d\theta d\phi\, R^s\bigl[ R^4 (\partial_R\partial_R\varphi)^2+R^2\sigma^{AB}\left(D_A\partial_R\varphi\right) \left(D_B\partial_R\varphi\right) \nonumber\\
   &\hspace{4.5cm}+\sigma^{AB}\sigma^{CD} \left(D_A D_C\varphi\right)\left( D_B D_D\varphi\right) \bigr]R^2 .&
\end{align} 
where have used the uncompactified radial coordinate $R:=2\rho/(1-\rho^2)$ to allow for a straightforward comparison with the definitions of \cite{Holzegel:2011qj}. In practice, it is convenient to compute the integral in Cartesian coordinates $(x,y,z)$ as this can be performed by simple and fast algorithms, since the numerical solution is known at equally spaced values of the Cartesian coordinates.\footnote{The complete integrand in \eqref{eq:H0AdS}--\eqref{eq:H2AdS} for the Cartesian integration is obtained by re-writing the measure as $R^2 dR \sin\theta d\theta d\phi=\frac{R^2}{\rho^2} \rho^2 \frac{ dR}{d\rho} d\rho \sin\theta d\theta d\phi=\frac{R^2}{\rho^2} \frac{ dR}{d\rho} dx dy dz$.}

In our simulations we also compute the following ``energy'' norms at $t=0$,
\begin{align}
  \label{eq:E1AdS}
\hspace{-0.15cm}E_1[\varphi]:=&\left|\left|\varphi\right|\right|^2_{H_{\text{AdS}}^{1,0}}(0)+\left|\left|\partial_t\varphi\right|\right|^2_{H_{\text{AdS}}^{0,-2}}(0) ,&\\
  \label{eq:E2AdS}
\hspace{-0.15cm}E_2[\varphi]:=&\left|\left|\varphi\right|\right|^2_{H_{\text{AdS}}^{2,0}}(0)+\left|\left|\partial_t\varphi\right|\right|^2_{H_{\text{AdS}}^{1,0}}(0)+\sum_{i=1}^3 \left|\left|m_i(\varphi)\right|\right|^2_{H_{\text{AdS}}^{1,0}}(0)+\left|\left|\partial_t\partial_t\varphi\right|\right|^2_{H_{\text{AdS}}^{0,-2}}(0),&
\end{align}
defined similarly to \cite{Holzegel:2011uu,Holzegel:2013kna} after replacing their coordinates with our $(t,R,\theta,\phi)$, and where $m_1=\frac{\partial}{\partial \phi}$, $m_2=-\sin\phi\frac{\partial}{\partial\theta}-\cot\theta\cos\phi \frac{\partial}{\partial \phi}$, $m_3=\cos\phi\frac{\partial}{\partial\theta}-\cot\theta\sin\phi \frac{\partial}{\partial \phi}$ are the standard generators of $SO(3)$ rotations.

\section{Kerr-AdS}
\label{sec:appKerrAdS}

In this appendix we review the relevant properties (for our purposes) of the sub-extremal Kerr-AdS spacetime, found by Carter in \cite{Carter:1968ks}.
To avoid confusion, when referring to quantities that are defined elsewhere in the article for a general asymptotically AdS spacetime, we denote their pure Kerr-AdS values by the symbol $\hat{\phantom a}$.

In Boyer-Lindquist (BL) coordinates $(\tau,r,\Theta,\Phi)\in(-\infty,+\infty)\times(r_+,+\infty)\times(0,\pi)\times(0,2\pi)$, the Kerr-AdS metric is given by
\begin{equation}
\label{eq:KerrAdS1}
\hat{g}=-\frac{\Delta}{\Sigma^2}\biggl(d\tau-\frac{a}{\Xi}\sin^2\Theta d\Phi\biggr)^2+
\frac{\Sigma^2}{\Delta}dr^2+\frac{\Sigma^2}{\Delta_\Theta}d\Theta^2+\frac{\Delta_\Theta}{\Sigma^2}\sin^2\Theta\biggl(a d\tau-\frac{r^2+a^2}{\Xi}d\Phi\biggr)^2,
\end{equation}
where
\begin{eqnarray}
\label{eq:fnrule}
&&\Delta=(r^2+a^2)(1+r^2)-2Mr\,, \quad\quad\Delta_\Theta=1-a^2\cos^2\Theta\,,\nonumber\\
&&\Sigma^2=r^2+a^2\cos^2\Theta\,, \quad\quad\Xi=1-a^2,
\end{eqnarray}
The largest real root of the equation $\Delta(r)=0$, denoted by $r_+$, corresponds to location of the outer event horizon.
The sub-extremal range is defined by requiring that the parameters $M,a$ satisfy $M> M_{\text{extr}}(a)$, where the critical mass parameter $M_{\text{extr}}(a)$ is given by
\begin{equation}
\label{eq:KerrAdScondM}
M_{\text{extr}}(a)={\frac{1}{3 \sqrt{6}} \left(2 \left(a^2+1\right)+\sqrt{12a^2+\left(a^2+1\right)^2}\right)
\sqrt{-(a^2+1)+\sqrt{12a^2+\left(a^2+1\right)^2}}}.
\end{equation}
We must also have $a^2<1$ for \eqref{eq:KerrAdS1} to be regular. 
If $a<0$, the redefinition $\Phi\to -\Phi$ leads to the metric \eqref{eq:KerrAdScondM} with parameter $-a>0$. 
Therefore, we can restrict to $a\geq 0$ without loss of generality. 
The cases $a=0$ and $M=a=0$ correspond to Schwarzschild-AdS and pure AdS respectively. 
Kerr-AdS is stationary and axisymmetric: 
$\frac{\partial}{\partial \tau}$ is a stationary Killing vector field and $\frac{\partial}{\partial\Phi}$ is an axial Killing vector field.

Despite allowing to write the Kerr-AdS metric in a relatively simple form, BL coordinates are not well-suited to characterise the near-boundary geometry. The reason is that the large $r$ limit of \eqref{eq:KerrAdS1} depends on $a$, and therefore Kerr-AdS in BL coordinates does not manifestly approach pure AdS. 
As a consequence, $\tau$ and $\Phi$ are not the time and azimuthal angular coordinate measured by stationary observers at the boundary.

These complications do not arise if we use coordinates that are adapted to the asymptotic geometry.
For the Kerr-AdS spacetime, one such frame was originally proposed by Hawking, Hunter and Taylor-Robinson (HHT) in \cite{Hawking:1998kw}. 
It is given by $x^{\dot\alpha}=(\mathcal{T},\mathcal{R},\chi,\psi)\in(-\infty,+\infty)\times(\mathcal{R}_+(\chi),+\infty)\times(0,\pi)\times(0,2\pi)$ defined in terms of BL coordinates by 
\begin{equation}
\label{eq:nonrotnonhorpen}
\begin{aligned}
\mathcal{T}&=\tau,\\
\mathcal{R}^2\cos^2\chi&=r^2\cos^2\Theta,\\
\Xi\, \mathcal{R}^2\sin^2\chi&=(r^2+a^2)\sin^2\Theta,\\
\psi&=\Phi+at.
\end{aligned}
\end{equation}
Explicitly, 
\begin{equation}
\label{eq:nonrotnonhorpenexpl}
\begin{aligned}
\mathcal{T}&=\tau,\\
\mathcal{R}&=\sqrt{r^2+\frac{a^2 \left(r^2+1\right) \sin ^2\Theta }{1-a^2}},\\
\chi&=
\begin{cases}
 \arcsin\left( \sqrt{\frac{a^2+r^2}{a^2 \left(r^2+1\right) \sin
  ^2\Theta +\left(1-a^2\right) r^2}}\sin \Theta\right) & 0 <\Theta\leq \frac{\pi}{2} \\
 \pi -\arcsin\left( \sqrt{\frac{a^2+r^2}{a^2 \left(r^2+1\right) \sin
  ^2\Theta +\left(1-a^2\right) r^2}}\sin \Theta\right) & \frac{\pi}{2} <\Theta<\pi
\end{cases},  \\
\psi&=\Phi+at\,,
\end{aligned}
\end{equation}
and the inverse relations are
\vspace{-0.3cm}
\begin{align}
\label{eq:nonrotnonhorpenexplinv}
\hspace{-0.5cm}&\tau=\mathcal{T},\nonumber \\
\hspace{-0.5cm}&r=\frac{1}{\sqrt{2}}\sqrt{\mathcal{R}^2 (1-a^2 \sin ^2\chi) -a^2+\sqrt{a^4 \mathcal{R}^4 \sin ^4\chi +2 a^2 \mathcal{R}^2 \left(a^2-\mathcal{R}^2-2\right) \sin ^2\chi
  +\left(a^2+\mathcal{R}^2\right)^2}},\nonumber \\
\hspace{-0.5cm}&\Theta=
\begin{cases}
 \arcsin\left(\frac{\sqrt{\mathcal{R}^2 (1-a^2 \sin ^2\chi) +a^2-\sqrt{a^4 \mathcal{R}^4 \sin ^4\chi +2 a^2 \mathcal{R}^2 \left(a^2-\mathcal{R}^2-2\right) \sin ^2\chi
  +\left(a^2+\mathcal{R}^2\right)^2}}}{\sqrt{2} a}\right) & 0 <\chi\leq \frac{\pi}{2} \nonumber \\
 \pi - \arcsin\left(\frac{\sqrt{\mathcal{R}^2 (1-a^2 \sin ^2\chi) +a^2-\sqrt{a^4 \mathcal{R}^4 \sin ^4\chi +2 a^2 \mathcal{R}^2 \left(a^2-\mathcal{R}^2-2\right) \sin ^2\chi
  +\left(a^2+\mathcal{R}^2\right)^2}}}{\sqrt{2} a}\right) & \frac{\pi}{2} <\chi<\pi
\end{cases}, \nonumber \\
\hspace{-0.5cm}&\Phi=\psi-a\mathcal{T}.
\end{align}
The location of the event horizon in these coordinates is given by
\begin{equation}
\label{eq:Rpluschi}
\mathcal{R}_+(\chi)=r_+ \sqrt{\frac{2(a^2+r_+^2)}{a^2 \left(r_+^2+1\right)
  \cos 2 \chi -\left(a^2-2\right)r_+^2+a^2}}.
\end{equation}

For large $\mathcal{R}$, the Kerr-AdS metric in HTT coordinates asymptotes to the pure AdS metric in the standard form
\begin{equation}
\label{eqn:ads}
g^{\text{AdS}}= -\left(1+\mathcal{R}^2\right) d\mathcal{T}^2 + \frac{1}{1+\mathcal{R}^2} d\mathcal{R}^2 +\mathcal{R}^2 d{\Omega}^2 \,
\end{equation}
with $d{\Omega}^2=d\chi^2+\sin^2\chi d\psi^2$. 
More precisely, the components of the deviation from pure AdS, $h:=\hat{g}-g^{\text{AdS}}$, satisfy
\begin{align}
\label{eq:HHth}
&h_{\mathcal{T}\mathcal{T}}=O(\mathcal{R}^{-1}), &h_{\mathcal{T}\mathcal{R}}=0, &\qquad h_{\mathcal{T}\chi}=0, &h_{\mathcal{T}\psi}=O(\mathcal{R}^{-1}), \nonumber\\
&h_{ \mathcal{R}\mathcal{R}}=O(\mathcal{R}^{-5}), &h_{ \mathcal{R}\chi}=O(\mathcal{R}^{-4}), &\qquad h_{\mathcal{R}\psi}=0,\nonumber\\
&h_{ \chi\chi}=O(\mathcal{R}^{-3}), &h_{\chi\psi}=0, \nonumber\\
&h_{\psi\psi}=O(\mathcal{R}^{-1}). 
\end{align}
In HHT frame, the angular velocity of the spacetime is given by $\Omega=-\hat{g}_{\mathcal{T}\psi}/\hat{g}_{\psi\psi}$.
At the horizon, we have
\begin{equation}
\label{eq:HHTangvelhor}
 \Omega_H=\frac{a(1+r_+^2)}{r^2_+ +a^2}\,,
\end{equation}
while the angular velocity at the boundary vanishes, i.e., $ \Omega_{\infty}=0$. The event horizon is a Killing horizon of the Killing vector $\xi=\frac{\partial}{\partial \mathcal{T}}+ \Omega_H\frac{\partial}{\partial \psi}$. Its surface gravity is given by
\begin{equation}
\label{eq;surgra}
\kappa=\frac{r_+(1+r_+^2)}{r_+^2+a^2}-\frac{1-r_+^2}{2r_+}\,,
\end{equation}
and the horizon area is 
\begin{equation}
\label{eq;areahor}
A_H=\frac{4\pi (r_+^2+a^2)}{(1-a^2)}.
\end{equation}

\subsection{Dual CFT observables}
\label{sec:setKAdS}

Let us define the compactified radial coordinate by $\zeta=\left(-1+\sqrt{1+\mathcal{R}^2}\right)/\mathcal{R}$ with $\zeta\in(0,1)$. It is straightforward to verify, using \eqref{eq:HHth}, that the coordinates $(\mathcal{T},\zeta,\chi,\psi)$ are asymptotically spherical, according to the definition of Appendix~\ref{sec:CFTquants}.
Hence, we can use them to obtain the boundary observables for the exact Kerr-AdS spacetime.
Notice that, in the bulk, $(\mathcal{T},\zeta,\chi,\psi)$ clearly differ from the coordinates $(t,\rho,\theta,\phi)$ on Kerr-AdS defined in Section~\ref{sec:horpenbackgr} (for instance, the former set is not horizon-penetrating, while the latter is).
However, these two sets are equivalent near the AdS boundary, $\zeta\to 1$ (or $\rho\to 1$).
This means that the Kerr-AdS boundary quantities take the same form regardless of the set that we choose.
In the rest of this appendix, we write the boundary quantities in $(t,\rho,\theta,\phi)$ coordinates to make contact with the notation of Appendix~\ref{sec:CFTquants}.

We have $\hat{\bar\varphi}_{(1)}=0$ and
\begin{align}
\label{eq:gbar1kads}
&\hat{\bar g}_{(1)tt}=\frac{8 \sqrt{2} M}{\left(2-a^2+a^2 \cos 2 \theta \right)^{5/2}}, \quad \hat{\bar g}_{(1)t\rho}=\frac{8 M}{\left(2-a^2+a^2 \cos 2 \theta \right)^2}, \quad \hat{\bar g}_{(1)t\theta}=0, \nonumber\\ &\hat{\bar g}_{(1)t\phi}=-\frac{8 \sqrt{2} a M \sin ^2\theta}{\left(2-a^2+a^2 \cos 2 \theta \right)^{5/2}}, \nonumber\\
&\hat{\bar g}_{(1)\rho\rho}=\frac{4 \sqrt{2} M}{\left(2-a^2+a^2 \cos 2 \theta\right)^{3/2}}, \quad \hat{\bar g}_{(1)\rho\theta}=0, \qquad \hat{\bar g}_{(1)\rho\phi}=-\frac{8 a M \sin^2 \theta }{ \left(2-a^2+a^2 \cos 2 \theta\right)^2},\nonumber\\
&\hat{\bar g}_{(1)\theta\theta}=0, \quad \hat{\bar g}_{(1)\theta\phi}=0, \nonumber\\
&\hat{\bar g}_{(1)\phi\phi}=\frac{8 \sqrt{2} a^2 M \sin ^4\theta}{\left(2-a^2+a^2 \cos 2 \theta\right)^{5/2}}. 
\end{align}

From these, we obtain the boundary observables with the formulae in Appendix~\ref{sec:CFTquants}. The expectation value of the boundary scalar operator is trivial, $\langle \mathcal{\hat O}\rangle_{\text{CFT}}=0$, and boundary energy-momentum tensor is given by
 \begin{equation}
\label{eq:set_kads_explicit}
\begin{aligned}
\langle \hat T_{tt}\rangle_{\text{CFT}}&=\frac{M \left(4+a^2-a^2 \cos 2 \theta \right)}{2 \sqrt{2} \pi \left(2-a^2+a^2 \cos 2
  \theta\right)^{5/2}},  \\
\langle \hat T_{t\theta}\rangle_{\text{CFT}}&=0, \\
\langle \hat T_{t\phi}\rangle_{\text{CFT}}&=-\frac{3 a M \sin ^2\theta }{\sqrt{2} \pi \left(2-a^2+a^2 \cos 2 \theta
  \right)^{5/2}},  \\
\langle \hat T_{\theta\theta}\rangle_{\text{CFT}}&=\frac{M}{2 \sqrt{2} \pi \left(2-a^2+a^2 \cos 2 \theta \right)^{3/2}},  \\
\langle \hat T_{\theta\phi}\rangle_{\text{CFT}}&=0,  \\
\langle \hat T_{\phi\phi}\rangle_{\text{CFT}}&=\sin ^2\theta \frac{M \left(1+a^2-a^2 \cos 2 \theta \right)}{\sqrt{2} \pi 
  \left(2-a^2+a^2 \cos 2 \theta \right)^{5/2}}.
\end{aligned}
\end{equation}
It is also straightforward to verify that $\langle \hat T_{ab}\rangle_{\text{CFT}}$ is traceless and conserved.
The boundary energy density is given by
\begin{equation}
  \label{eq:CFTendens}
  \hat \epsilon_{\text{CFT}}=\frac{M}{\sqrt{2}\pi\left(2-a^2+a^2\cos2\theta\right)^{3/2}}.
\end{equation}
and boundary anisotropy vanishes, $\widehat{\Delta p}_{\text{CFT}}=0$.

The generator of asymptotic time-translation is given by $k=\frac{\partial}{\partial t}$ and the generators of asymptotic SO(3) rotations are given by $m_1=\frac{\partial}{\partial \phi}$, $m_2=-\sin\phi\frac{\partial}{\partial\theta}-\cot\theta\cos\phi \frac{\partial}{\partial \phi}$, $m_3=\cos\phi\frac{\partial}{\partial\theta}-\cot\theta\sin\phi \frac{\partial}{\partial \phi}$.\footnote{Given the discussion at the beginning of this appendix, it should be clear that the vector field $\frac{\partial}{\partial \mathcal{T}}$ differs from $\frac{\partial}{\partial t}$ in the bulk. However, the two are equivalent at the AdS boundary. The same applies to the other pairs of vector fields, $\frac{\partial}{\partial \psi}$ and $m_1$, $-\sin\psi\frac{\partial}{\partial\chi}-\cot\chi\cos\psi \frac{\partial}{\partial \psi}$ and $m_2$, $\cos\psi\frac{\partial}{\partial\chi}-\cot\chi\sin\psi \frac{\partial}{\partial \psi}$ and $m_3$. For completeness, we note that, in terms of BL coordinates, we have $\frac{\partial}{\partial \mathcal{T}}=\frac{\partial}{\partial \tau}-a\frac{\partial}{\partial \Phi}$ and $\frac{\partial}{\partial \psi}=\frac{\partial}{\partial \Phi}$.}
The conserved charges associated with these asymptotic Killing vector fields, namely the energy, $\hat E$, and angular momenta, $\hat J_1$, $\hat J_2$, $\hat J_3$ (see \eqref{eq:AdSmasscalc} and \eqref{eq:AdSangmom}), are given by
\begin{equation}
\label{eq:kerradsconscharge2}
\hat E=\frac{M}{\Xi^2}, \quad \quad \hat J_1=\frac{aM}{\Xi^2}, \quad \quad \hat J_2=\hat J_3=0.
\end{equation}
Since $\hat J_2$ and $\hat J_3$ vanish, it is common to refer to $\hat J_1$ as the angular momentum of the Kerr-AdS spacetime and denote it by $\hat J$.

It was noted in \cite{Caldarelli:1999xj,Gibbons:2004ai} that the physical notions of energy, angular momentum and angular velocity in Kerr-AdS are precisely those naturally identified by working in any frame adapted to the near-boundary geometry.
In particular, Ref. \cite{Gibbons:2004ai} points out that, under a small variation of the parameters $M,a$ of the Kerr-AdS spacetime, the corresponding variations $\delta A$, $\delta \hat E$ and $\delta \hat J$ satisfy the first law of black hole mechanics, $\delta \hat E=\frac{\kappa}{8\pi}\delta A+\Omega_H\delta \hat J$, only if $\Omega_H$ is the angular velocity of the horizon given by \eqref{eq:HHTangvelhor}, and $\hat E$ and $\hat J$ are the energy and angular momentum given by \eqref{eq:kerradsconscharge2}.

\subsection{Spheroidal harmonics}
\label{subsec:sphharm}

We here review the definition and the basic properties of Kerr-AdS spheroidal harmonics, employed in the discussion of Section~\ref{sec:potbar}. 
More details can be found in \cite{Dafermos:2010hb,Holzegel:2011uu}.

As in Section~\ref{sec:potbar}, given a Kerr-AdS spacetime with parameters $r_+,a$ in Boyer-Lindquist coordinates $(\tau,r,\Theta,\Phi)$, we consider a scalar field of mass $\mu$, satisfying the BF bound $\mu^2\geq -\frac{9}{4}$, with modes of the form $e^{-i\omega \tau} \tilde f^{(\omega)}(r,\Theta,\Phi)$ for a given frequency $\omega$. In general, this frequency is complex, although we restrict to $\omega \in \mathbb{R}$ in our study as explained in Appendix~\ref{sec:Kerr_effective_potential}.
$\tilde f^{(\omega)}(r,\Theta,\Phi)$ can be regarded as a function of $(\Theta,\Phi)$ on the oblate 2-sphere $S^2_{\tau,r}$ at fixed $\tau$ and $r$.
On each $S^2_{\tau,r}$ and for any pair of integers $l,m$ satisfying $l\geq0$ and $-l\leq m \leq l$, we can define the generalisation of spherical harmonics, called spheroidal harmonics and denoted by $S^{(\omega)}_{lm}(\Theta,\Phi)$, as the eigenfunctions of the operator
\begin{equation}
\label{eq:sphharmop}
  P_{\mu}^{(\omega)}=  \begin{cases*}
   P^{(\omega)}-\mu^2 a^2\sin^2\Theta & if $\mu^2 \leq 0$ \\
   P^{(\omega)}+\mu^2 a^2\cos^2\Theta & if $\mu^2 > 0$
  \end{cases*},
\end{equation}
where
\begin{equation}
  -P^{(\omega)}:=\frac{1}{\sin\Theta}\partial_\Theta\left(\Delta_\Theta\sin\Theta \partial_\Theta \right)+\frac{\Xi^2}{\Delta_\Theta}\frac{1}{\sin^2\Theta}\partial^2_\Phi+\Xi\frac{a^2\omega^2}{\Delta_\Theta}\cos^2\Theta-2 i a^3\omega \frac{\Xi}{\Delta_\Theta}\cos^2\Theta \partial_\Phi.
\end{equation}
The $\Phi$ dependence of $S^{(\omega)}_{lm}(\Theta,\Phi)$ is given by $S^{(\omega)}_{lm}(\Theta,\Phi)=s^{(\omega)}_{lm}(\Theta)e^{im\Phi}$.
We denote the corresponding eigenvalues, called separation constants, by $\lambda^{(\omega)}_{lm}$.
In the simple case of a Schwarzschild-AdS black hole ($a=0$), $-P_{\mu}^{(\omega)}$ reduces to the Laplacian of the unit round 2-sphere. Hence, the spheroidal harmonics $S^{(\omega)}_{lm}(\Theta,\Phi)$ reduce to the usual spherical harmonics $Y_{lm}(\Theta,\Phi)$, and the corresponding eigenvalues $\lambda_{lm}^{(\omega)}$ are simply $l(l+1)$.
However, in the rotating case $a\neq 0$, there are no explicit analytic expressions for $S_{lm}^{(\omega)}(\Theta,\Phi)$ and $\lambda^{(\omega)}_{lm}$. 
These quantities can be computed numerically (for example, with the methods of \cite{Uchikata:2009zz} and \cite{Cardoso:2013pza}).

Spheroidal harmonics $S^{(\omega)}_{lm}(\Theta,\Phi)$ form a complete basis for smooth functions on $S^2_{\tau,r}$, i.e., any function $\tilde f^{(\omega)}(r,\Theta,\Phi)$ can be decomposed in spheroidal harmonics as
\begin{equation}
\label{eq:decsphharm}
\tilde f^{(\omega)}(r,\Theta,\Phi)=\sum_{lm} f_{lm}^{(\omega)}(r) S_{lm}^{(\omega)}(\Theta,\Phi).
\end{equation}
Moreover, $S^{(\omega)}_{lm}(\Theta,\Phi)$ are orthonormal, i.e.,
\begin{equation}
\label{eq:orthonormsphharm}
\int_{0}^{\pi} d\Theta \int_{0}^{2\pi}d\Phi\, \sqrt{\hat\sigma}\, \big(S_{l'm'}^{(\omega)}(\Theta,\Phi)\big)^\star S_{lm}^{(\omega)}(\Theta,\Phi)=\delta_{ll'}\delta_{mm'},
\end{equation}
where $\hat\sigma=\sin^2\Theta$ is the determinant of the metric on $S^2_{\tau,r}$ ($\star$ denotes complex conjugation).

The result used in Section~\ref{subsec:potbarkerrads} is 
\begin{equation}
\label{eq:estlambda}
 \lambda_{lm}^{(\omega)}\geq \Xi^2 l(l+1)-a^2\omega^2.
\end{equation}
We prove this by following the proof of Lemma 5.1 in \cite{Holzegel:2011uu}.
Consider the operator $P_{\mu}^{(\omega)}+a^2\omega^2$, whose eigenvalues are $\lambda_{lm}^{(\omega)}+a^2\omega^2$.
Let us define the inner product on $S^2_{\tau,r}$,
\begin{equation}
  \label{eq:inprod}
  \left(f,g\right):=\int_{0}^{\pi} \int_{0}^{2\pi} d\Theta \,d\Phi\, \sqrt{\hat\sigma}\, f^\star g 
\end{equation}
for any two functions $f,g$ on $S^2_{\tau,r}$. 
Consider $(S^{(\omega)}_{lm},[P_{\mu}^{(\omega)}+a^2\omega^2]S^{(\omega)}_{lm})$ for an arbitrary spheroidal harmonic $S^{(\omega)}_{lm}$. Since the mass term of $P_{\mu}^{(\omega)}$ is non-negative, we have
\begin{equation}
  \big(S^{(\omega)}_{lm},\left[P_{\mu}^{(\omega)}+a^2\omega^2\right]S^{(\omega)}_{lm}\big) \geq \big(S^{(\omega)}_{lm},\left[P^{(\omega)}+a^2\omega^2\right]S^{(\omega)}_{lm}\big)\,.
\end{equation}
Using $S^{(\omega)}_{lm}=s^{(\omega)}_{lm} e^{im\Phi}$, we can write
\begin{align}
\big(P^{(\omega)}+a^2\omega^2\big)S^{(\omega)}_{lm}=\big(\tilde P^{(\omega)} +P_c^{(\omega)}\big)S^{(\omega)}_{lm},
\end{align}
where in the last equality we have defined
\begin{align}
 \tilde P^{(\omega)} S^{(\omega)}_{lm}&:=-\frac{1}{\sin\Theta}\partial_\Theta\left(\Delta_\Theta\sin\Theta\partial_\Theta S^{(\omega)}_{lm}\right)-\frac{\Xi^2}{\Delta_\Theta \sin^2\Theta}\partial_\Phi^2 S^{(\omega)}_{lm} \,,\\
 P_c^{(\omega)}&:= \frac{1}{\Delta_\Theta}\left(a \omega \sin\Theta-m\Xi a^2\frac{\cos^2\Theta}{\sin\Theta}\right)^2+\Xi^2 m^2 a^2 \frac{\cos^2\Theta}{\sin^2\Theta}\,. 
\end{align}
Since $P_c^{(\omega)}\geq 0$, we have
\begin{equation}
  \label{eq:Pcineq}
\big(S^{(\omega)}_{lm},\left[P_{\mu}^{(\omega)}+a^2\omega^2\right]S^{(\omega)}_{lm}\big) \geq \big(S^{(\omega)}_{lm},\tilde P^{(\omega)}S^{(\omega)}_{lm}\big).
\end{equation}
The right-hand side is given by 
\begin{align}
 &\big(S^{(\omega)}_{lm},\tilde P^{(\omega)}S^{(\omega)}_{lm}\big)=\nonumber\\
 & -\int_{0}^{\pi} \int_{0}^{2\pi} d\Theta\, d\Phi\Big(\partial_\Theta\big(\Delta_\Theta\sin\Theta\partial_\Theta S^{(\omega)}_{lm}\big)\Big) S^{(\omega)\star}_{lm} 
 -\int_{0}^{\pi} \int_{0}^{2\pi}d\Theta\, d\Phi \frac{\Xi^2}{\Delta_\Theta}\frac{1}{\sin\Theta}\big(\partial_\Phi^2 S^{(\omega)}_{lm}\big) S^{(\omega)\star}_{lm} \,.
\end{align}
Integrating by parts the first term in $d\Theta$ and the second term in $d\Phi$, we obtain
\begin{equation}
  \big(S^{(\omega)}_{lm},\tilde P^{(\omega)}S^{(\omega)}_{lm}\big)=\int_{0}^{\pi} \int_{0}^{2\pi}d\Theta d\Phi \left(\Delta_\Theta \lvert\partial_\Theta S^{(\omega)}_{lm}\rvert^2 +\Xi^2\frac{1}{\sin\Theta}\lvert\partial_\Phi S^{(\omega)}_{lm}\rvert^2\right)\,.
\end{equation}
Since $a\in[0,1)$, $\cos^2\Theta \in [0,1),\sin\Theta\in (0,1]$, we have
\begin{equation}
  \Delta_\Theta=1-a^2\cos^2\Theta\geq 1-a^2 \geq (1-a^2)^2\geq (1-a^2)^2 \sin\Theta =\Xi^2 \sin\Theta\,.
\end{equation}
Hence, 
\begin{equation}
\begin{split}
  \big(S^{(\omega)}_{lm},\tilde P^{(\omega)}S^{(\omega)}_{lm}\big)&\geq \Xi^2 \int_{0}^{\pi} \int_{0}^{2\pi} d\Theta\, d\Phi\left( \lvert\partial_\Theta S^{(\omega)}_{lm}\rvert^2 +\frac{1}{\sin^2\Theta}\lvert\partial_\Phi S^{(\omega)}_{lm}\rvert^2\right) \sin\Theta \\
  &\hspace{-1cm}=\Xi^2 \int_{0}^{\pi} \int_{0}^{2\pi}d\Theta\, d\Phi \left[ -\frac{1}{\sin\Theta} \partial_\Theta \left( \sin\Theta \partial_\Theta S^{(\omega)}_{lm}\right) -\frac{1}{\sin^2\Theta}\partial_\Phi ^2 S^{(\omega)}_{lm}\right] S^{(\omega)\star}_{lm} \sin\Theta \\
  &\hspace{-1cm}=\Xi^2 \big(S^{(\omega)}_{lm},\Delta S^{(\omega)}_{lm}\big)\,,
\end{split}
\end{equation}
where the second line is obtained from the first line by integrating by parts the first term in $d\Theta$ and the second term in $d\Phi$, and the third line is obtained by simply using the definitions of inner product and Laplacian $\Delta$.
Combined with \eqref{eq:Pcineq}, this gives
\begin{equation}
  \big(S^{(\omega)}_{lm},\left[P_{\mu}^{(\omega)}+a^2\omega^2\right]S^{(\omega)}_{lm}\big) \geq \big(S^{(\omega)}_{lm},\Xi^2 \Delta S^{(\omega)}_{lm}\big)\,.
\end{equation} 
Since the spheroidal harmonics $S^{(\omega)}_{lm}$ form a basis for functions on $S^2_{\tau,r}$, the last inequality implies that the operator $P_{\mu}^{(\omega)}+a^2\omega^2-\Xi^2 \Delta$ is positive semi-definite.
Therefore, its eigenvalues, $\lambda_{lm}^{(\omega)}+a^2\omega^2-\Xi^2 l(l+1)$, must be non-negative.
This proves \eqref{eq:estlambda}.

\section{Approximate effective potential for Kerr-AdS}
\label{sec:Kerr_effective_potential}

The qualitative picture of the potential barrier for the entire range of sub-extremal Kerr-AdS black holes displayed in Section~\ref{subsec:potbarkerrads} is obtained by considering the approximate version of the eigenvalue problem, as we explain in this appendix.\footnote{Rigorous proofs of properties of the potential are provided in \cite{Holzegel:2011uu} under certain restrictions on $a$ and $\mu$, and in \cite{Dias:2012tq} using the WKB approximation for large $l$.}

To begin, we restrict the original problem of a scalar wave travelling in the entire black hole spacetime to a region outside of the black hole, thus turning it into a ``wave in a box'' problem that can only admit real eigenvalues $\omega$.
For certain Kerr-AdS parameters $(r_+,a)$, $(l,m)$ and $\omega\in \mathbb{R}$, the potential $V_{lm}^{(\omega)}$ has a local maximum $V_{lm,\text{max}}^{(\omega)}$ at $r=r^{(\omega)}_{lm,\text{max}}$ and a local minimum $V_{lm,\text{min}}^{(\omega)}$ at $r=r^{(\omega)}_{lm,\text{min}}$, with $r_+<r^{(\omega)}_{lm,\text{max}}<r^{(\omega)}_{lm,\text{min}}$.
Considering only modes $(l,m,\omega)$ with $V_{lm,\text{min}}^{(\omega)}\leq \omega^2<V_{lm,\text{max}}^{(\omega)}$, i.e., modes that ``feel'' a potential barrier, the boundary conditions that we impose in this ``wave in a box'' problem are the usual reflective condition at the AdS boundary, $u^{(\omega)}_{lm}\to 0$ at $r\to+\infty$, and a Dirichlet boundary condition, $u^{(\omega)}_{lm}=0$ at $r=r^{(\omega)}_{lm,\text{max}}$. With these boundary conditions, the problem only admits a discrete spectrum of real eigenvalues $\omega$ (see Proposition 4 in \cite{Holzegel:2012wt}) and hence the solutions can neither grow nor decay in time.

Clearly, the approximate problem is significantly different from the original one. Nevertheless, for large $l$, the resulting real frequencies are close enough to the complex quasi-normal frequencies so that one can plug these values of $\omega$ in $V_{lm}^{(\omega)}$ and obtain a rather accurate picture of the potential.
To understand this, we note that solutions to the axisymmetric (i.e., $m=0$) ``wave in a box'' problem have been used to construct the so-called quasimodes, which are smooth functions for all $r>r_+$ that solve the Klein-Gordon equation on Kerr-AdS up to an error that decays exponentially in $l$ \cite{Holzegel:2013kna}.\footnote{In fact, real quasimode frequencies have been shown to approximate complex quasi-normal frequencies for a number of eigenvalue problems, such as those studied in \cite{cmp/1104286118,Tang1998FromQT,10.1215/S0012-7094-99-09903-9,Gannot:2012pb,Gannot_2015}.}
Although this construction has not been explicitly carried out yet for modes with $m\neq 0$, it should also work; in particular, for modes with $l=m$. Indeed, in the large $l$ limit, \cite{Dias:2012tq} shows that the $l=m$ case is the one for which the potential barrier is the highest and the widest, making this case the closest one to the simplified ``wave in a box'' problem.
We note that approximating the quasi-normal frequencies with the real frequencies of the ``wave in a box'' problem is expected to be less accurate at small $l$. 

Next, we treat the discrete spectrum of real frequencies $\omega$ of the approximate problem as continuous. 
The motivation for this is two-fold. 
First, it is straightforward to verify that the shape of the potential and its qualitative features are not sensitive to small changes in $\omega$ at fixed $r_+,a,l,m$. 
Second, the results of \cite{Holzegel:2013kna} show that the discrete real frequencies are close to each other for sufficiently large $l$ and $m=0$.
Our approximation is, in essence, an extension of this property of the spectrum to all values of $l,m$. Finally, for simplicity, we do not use the exact values of $\lambda_{lm}^{(\omega)}$ (which can be computed numerically as in \cite{Uchikata:2009zz} and \cite{Cardoso:2013pza}). 
Instead, we use the inequality \eqref{eq:estlambda} to obtain an analytic lower bound on $V_{lm}^{(\omega)}$:
\begin{equation}
  V_{lm}^{(\omega)}\geq \tilde{V}_{lm}^{(\omega)}:=V_{+,lm}^{(\omega)}+\tilde V_{0,lm}^{(\omega)}
\end{equation}
with
\begin{equation}
  \tilde V_{0,lm}^{(\omega)}:=\frac{\Delta\, \Xi^2 \,l(l+1)-\Xi^2\, a^2\, m^2-2\, m\,\omega \,a\, \Xi\left(\Delta-\left(r^2+a^2\right)\right)}{\left(r^2+a^2\right)^2}\,.
\end{equation}

Summarizing, we can get a reasonably accurate description of the full potential $V_{l=m}^{(\omega)}$ by considering the much simpler quantity $\tilde{V}_{l=m}^{(\omega)}$ for real, continuous values of $\omega$.
Although this approximation might seem very crude and its validity is not justified for small values of $l$, the resulting picture is consistent with the qualitative behaviour of modes in the linear and fully non-linear simulations that we have performed in Section~\ref{sec:res}.

\section{Convergence tests}
\label{sec:conv}

In this appendix we present some numerical tests to assess the accuracy and convergence of our simulations. The code that we have used is essentially the same as in \cite{Bantilan:2020xas}, and more thorough convergence tests can be found in this reference. 

\begin{figure}[t!]
\vspace{0.5cm}
    \centering
    \includegraphics[scale=0.7]{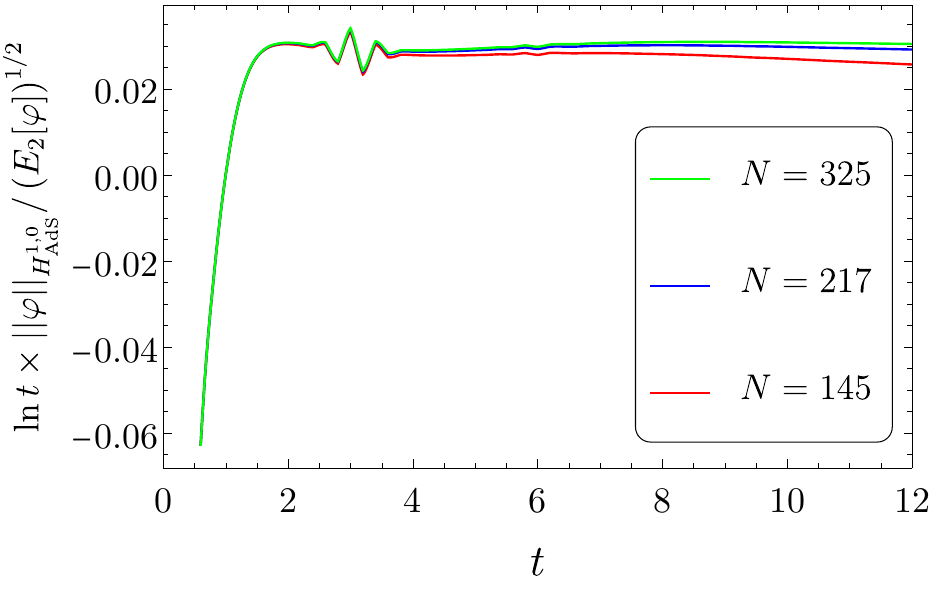}
	\caption{Inverse logarithmic decay of $||\varphi||_{H^{1,0}_{\text{AdS}}}$ for unboosted initial data for three different resolutions.}
  \label{fig:convergence_logdecay}
\end{figure}

\begin{figure}[t!]
    \centering
    \includegraphics[scale=0.45]{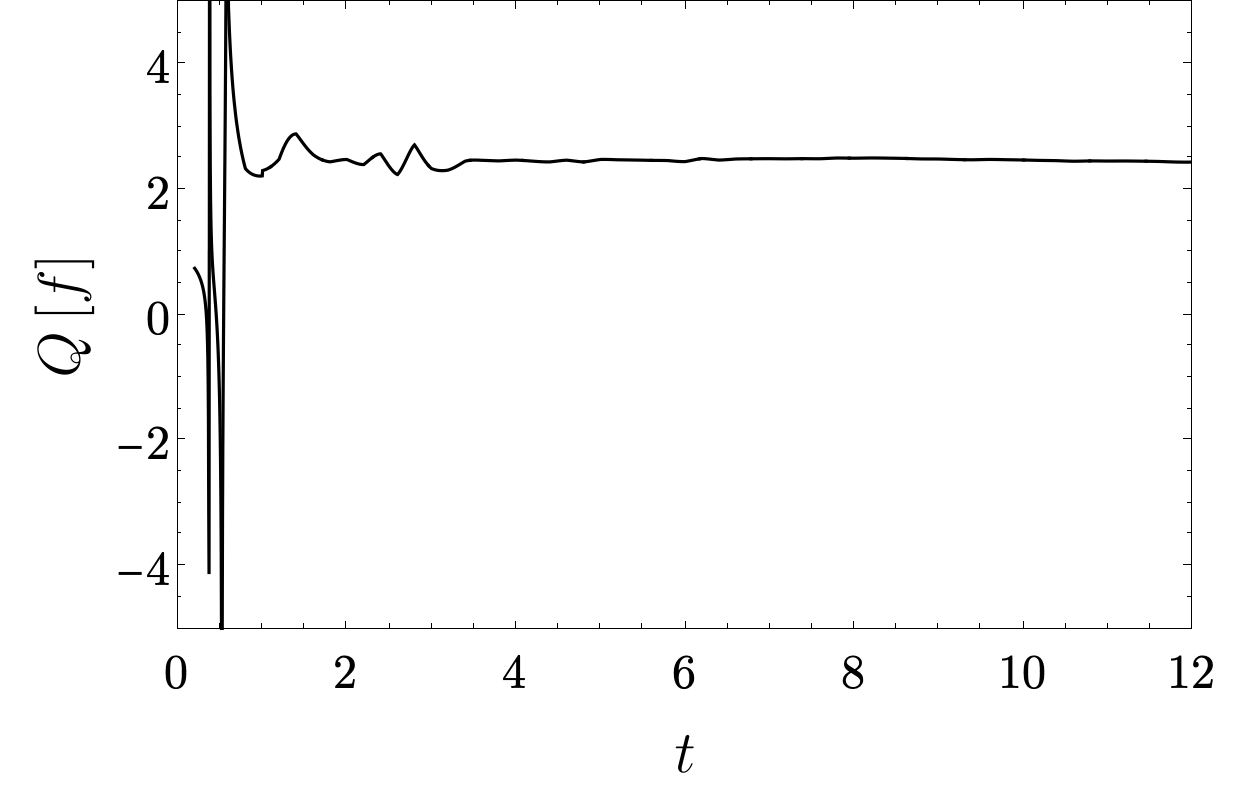}
	\caption{Convergence factor obtained from the data in Fig.~\ref{fig:convergence_logdecay}.}
	\label{fig:convegence_factor}
\end{figure}

First we test convergence; we focus on the case without backreaction because of the prohibitive cost of the simulations in the fully non-linear case. We consider Gaussian initial data (without boost) for the scalar field, with the same parameters as in the main text, on a Kerr-AdS background with $M=0.4$ and $a=0.2$. We evolve this initial data for different resolutions and Fig.~\ref{fig:convergence_logdecay} shows the logarithmic decay in each case. Using these simulations, we compute the convergence factor, 
\begin{equation}
  Q\left[f\right]:=\frac{1}{\ln (3/2)} \ln\left(\left|\frac{f_{N=145}-f_{N=217}}{f_{N=217}-f_{N=325}}\right|\right)
\end{equation}
as a function of time, where $f=\ln t \times ||\varphi||_{H^{1,0}_{\text{AdS}}}/(E_2[\varphi])^{\frac{1}{2}}$ and $f_N$ denotes the numerical values of $f$ in the simulation with $N$ points along each direction of the Cartesian grid. The result is shown in Fig.~\ref{fig:convegence_factor}. Given the discretization scheme used in our code, one should expect second order convergence, and hence $Q\left[f\right]=2$. Fig.~\ref{fig:convegence_factor} confirms that, after an initial rapid period of adjustment, our code converges at the expected order throughout the rest of the simulation.

\begin{figure}[t!]
    \centering
    \includegraphics[scale=0.6]{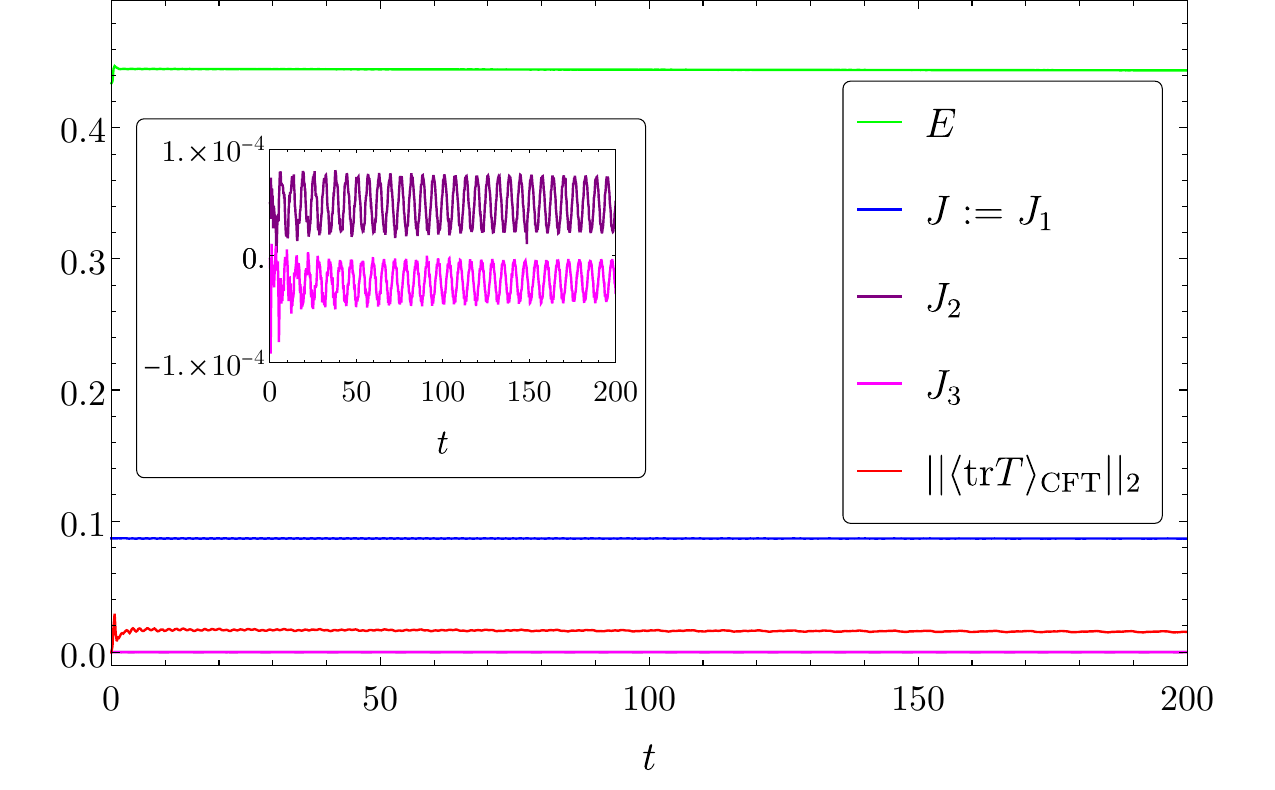}
	\caption{Evolution of the conserved charges and the $L^2$ norm of the trace of the boundary energy-momentum tensor of the dual CFT.}
    \label{fig:cons_charges}
\end{figure}

A non-trivial check of the accuracy of our numerical solution to the non-linear problem is to verify that the charges of Appendix~\ref{sec:CFTquants}, conserved for an exact solution of the equations of motion~\cite{Fischetti:2012rd}, are truly conserved in practice. In Fig.~\ref{fig:cons_charges} we display the total energy $E$ and angular momentum $J$ as functions of time. This figure shows that these quantities do stay constant throughout the evolution after a brief initial period of gauge adjustment, during which the constraint violations are exponentially damped. We also display the evolution of the other components of the angular momentum, which should be zero in the continuum limit; in our simulations they are $O(10^{-4})$ at all times. Finally, we show the evolution of the $L^2$ norm of the trace of the boundary energy-momentum tensor, $||\langle \text{tr}T\rangle_{CFT}||_2$, which should also be zero in the continuum limit; this quantity is of a few percent at all times. We point out that the local values of $|\langle \text{tr}T\rangle_{CFT}|$ are $O(10^{-4})$, so much smaller than the $L^2$ norm of this quantity. These tests confirm the correctness of our simulations.

\bibliographystyle{JHEP}
\bibliography{main}

\end{document}